\documentclass[letterpaper, secnumarabic, amssymb, floatfix, superscriptaddress, aps, prl, longbibliography, nobibnotes, reprint, 10pt]{revtex4-2}

\usepackage{etoolbox}
\usepackage{amsmath}
\usepackage{amsfonts}
\usepackage{graphicx}
\usepackage{braket}
\usepackage{booktabs}
\usepackage{bbm}
\usepackage{float}
\usepackage{nicefrac}
\usepackage{bm}
\usepackage{hhline}
\usepackage{multirow}
\usepackage{lipsum}
\usepackage[percent]{overpic}
\usepackage{subcaption}
\usepackage[dvipsnames]{xcolor}
\usepackage{xcite}
\usepackage{xr}

\usepackage{silence}
\usepackage[hidelinks]{hyperref}

\newcommand{\angstrom}{\mbox{\normalfont\AA}}

\makeatletter
\renewcommand{\@makecaption}[2]{%
  \vskip\abovecaptionskip
  \small
  \parindent 0pt
  \parfillskip=0pt plus 1fil   
  \setlength{\leftskip}{0pt}
  \setlength{\rightskip}{0pt}
  {#1.} #2\par          
  \vskip\belowcaptionskip
}
\makeatother

\begin{document}

\title{Ultrafast Spontaneous Exciton Dissociation via Phonon Emission in \texorpdfstring{BiVO$_4$}{BiVO4}}

\author{Stephen E. Gant}%
\affiliation{Department of Physics, University of California Berkeley, Berkeley, California 94720, USA}

\author{Antonios M. Alvertis}%
\affiliation{KBR, Inc., NASA Ames Research Center, Moffett Field, California 94035, United States}

\author{Christopher J. N. Coveney}
\affiliation{Department of Physics, University of Oxford, Oxford OX1 3PJ, United Kingdom}

\author{Jonah B. Haber}%
\affiliation{Materials Science and Engineering, Stanford University, Stanford, CA 94305, USA}

\author{Marina R. Filip}%
\affiliation{ Department of Physics, University of Oxford, Oxford OX1 3PJ, United Kingdom}

\author{Jeffrey B. Neaton}%
\affiliation{Department of Physics, University of California Berkeley, Berkeley, California 94720, USA}
\affiliation{Molecular Foundry, Lawrence Berkeley National Laboratory, Berkeley, California 94720, USA}
\affiliation{Kavli Energy NanoScience Institute at Berkeley, Berkeley, California 94720, USA}

\begin{abstract}
Monoclinic bismuth vanadate (m-BiVO\textsubscript{4}) is a promising indirect band gap semiconductor for photoelectrochemical water splitting, yet the characteristics of its low-lying photoexcitations, or excitons, remain poorly understood.
Here, we use an \textit{ab initio} Bethe-Salpeter equation approach that incorporates phonon screening to compute the nature and lifetimes of the low-lying excitons of m-BiVO\textsubscript{4}.
Our calculations indicate that at $0$~K, the lowest-lying exciton energy exceeds the indirect band gap, enabling spontaneous dissociation into free carriers via phonon emission within picoseconds.
At $300$~K, both phonon emission and absorption effects reduce this timescale to only a few femtoseconds. 
Phonon screening also greatly reduces the binding energy of the lowest-lying exciton, leading to an optical absorption spectrum that better reproduces experimental measurements.
Overall, our findings establish the general conditions under which phonon emission-driven exciton dissociation can occur in indirect gap semiconductors, and they emphasize the critical role phonon screening can play in predictive calculations of photophysical properties of complex materials.
\end{abstract}

\maketitle
Monoclinic bismuth vanadate (m-BiVO\textsubscript{4}) is a prominent candidate photocatalyst for solar water splitting applications due to its visible-range light absorption, favorable band edge energies versus water reduction and oxidation potentials, and its stability under operating conditions \cite{park_progress_2013, hisatomi_recent_2014, sivula_semiconducting_2016, wang_crystal_2019, gaikwad_emerging_2022, yi_bivo4_2024, arunachalam_surface_2024, wang_bivo4_2025}.
Notably, m-BiVO\textsubscript{4}-based devices have achieved some of the highest measured solar-to-hydrogen conversion efficiencies~\cite{pan_boosting_2018,huang_317_2021}.
A wealth of calculations have been performed to understand the electronic structure and optical properties of m-BiVO\textsubscript{4}.
These theoretical investigations have computed its band structure\cite{walsh_band_2009, zhao_electronic_2011, yang_theoretical_2013, ding_comparative_2014, kim_simultaneous_2015, wiktor_comprehensive_2017}, its optical absorption spectrum  \cite{zhao_electronic_2011, das_investigation_2017, wiktor_comprehensive_2017, steinitz-eliyahu_mixed_2022, ohad_optical_2023}, its defects \cite{newhouse_combinatorial_2018, osterbacka_influence_2021, osterbacka_charge_2022, osterbacka_spontaneous_2024} and polarons \cite{kweon_electron_2014, wiktor_role_2018, cooper_physical_2018, wiktor_electron_2019, ambrosio_strong_2019, liu_hole_2020, moslinger_competing_2025}, the magnitude of phonon contributions to electronic states \cite{wiktor_comprehensive_2017}, and its properties at surfaces and interfaces \cite{crespo-otero_variation_2015, ambrosio_phdependent_2018, ambrosio_phdependent_2018a, steinitz-eliyahu_mixed_2022, qi_unraveling_2022}.

Nonetheless, the nature of the photoexcitations, or excitons, of m-BiVO\textsubscript{4} and their dissociation into free carriers---an important energy transduction step \cite{nelson_physics_2003, chen_photoelectrochemical_2013, hisatomi_recent_2014, yi_bivo4_2024}---remains less well understood.
In particular, prior calculations that neglect the effects of lattice vibrations, or phonons, predict a strong in-gap excitonic peak \cite{das_investigation_2017, wiktor_comprehensive_2017, steinitz-eliyahu_mixed_2022, ohad_optical_2023}  that is inconsistent with measured absorption spectra at room temperature \cite{cooper_indirect_2015}; these calculations also predict a relatively large exciton binding energy of $\mathord{\sim}0.1$~eV \cite{wiktor_comprehensive_2017,ohad_optical_2023}, a value that in principle could hinder the rapid dissociation of excitons into free carriers. 
However, the large reported effects of electron-phonon coupling on the band gap of m-BiVO\textsubscript{4} \cite{wiktor_comprehensive_2017}, as well as its large static dielectric constant \cite{zhou_phase_2011, zhou_phase_2011a, gu_new_2015}, suggest that phonons may play a critical role in screening excitons.

In this Letter, we compute the properties of low-lying excitons in m-BiVO\textsubscript{4} at finite temperatures using a state-of-the-art \textit{ab initio} Bethe-Salpeter equation (BSE) approach that includes dynamical screening due to phonons at lowest order in the electron-phonon interaction~\cite{filip_phonon_2021, alvertis_phonon_2024, coveney_rearrangement_2024, lee_phonon_2024}.
We find that phonon screening significantly reduces the predicted binding energies of the lowest-lying zero center-of-mass momentum exciton compared to the clamped-ion limit.
Furthermore, our calculations predict a large linewidth at room temperature associated with the ultrafast dissociation of this exciton into free electron and hole carriers via phonon emission and absorption.
We demonstrate that the indirect band gap of m-BiVO\textsubscript{4} enables exciton dissociation via phonon emission---a spontaneous process that can occur even at zero temperature.
Finally, our calculations incorporating phonon effects bring the room-temperature linear absorption spectrum into better agreement with experimental measurements \cite{cooper_indirect_2015}.
Our findings reveal that phonon emission-driven exciton dissociation can enable the rapid photogeneration of free carriers even at low temperatures, with broad implications for other indirect band gap semiconductors exhibiting strong electron-phonon coupling.

In the standard \textit{ab initio} $GW$-BSE approach \cite{hybertsen_electron_1986, benedict_optical_1998, albrecht_initio_1998, rohlfing_electronhole_1998, rohlfing_electronhole_2000, onida_electronic_2002}, the BSE can be written in reciprocal space within the Tamm-Dancoff approximation \cite{onida_electronic_2002} as
\begin{equation}
\label{eq:bse_def}
    \Delta_{\rm c\mathbf{k}v\mathbf{k}} A^S_{\rm cv\mathbf{k}}
    +\sum_{\rm c'v'\bm{k}'}K_{\rm cv\mathbf{k}, c'v'\mathbf{k}'}(\Omega)A^S_{\rm c'v'\mathbf{k}'}
    =\Omega^S A^S_{\rm cv\mathbf{k}},
\end{equation}
where $\Delta_{\rm c\bm{k}v'\mathbf{k}'}\,{=}\,E_{\rm c\mathbf{k}}\,{-}\,E_{\rm v'\mathbf{k}'}$ are the single-particle energy differences (usually computed in the $GW$ approximation \cite{hedin_new_1965, hybertsen_firstprinciples_1985, hybertsen_electron_1986}) between occupied ($\rm v$) and unoccupied ($\rm c$) states at the reciprocal lattice points $\rm \mathbf{k}$ and $\rm \mathbf{k}'$ respectively.
$K_{\rm cv\mathbf{k},c'v'\mathbf{k}'}(\Omega)\,{=}\braket{{\rm cv\mathbf{k}}|K(\Omega)|{\rm c'v'\mathbf{k}'}}$ is the dynamical two-particle interaction kernel in the electron-hole basis, and $\Omega^S$ and $A^S_{\rm cv\mathbf{k}}$ are the eigenvalue and eigenvector solutions with corresponding index $S$. 
In Eq.~\ref{eq:bse_def}, we restrict our focus to zero center-of-mass momentum (${\rm\mathbf{Q}\,{=}\,0}$) excitons, as they are the most relevant photoexcitations for optical absorption.
Furthermore, we neglect spin-orbit coupling effects, noting they are minimal at the band edges in m-BiVO\textsubscript{4} (see SM S1.1.2 \cite{sm_see_}).
For singlets, the interaction kernel $K$ in the standard BSE includes bare exchange $K^{\rm X}$ and screened direct interaction $K^{\rm D}$ terms \cite{rohlfing_electronhole_2000,onida_electronic_2002}.

In this work, we include the additional screening term $K^{\rm ph}$ in the BSE kernel;
it encodes, to lowest order in the electron-phonon interaction, the phonon-mediated screened interaction $W^{\rm ph}$ between electron-hole pairs \cite{baym_fieldtheoretic_1961, hedin_effects_1970, filip_phonon_2021, alvertis_phonon_2024}.
$K^{\rm ph}$ depends on both phonon occupations and temperature, and it can be expressed in the clamped-ion exciton basis ($A^S_{\rm cv\mathbf{k}}$) for the emission ($-$) and absorption ($+$) channels as \cite{alvertis_phonon_2024}
\begin{equation}
\label{eq:Kph_S}
\begin{aligned}
    &K^{\rm ph}_{SS'}(\Omega,T)=\kern-0.4em\sum_{\substack{\rm \pm \nu cv\mathbf{k}\\\rm c'v'\mathbf{k}'}}\kern-0.4em
    {A^S_{\rm cv\mathbf{k}}}^*A^{S'}_{\rm c'v'\mathbf{k}'}
    g_{\rm c\mathbf{k}c'\mathbf{k}'\nu}
    g^*_{\rm v\mathbf{k}v'\mathbf{k}'\nu}\mathcal{N}_B^\pm\\
    &\;\;\;\;
    \left[
    \frac{1}{\Omega-\Delta_{\rm c\mathbf{k}v'\mathbf{k}'}\pm\omega_{\rm \mathbf{k}-\mathbf{k}',\nu}+i\eta}
    +\left(\rm{cv\mathbf{k}\leftrightarrow c'v'\mathbf{k}'}\right)
    \right].
\end{aligned}
\end{equation}

In Eq.~\ref{eq:Kph_S}, the second term is the same as the first but with the permutation of the $\rm{cv\mathbf{k}}$ indices with their primes.
$\omega_{\rm \mathbf{k}-\mathbf{k}',\nu}$ is a phonon frequency for the mode $\rm \nu$ and momentum $\rm \mathbf{k}\,{-}\,\mathbf{k}'$, the electron-phonon coupling matrix element associated with this phonon is $g_{\rm n\mathbf{k}m\mathbf{k}'\nu}\,{=}\braket{{\rm n\mathbf{k}}|g_{\rm \mathbf{k}-\mathbf{k}',\nu}|{\rm m\mathbf{k}'}}$  \cite{giustino_electronphonon_2017}, $\mathcal{N}_B^\pm\,{=}\,{\pm}\,\frac{1}{2}\,{-}\,\frac{1}{2}\,{-}\,n_B$ is the effective Bose-Einstein occupation factor where $n_B$ is the standard Bose occupation of a phonon of energy $\omega$ at temperature $T$, and $\eta$ is a small positive number needed for numerical convergence
(SM S1.4.3 and S2 \cite{sm_see_}).
In Eq.~\ref{eq:Kph_S}, we have omitted the thermal occupation terms of the electron-hole states since, for the temperatures considered in this work, they are much smaller than those of the phonons \cite{alvertis_phonon_2024}.

Unlike for $K^{\rm D}$, dynamical effects for $K^{\rm ph}$ are non-negligible because the energy scales of phonons and exciton binding energies are comparable \cite{rohlfing_electronhole_2000, onida_electronic_2002, filip_phonon_2021}; this has two important implications.
First, it becomes possible for the imaginary part of $K^{\rm ph}$ to be non-zero, indicating finite exciton lifetimes due to dissociation into free electrons and holes \cite{coveney_rearrangement_2024}. 
Second, the real part of $K^{\rm ph}$ can vary significantly with respect to the exciton energy $\Omega$, and the phonon-screened exciton energy must then be solved for self-consistently. 

Except for the evaluation of the optical absorption spectra, we retain only the diagonal components of $K^{\rm ph}_{SS'}$ when evaluating the change to the exciton energy;
this approximation is discussed and justified in Appendix~\ref{sec:N_S_diag} and Ref.~\cite{alvertis_phonon_2024}.
The self-consistent equation for the exciton energy in the diagonal approximation is then
\begin{equation}
\label{eq:omega_til_def}
    \tilde{\Omega}^S(T)=\Omega_0^S+K^{\rm ph}_{S,S}(\text{Re}\{\tilde{\Omega}^S\},T),
\end{equation}
where $\Omega_0^S$ is the clamped-ion exciton energy.  
Appendix~\ref{sec:N_S_diag} shows diagonalizing the BSE with non-zero off-diagonal elements in $K^{\rm ph}_{SS'}$ for the first $70$ lowest energy $S$ states affects results by only a few meV.

As discussed in Refs.~\cite{alvertis_phonon_2024, coveney_rearrangement_2024}, both the real and imaginary parts of $\tilde{\Omega}^S(T)$ have physical significance. The real part corresponds to the exciton energy including phonon screening, while the imaginary part is related to the linewidth $\gamma^S$ associated with the exciton dissociating into a free electron-hole pair via the emission ($-$) or absorption ($+$) of a phonon according to the Fermi's Golden rule-like expression:
\begin{equation}
\label{eq:lifetime_def}
\begin{aligned}
    &\gamma^{S}(T)=2\pi\kern-0.4em\sum_{\substack{\rm \pm \nu cv\mathbf{k}\\\rm c'v'\mathbf{k}'}}\kern-0.4em
    {A^{S}_{\rm cv\mathbf{k}}}^* A^{S}_{\rm c'v'\mathbf{k}'}
    g_{\rm c\mathbf{k}c'\mathbf{k}'\nu}
    g^*_{\rm v\mathbf{k}v'\mathbf{k}'\nu}
    \left|\mathcal{N}_B^\pm\right|\\
    &\;\;
    \left[
    \delta\left(\tilde{\Omega}^S-\Delta_{\rm c\mathbf{k}v'\mathbf{k}'}\pm\omega_{\rm \mathbf{k}-\mathbf{k}',\nu}\right)
    +\left(\rm{cv\mathbf{k}\leftrightarrow c'v'\mathbf{k}'}\right)
    \right].
\end{aligned}
\end{equation}
The dissociation lifetime is then $\tau^S\,{=}\,(\gamma^S)^{-1}$.

\begin{figure}[htbp!]
    \centering
    \includegraphics[width=\linewidth]{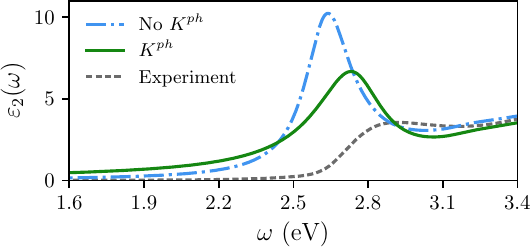}
    \\[-0.2cm]
    \caption{Computed optical absorption spectra with and without $K^{\rm ph}$ with experiment~\cite{cooper_indirect_2015} as reference. Corrections from $K^{\rm ph}$ reduce the first exciton peak height and blue-shift it, leading to an onset with closer agreement to experiment.}
    \label{fig:spectra}
\end{figure}

Following Refs.~\cite{filip_phonon_2021, coveney_rearrangement_2024,alvertis_phonon_2024}, Eqs.~\ref{eq:Kph_S}, \ref{eq:omega_til_def}, and \ref{eq:lifetime_def} neglect electron-phonon self-energy corrections to the single-particle energies (Fan-Migdal and Debye-Waller self-energies \cite{giustino_electronphonon_2017}). Here, as summarized in SM S3.2 \cite{sm_see_}, we find that dressing the single-particle eigenvalues used in $K^{\rm ph}$ with electron-phonon self-energy corrections does not significantly alter the real part of $K^{\rm ph}$, and the imaginary part of $K^{\rm ph}$ continues to exhibit ultrafast room-temperature dissociation via phonon emission but at a rate that is $\mathrm{\sim}2\times$ times slower. We also observe that the overall effects of electron-phonon renormalization on the lowest-lying exciton energy are well approximated by incorporation of phonon screening via Eq.~\ref{eq:Kph_S} and a scissor-shift of solutions to Eq.~\ref{eq:omega_til_def} that neglect electron-phonon self-energy corrections to the single-particle energies in its kernel; this approximation is employed in Fig.~\ref{fig:spectra}.
Further justifying this approach, it is discussed in Refs.~\cite{filip_phonon_2021, coveney_rearrangement_2024,alvertis_phonon_2024,mahanti_effective_1972, pollmann_effective_1977} and SM S3.1 \cite{sm_see_} that higher-order electron-phonon effects are expected to cancel the effects of these neglected single particle corrections on the exciton binding energy.

Both phonon emission- and absorption-driven processes contribute to the imaginary part of $K^{\rm ph}$ provided that energy and momentum conservation are satisfied.
While dissociation via phonon absorption requires a non-zero phonon population that occurs at finite temperatures, dissociation via phonon emission can occur at zero as well as finite temperatures.
In direct gap materials, absorption-driven dissociation of an exciton with energy $\Omega^S$ is energetically forbidden unless $\omega\geq E_{\rm g}^{\text{dir}}-\Omega^S$, where $E_{\rm g}^{\text{dir}}$ is the lowest energy direct single-particle gap and $\omega$ is the frequency of a phonon.
In indirect gap materials, this constraint is $\omega\geq E_{\rm g}^{\text{ind}}-\Omega^S$, where $E_{\rm g}^{\text{ind}}$ is the lowest indirect gap.
For phonon emission-driven dissociation in direct gap systems, the condition $E_{\rm g}^{\text{dir}}\leq\Omega^S$ must hold, typically prohibiting this channel for the lowest bound exciton (but allowing it for resonant excitons).
For indirect gap systems, the constraint for dissociation via phonon emission is $E_{\rm g}^{\text{ind}}\leq\Omega^S$. We refer to Appendix~\ref{sec:diss_cond} and SM S4 \cite{sm_see_} for more details on this.

In m-BiVO\textsubscript{4}, the conditions for both emission and absorption driven exciton dissociation are met for the lowest-lying exciton after solving Eq.~\ref{eq:omega_til_def}.
While the constraint for emission-driven exciton dissociation can be satisfied in other indirect gap semiconductors, the density of final free electron and hole states into which an exciton can scatter into is not guaranteed to be large.
However, because m-BiVO\textsubscript{4} is an indirect gap semiconductor with relatively flat bands that have lower/higher-energy conduction/valence states to scatter into that are close to the direct gap, a large rate of phonon-emission driven dissociation is possible.
Moreover, because this process in indirect gap systems scatters the electron and hole to valence and conduction band extrema away from the lowest direct gap, the associated matrix element of this process can be small because $A_{\rm cv\mathbf{k}}$ is peaked around the $\rm \mathbf{k}$ points near the direct gap.


\begin{figure}[htbp!]
    \centering
    \includegraphics[width=\linewidth]{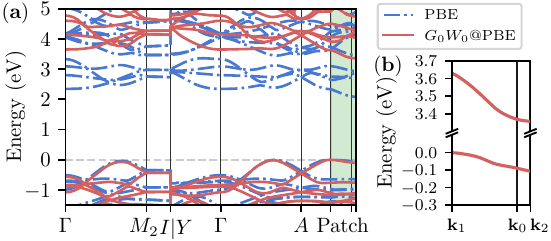}
    \\[-0.2cm]
    \caption{Electronic band structure of m-BiVO\textsubscript{4}. (\textbf{a}) PBE and $G_0W_0$@PBE bands along high symmetry paths as well as within the patch in which the BSE is solved. (\textbf{b}) The same path as in the green box in (a) shows the valence band maximum, direct gap, and the conduction band minimum. In crystal coordinates, the labeled points are $\rm \mathbf{k}_1=[\nicefrac{1}{24},\nicefrac{-1}{24},\nicefrac{9}{32}]$, $\rm \mathbf{k}_0=[\nicefrac{1}{6},\nicefrac{-1}{6},\nicefrac{1}{8}]$, and $\rm \mathbf{k}_2=[\nicefrac{1}{6},\nicefrac{-5}{24},\nicefrac{3}{32}]$.}
    \label{fig:combined_bands}
\end{figure}

To investigate this, we perform \textit{ab initio} BSE calculations with $K^{\rm ph}$ for m-BiVO\textsubscript{4}, evaluating temperature-dependent dissociation rates and renormalized exciton binding energies.
Our \textit{ab initio} calculation of the BSE with phonon screening begins with density functional theory (DFT) calculations using the PBE exchange-correlation functional \cite{perdew_generalized_1996} to generate the Kohn-Sham eigensystem, and we then use density functional perturbation theory (DFPT) to calculate the phonon eigensystem and electron-phonon coupling constants.
These quantities are subsequently used as input to the $GW$ and BSE calculations in this work (SM S1 \cite{sm_see_}).

The quasiparticle band structures computed with DFT-PBE and $G_0W_0$@PBE are shown in Fig.~\ref{fig:combined_bands} and are consistent with prior work \cite{wiktor_comprehensive_2017, ohad_optical_2023}.
Notably, our $G_0W_0$@PBE direct band gap is $3.45$~eV, a significant increase over the DFT-PBE result of $2.13$~eV but considerably larger than the measured direct gap of $2.7$~eV \cite{cooper_indirect_2015}.
This difference can be rationalized by a large renormalization of the electronic band structure via electron-phonon interactions \cite{wiktor_comprehensive_2017}.
The band gap is indirect, and interestingly, the valence band maximum, conduction band minimum, nor the smallest direct gap are located along the high-symmetry paths shown in Fig.~\ref{fig:combined_bands}(a).
Instead, the smallest direct gap lies at a lower symmetry point near $\Gamma$ and $A$ in the irreducible wedge of Brillouin zone (BZ).
Near this point there are also higher- and lower-energy valence and conduction band states which give a $G_0W_0$@PBE indirect band gap of $3.35$~eV; Fig.~\ref{fig:combined_bands}(b) contains the path which shows both of these direct and indirect gaps.
These values are close to but slightly smaller than eigenvalues along the high symmetry paths. Our results indicate a direct-indirect gap difference of $0.10$~eV, in agreement with the measured direct-indirect gap difference of $\mathord{\sim} 0.1$-$0.2$~eV \cite{cooper_indirect_2015}.
We also find this trend holds when using LDA \cite{kohn_selfconsistent_1965} as a starting point (SM S1.5 \cite{sm_see_}).

Next, we solve the \textit{ab initio} BSE in a small volume, or \textit{patch}, in the BZ on a dense $\rm \mathbf{k}$-grid centered at $\rm \mathbf{k}_0$, the $\rm \mathbf{k}$-point with the smallest $G_0W_0$@PBE direct gap \cite{alvertis_importance_2023} (Fig.~\ref{fig:combined_bands}(b)), and where the lowest-lying exciton wavefunction is centered in reciprocal space.
Initially neglecting $K^{\rm ph}$, i.e. the clamped-ion limit, we find the lowest optical excitation energy $\Omega^S_0$ is $3.35$~eV, implying $E_{\rm g}^d-\Omega^S_0\,{=}\,0.10$~eV, a value consistent with prior work \cite{wiktor_comprehensive_2017, ohad_optical_2023} and the same as our $G_0W_0$@PBE direct-indirect gap difference.
In this work, we define the exciton binding energy as
$E_B^S\,{=}\,\sum_{\rm cv\mathbf{k}}\left|A^S_{\rm cv\mathbf{k}}\right|^2\left(E_{\rm c\mathbf{k}}-E_{\rm v\mathbf{k}}\right)-\Omega^S$,
following e.g.~\cite{champagne_strongly_2024}, where, instead of referencing the lowest direct gap, the reference is a weighted average of the energy of each direct transition contributing to the exciton wavefunction.
This definition yields a larger binding energy of $150$~meV, and is expected to be more representative of the strength of the electron-hole interaction.
We emphasize that this definition of $E_B$ has no effect on our computed linewidths.

\begin{figure}[htbp!]
    \centering
    \includegraphics[width=\linewidth]{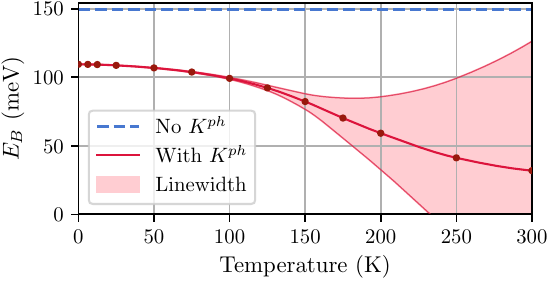}
    \\[-0.2cm]
    \caption{Computed binding energies $E_B$ of the first exciton vs. temperature. In blue is the ion-clamped \textit{ab intio} BSE value. The results from solving Eq.~\ref{eq:omega_til_def} are plotted in red; the linewidth from the imaginary part of $\tilde{\Omega}^S$ is overlaid as well.}
    \label{fig:ebe}
\end{figure}

Then, we consider the effects of phonon screening by including $K^{\rm ph}$. Our calculation of the orientationally averaged static dielectric constant $\epsilon_0$ yields a large phonon contribution, with $\epsilon_\infty\,{=}\,7.2$ and $\epsilon_0\,{=}\,77$. Following Refs.~\cite{filip_phonon_2021, alvertis_phonon_2024}, the large lattice contribution, combined with the clamped-ion exciton binding energy and typical zone-center phonon frequencies in range of $\mathrm{\sim}10-100$~meV, suggests a large binding energy renormalization due to phonon screening.
Indeed, after re-evaluating the BSE to include $K^{\rm ph}$ via Eqs.~\ref{eq:Kph_S} and \ref{eq:omega_til_def}, we obtain an exciton energy of $\mathrm{Re}\{\tilde{\Omega}^S\}\,{=}\, 3.31$~eV at zero temperature, corresponding to a $0.04$~eV (or a $27\%$) reduction in binding energy versus the clamped-ion result.
Fig.~\ref{fig:ebe} shows that this phonon-induced screening becomes especially pronounced at higher temperatures.
Notably, at room temperature, $\mathrm{Re}\{\tilde{\Omega}^S\}$ increases by $117$~meV relative to $\Omega^S_0$, indicating a substantial $\mathord{\sim}80\%$ reduction of the exciton binding energy, and consistent with polar phonons possessing energies low enough to be activated with temperatures on the order of $k_{B}T\,{\sim}\,20$~meV.
This large reduction also implies the exciton energy now lies in the continuum above the direct gap, opening up the possibility of direct gap ($|{\rm\mathbf{q}}|\sim 0$) emission-driven dissociation at these temperatures as shown in SM S5 \cite{sm_see_}.

The large reduction in $E_B$ via phonon screening predicted here can be rationalized by considering the relatively low frequencies of many of the $15$ zone-center polar phonons, all of which are in the range of $8$-$107$~meV.
In particular, the lower frequency phonon modes contribute comparably with the highest LO mode to the overall magnitude of the real part of $K^{\rm ph}$ (SM S6 \cite{sm_see_}).
The large impact of phonon screening on the exciton binding energy is consistent with prior work demonstrating large electron-phonon coupling and the concomitant significant reduction of the fundamental band gap \cite{wiktor_comprehensive_2017}. We also compute the electron-phonon renormalization of the fundamental band gap to be similarly large in SM S3.2 \cite{sm_see_}.


\begin{figure}[htbp]
    \centering
    \includegraphics[width=\linewidth]{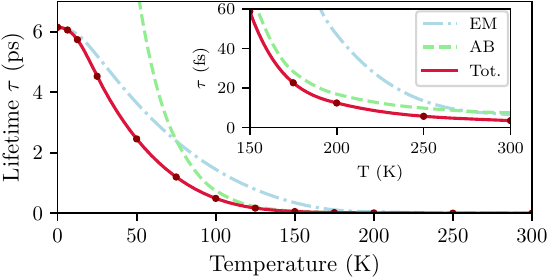}
    \\[-0.2cm]
    \caption{Computed lifetimes for the first exciton vs. temperature. At $0$~K, the lifetime is entirely emission-driven, and at finite temperatures the absorption channel becomes active.}
    \label{fig:lifetimes}
\end{figure}

Fig.~\ref{fig:ebe} shows the results of our \textit{ab initio} calculations of the linewidth corresponding to phonon-driven exciton dissociation into free electrons and holes for the first exciton; it increases significantly with temperature, indicating a rapid decrease in the exciton lifetime. This trend is further illustrated in Fig.~\ref{fig:lifetimes} which plots the lifetime of the lowest-lying exciton as a function of temperature. Remarkably, even at zero temperature, the exciton has a finite and relatively short lifetime of $6$~ps, a consequence of the favorable conditions for phonon emission-driven exciton dissociation in m-BiVO\textsubscript{4}. This is in contrast to direct gap semiconductors, where non-resonant excitons can only dissociate at finite temperatures via phonon absorption.

With increasing temperature, the dissociation lifetime decreases dramatically due to contributions from both phonon emission and absorption processes. By $100$~K, it falls below $1$~ps, and beyond $200$~K it reaches the single-digit femtosecond regime with a computed room temperature value of $3$~fs. This is markedly smaller than measured and computed dissociation lifetimes for other semiconductors such as GaN \cite{alvertis_phonon_2024}, CH$_3$NH$_3$PbI$_3$ \cite{jha_direct_2018}, or GaSe \cite{allerbeck_probing_2021}, which range between $\mathord{\sim}50$ and $\mathord{\sim}100$~fs. 

For comparison, we estimate the timescales of two other major exciton scattering processes that can compete with exciton dissociation in m-BiVO\textsubscript{4}: the bulk radiative lifetime due to electron-hole recombination and phonon-driven scattering between one bound exciton state and another.
Following Ref.~\cite{coveney_rearrangement_2024}, we estimate the scattering rate of the lowest lying exciton to other higher lying exciton states via phonon absorption, and obtain a timescale at $300$~K of $1$~ps, a value that is comparable to, if not somewhat larger than, exciton-exciton scattering lifetimes computed for other systems in literature \cite{amit_ultrafast_2023,chan_exciton_2023,cohen_phonondriven_2024,coveney_rearrangement_2024}.
Following Ref.~\cite{chen_initio_2019}, we also estimate the radiative recombination lifetime of m-BiVO\textsubscript{4} at $300$~K to be $1$~ns, a timescale that is much greater than both exciton scattering and dissociation and in line with previously reported measurements in bulk semiconductors at room temperature \cite{im_radiative_1997, sun_observation_2014}.
Appendix~\ref{sec:decay_processes} contains the details of these calculations.
Taken together, these results suggest that despite clamped-ion calculations suggesting m-BiVO\textsubscript{4} has a strongly bound in-gap exciton, once phonon screening is taken into account, this binding energy becomes much smaller. Moreover, it strongly suggests that the dissociation of photo-generated excitons into free electrons and holes in m-BiVO\textsubscript{4} is a rapid and dominant process at and below room temperatures.

The effects of electron-phonon interactions in m-BiVO\textsubscript{4} have a direct impact on its absorption spectrum.
To obtain the phonon-screened line-shape of the first bright exciton---which dominates the spectrum onset in the clamped-ion limit---we solve Eq.~\ref{eq:S_basis_eig}, including the off-diagonal elements of $K^{\rm ph}_{SS'}$ for the lowest 70 clamped-ion $S$ states.
Appendix~\ref{sec:abs_spec} discusses the details of how this modifies the spectrum.
In short, the complex eigenvalues as well as the left and right eigenvectors from diagonalizing the Hamiltonian in Eq.~\ref{eq:S_basis_eig} result in an asymmetric Lorentzian line-shape, a result originally established by Toyozawa~\cite{toyozawa_theory_1958,toyozawa_interband_1964}.
This new line-shape, computed in the patch, is then applied to the first bright peak in the full-grid clamped-ion spectrum.
Additionally, each BSE eigenvalue $\tilde{\Omega}^{\tilde{S}}$ is red-shifted by $\Delta_{\text{scissor}}\,{=}\,0.633$~eV; this correction combines the finite-temperature electron-phonon correction to the direct gap with an adjustment for the $G_{0}W_{0}$@PBE starting-point error.
In Fig.~\ref{fig:spectra} we compare the directionally averaged optical absorption spectrum at 300~K computed with and without phonon screening; for reference we also include the experimentally measured spectrum~\cite{cooper_indirect_2015}.
Relative to clamped-ion calculations, $K^{\rm ph}$ blue-shifts the first exciton peak and reduces its height by mixing its wavefunction with other states and by increasing its broadening.
Overall, these changes result in a spectrum with an exciton peak more similar to experiment, though further broadening from additional scattering channels would likely bring the spectrum into even closer accord with experiment.

In summary, in this work we compute from first principles the effects of phonon screening on the lowest-lying excitons in the monoclinic phase of BiVO\textsubscript{4}. 
We find the binding energy of the lowest lying exciton is reduced by $27\%$ at zero temperature and nearly $80\%$ at room temperature.
We also predict exciton dissociation into free carriers on ultrafast timescales. At zero temperature, the lifetime due solely to phonon emission is on the order of ps, while at room temperature the lifetime shortens to just a few fs.
These findings significantly reduce the discrepancy between experiment and previous clamped-ion calculations, which predict a relatively large exciton binding energy.
This rapid dissociation is estimated to outpace other bulk scattering and recombination channels, facilitating efficient free-carrier photogeneration in m-BiVO\textsubscript{4}.
Additionally, once phonon screening is accounted for, our calculated optical absorption spectrum shows improved agreement with experiment.
We expect ultrafast low-temperature exciton dissociation via phonon emission can be a general phenomenon in indirect gap materials with strong electron-phonon coupling especially when the valence and/or conduction band extrema lie near the direct gap in the Brillouin zone.
Overall, these findings underscore the key role phonon screening can play in accurate and predictive calculations of the optical properties of complex semiconductors with strong electron-phonon coupling.

\begin{acknowledgments}
We thank Zhenglu Li and Sijia Ke for insightful discussions. This work was primarily supported by the Theory of Materials program at Lawrence Berkeley National Laboratory, funded by the U.S. Department of Energy, Office of Science, Basic Energy Sciences, Materials Sciences and Engineering Division, under Contract No. DE-AC02-05CH11231. J.B.N. acknowledges partial support for discussions about solar fuels materials from the Liquid Sunlight Alliance, a DOE Energy Innovation Hub, supported by the U.S. Department of Energy, Office of Science, Office of Basic Energy Sciences, under Award Number DE-SC0021266. S.E.G. was partially supported by the Kavli Energy NanoScience Institute. A.M.A. is supported by the U.S.
Department of Energy, Office of Science, National
Quantum Information Science Research Centers, Co-
design Center for Quantum Advantage (C$^2$QA). C.J.N.C. and M. R. F acknowledge funding from the UK Engineering and Physical Sciences Research Council (UK-EPSRC). Computational resources were provided by the Texas Advanced Computing Center (TACC) using the Frontera system through the allocation DMR23008 as well as by the National Energy Research Scientific Computing Center (NERSC), DOE Office of Science User Facilities supported by the Office of Science of the US Department of Energy under Contract DE-AC02-05CH11231.
\end{acknowledgments}


\setcounter{secnumdepth}{2}
\appendix
\section{Phonon Emission-Driven Dissociation}
\label{sec:diss_cond}
\begin{figure}[htbp!]
    \centering
    \includegraphics[width=\linewidth]{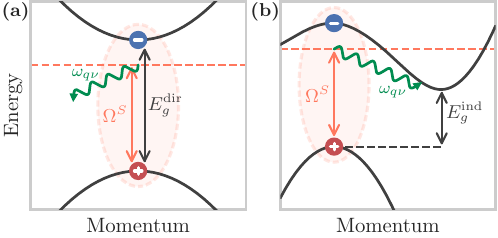}
    \caption{Dissociation via phonon emission in \textbf{(a)}--direct and \textbf{(b)}--indirect band gap systems.}
    \label{fig:indirect_dissociation}
\end{figure}

The basic conditions leading to a finite linewidth in Eq.~\ref{eq:lifetime_def} are determined by considering the arguments of the delta functions in the same expression.
Focusing on the emission channel $(-)$ of the first delta function, it must hold that $\omega_{\rm\mathbf{{k}}-\rm\mathbf{{k}}',\nu}
\,{=}\,\tilde{\Omega}^S-(E_{\rm{c}\rm\mathbf{{k}}}-E_{\rm{v}'\rm\mathbf{{k}}'})$.
Because $\omega_{\rm\mathbf{{k}}-\rm\mathbf{{k}}',\nu}$ is bounded from below by $0$ and from above by $\omega_{LO}^{\text max}$, there can only be emission from this channel if $\omega_{LO}^{\text max}\geq(E_{\rm{c}\rm\mathbf{{k}}}-E_{\rm{v}'\rm\mathbf{{k}}'})-\tilde{\Omega}^S\geq0$.
In the case of a direct gap system, unless $\tilde{\Omega}^S$ lies above the direct gap, $\tilde{\Omega}^S-(E_{\rm{c}\rm\mathbf{{k}}}-E_{\rm{v}'\rm\mathbf{{k}}'})$ will always be negative, meaning emission cannot occur as shown in Fig.~\ref{fig:indirect_dissociation}--(a).

If the band gap is indirect, however, phonon emission-driven dissociation becomes possible even for states where $\tilde{\Omega}^S$ is below the direct gap since $\tilde{\Omega}^S-(E_{\rm{c}\rm\mathbf{{k}}}-E_{\rm{v}'\rm\mathbf{{k}}'})$ can be positive as shown in Fig.~\ref{fig:indirect_dissociation}--(b).
The minimal value of this expression is $\tilde{\Omega}^S-E_g^{\text{ind}}$, where $E_g^{\text{ind}}$ is the indirect band gap, so as long as $\tilde{\Omega}^S\geq E_g^{\text{ind}}$, dissociation via phonon emission is allowed.
When the difference between $\tilde{\Omega}^S$ and $E_g^{\text{ind}}$ is small, only low-energy phonons can be emitted, but as the difference grows, the allowed phonon energies do as well.

It is worth noting that the above constraint is necessary but not sufficient for emission-driven dissociation. Phonon dispersions are not flat and they do not necessarily span the entire energy range between $0$ and $\omega_{LO}^{\text max}$. Ultimately a careful consideration of both energy and momentum conservation for each state must be applied, motivating the utility of \textit{ab initio} calculations. We also note that the same logic as above can be applied to analyze the other delta function in Eq.~\ref{eq:lifetime_def}, and the same constraints emerge.

\section{\texorpdfstring{$S$}{S} Basis Diagonal Approximation}
\label{sec:N_S_diag}
\begin{figure}[htbp!]
    \centering
    \includegraphics[width=\linewidth]{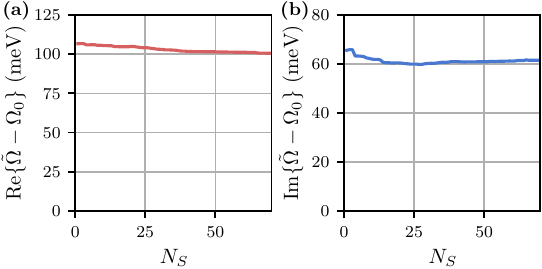}
    \caption{Convergence of the real--\textbf{(a)} and imaginary--\textbf{(b)} parts of the renormalization of the exciton energy vs the size of the non-diagonal $S$ basis ($N_S$) used to solve Eq.~\ref{eq:S_basis_eig} at $300$~K.}
    \label{fig:S_basis_conv}
\end{figure}
The resulting BSE with phonon screening can be expressed in the clamped-ion exciton basis as
\begin{equation}
    \label{eq:S_basis_eig}
    \sum\nolimits_{S'}\left[\Omega^S\delta_{SS'}+K^{\rm{ph}}_{S,S'}(\text{Re}\{\tilde{\Omega}^{\tilde{S}}\},T)\right]\tilde{A}_{S'}^{\tilde{S}}
    =\tilde{\Omega}^{\tilde{S}}\tilde{A}_{S}^{\tilde{S}}.
\end{equation}
Except when computing the optical absorption spectrum, we use the diagonal approximation to Eq.~\ref{eq:S_basis_eig}.
In this approximation, we set $K^{\rm ph}_{S,S'}\!\bigl(\text{Re}\{\tilde{\Omega}^S\},T\bigr)\,{=}\,0$ for $S\,{\neq}\,S'$, yielding Eq.~\ref{eq:omega_til_def}.
The validity of this approximation has been assessed to be accurate for the first bright in-gap excitons of a number of systems in prior work \cite{alvertis_phonon_2024}.
However, we also demonstrate its validity for BiVO\textsubscript{4} by explicitly solving Eq.~\ref{eq:S_basis_eig} at $T\,{=}\,300$ K for the first $70$ $S$ states in the clamped-ion basis.
Fig.~\ref{fig:S_basis_conv} shows the results of solving this as the number of non-zero non-diagonal S states $N_S$ increases.
The real and imaginary parts of the renormalization of the new first exciton energy vary by $6$~meV and $4$~meV respectively as $N_S$ increases from $1$ to $70$.
SM S7~\cite{sm_see_} also demonstrates that the exciton wavefunction itself for this lowest-lying state converges around $N_S=50$.

\section{Competing Decay Processes}
\label{sec:decay_processes}
For comparison, we also compute the rate of radiative recombination as well as the rate of scattering to other higher lying excitonic states via phonon absorption.
We find both the lifetimes associated with these processes to be significantly greater than that of dissociation at room temperature.
First, we compute the finite temperature radiative recombination lifetime of the lowest exciton following Ref.~\cite{chen_initio_2019} with the expression for the radiative recombination rate (in atomic units)
\begin{equation}
    \label{eq:rad_lifetime}
    \gamma_{\text{rad}}^S=\frac{2^5\sqrt{\pi^{3}\varepsilon_\infty}}{3 V}\frac{{d_S}^2}{N_k}{\frac{(\Omega^S\alpha)^3}{{\left(2Mk_B T\right)}^{3/2}}},
\end{equation}
where $\varepsilon_\infty$ is the clamped-ion macroscopic dielectric constant (computed via DFPT), $V$ is the unit cell volume, ${d_S}^2$ is the optical dipole matrix element (obtained by solving the BSE), $N_{\rm \mathbf{k}}$ is the number of ${\rm \mathbf{k}}$ points used when solving the BSE, $\alpha$ is the fine structure constant, and $M\,{=}\,m_e+m_h$ is the exciton total effective mass (computed in SM S3.1~\cite{sm_see_}).
The lifetime due to radiative recombination is then simply the inverse $\tau_{\text{rad}}^S\,{=}\,\left(\gamma_{\text{rad}}^S\right)^{-1}$.
We note that this expression assumes $\varepsilon_\infty$, $M$, and ${d_S}^2$ to be isotropic, which, strictly speaking, is not the case for BiVO\textsubscript{4}.
However, as an approximation, we directionally average each quantity when using Eq. \ref{eq:rad_lifetime}.
At $300$ K, this approach yields a lifetime of $1$ nanosecond.

Second, we compute the scattering rate of the lowest $\rm{\mathbf{Q}}\,{=}\,0,\; S\,{=}\,1$ exciton into other states using the model discussed at length in Ref.~\cite{coveney_rearrangement_2024}.
Specifically, a generalized Fr{\"o}hlich electron-phonon coupling~\cite{tubman_theory_2025} (assuming isotropic LO-TO splitting) is used, namely
\begin{equation}
    \label{eq:gen_frohlich}
    g^F_{\nu\rm\mathbf{q}}=\frac{i}{|\mathbf{q}|}\sqrt{\frac{2\pi}{N_k V \varepsilon_\infty}\frac{{\omega^\nu_{LO}}^2-{\omega^\nu_{TO}}^2}{\omega^\nu_{LO}}},
\end{equation}
with $\nu$ indexing over each LO-TO splitting of polar phonon modes, to construct the exciton-phonon coupling matrix element for an optically excited (effectively zero momentum) exciton of state $S$~\cite{antonius_theory_2022}
\begin{equation}
    \label{eq:ex-ph_coupling}
    G_{SS'\nu}(\bm{0},{\rm\mathbf{{q}}})=g^F_{\nu\rm\mathbf{q}}\sum\nolimits_{\rm\mathbf{{k}}}{A^{S,\rm{\mathbf{q}}}_{\rm\mathbf{{k}}}}^*\left(A^{S',\rm{\mathbf{0}}}_{\rm\mathbf{{k}}}-A^{S',\rm{\mathbf{0}}}_{\rm{\mathbf{k}}+\rm{\mathbf{q}}}\right).
\end{equation}
Here, $A^{S,\rm{\mathbf{q}}}_{\rm\mathbf{{k}}}$ is the wavefunction of an exciton with center-of-mass momentum $\rm{\mathbf{q}}$ at the reciprocal lattice point $\rm{\mathbf{k}}$. Due to the computational difficulty of calculating the finite center-of-mass momentum exciton wavefunctions needed to construct $G_{SS'\nu}$, we instead use hydrogenic wavefunctions for the $1s$, $2s$, $2p$, and $3s$ excitons, with the effective Bohr radius $a_0^{\text{eff}}$ obtained by fitting the BSE exciton wavefunction for the lowest excitation in SM S3.1 \cite{sm_see_}.
We also assume parabolic exciton dispersions using effective masses obtained from fitting the bands at the direct gap as done in SM S3.1 \cite{sm_see_}.
The scattering rate to other excitons is then calculated according to
\begin{equation}
    \label{eq:ex-ex_scattering_rate}
    \gamma^S_{\text{ex-ex}}=2\pi\sum_{S'\rm\mathbf{{q}}\nu}\left|G_{SS'\nu}(\bm{0},\rm\mathbf{{q}})\right|^2 \delta(\Omega^S-\Omega^{S'}_{\rm\mathbf{{q}}}+\omega_{\rm\mathbf{{q}}\nu})
\end{equation}
with the corresponding lifetime obtained via $\tau_{\text{ex-ex}}^S\,{=}\,\left(\gamma_{\text{ex-ex}}^S\right)^{-1}$.
At $300$ K we calculate this lifetime to be $1$ picosecond.

\section{Absorption Spectrum Renormalization}
\label{sec:abs_spec}
In order to obtain a renormalized optical absorption spectrum, we compute the complex eigenvalues and the left ($\tilde{B}^{\tilde{S}}_S$) and right ($\tilde{A}^{\tilde{S}}_{S}$) eigenvectors from the BSE Hamiltonian in Eq.~\ref{eq:S_basis_eig}, treating $K^{\rm{ph}}_{SS'}$ as non-diagonal for the first $70$ clamped-ion states (see Appendix~\ref{sec:N_S_diag} and SM S7~\cite{sm_see_} for convergence).
Because the Hamiltonian in Eq.~\ref{eq:S_basis_eig} is non-Hermitian, the left and right eigenvectors are not Hermitian conjugates, but they are normalized to satisfy bi-orthonormality ($\sum_S\tilde{B}^{\tilde{S}}_S\,\tilde{A}^{\tilde{S}'}_{S}\,{=}\,\delta_{\tilde{S}\tilde{S}'}$).
The square of the clamped-ion dipole matrix element~\cite{onida_electronic_2002,rohlfing_electronhole_2000} along the polarization direction $i$, $r^i_S\,{=}\,\sum_{\rm cv\mathbf{k}} \left(A_{\rm cv\mathbf{k}}^S\right)^*\braket{\rm{v\mathbf{k}}|r_i|\rm{c\mathbf{k}}}$,
is accordingly transformed~\cite{albrecht_initio_1998, onida_electronic_2002, chan_exciton_2023} to give
\begin{equation}
    D^i_{\tilde{S}}=\sum\nolimits_{SS'}r_S^{i*}\tilde{A}_S^{\tilde{S}}\;\,\tilde{B}^{\tilde{S}}_{S'}\,r_{S'}^i.
\end{equation}
The renormalized optical spectrum with only broadening from $K^{\rm ph}$ dissociation channels is then
\begin{equation}
\label{eq:eps_2_til_def}
\begin{aligned}
    \tilde{\varepsilon}_2^{\,i}(\omega)=8\pi^2&\sum_{\tilde{S}}
    \frac{\text{Re}\{D^i_{\tilde{S}}\}\;\text{Im}\{\tilde{\Omega}^{\tilde{S}}\}}{(\omega-\text{Re}\{\tilde{\Omega}^{\tilde{S}}\})^2+\text{Im}\{\tilde{\Omega}^{\tilde{S}}\}^2}\\
    &\;\,+\frac{\text{Im}\{D^i_{\tilde{S}}\}(\omega-\text{Re}\{\tilde{\Omega}^{\tilde{S}}\})}{(\omega-\text{Re}\{\tilde{\Omega}^{\tilde{S}}\})^2+\text{Im}\{\tilde{\Omega}^{\tilde{S}}\}^2}.
\end{aligned}
\end{equation}
This asymmetric form of the spectrum coming from $\text{Im}\{D^i_{\tilde{S}}\}$ due to exciton-phonon coupling is rarely calculated \textit{ab initio} (to our knowledge only Ref.~\cite{chan_exciton_2023} reports a comparable calculation), but has been long-established by Toyozawa~\cite{toyozawa_theory_1958, toyozawa_interband_1964}.
This definition of $\tilde{\varepsilon}_2(\omega)$ affects the strength of dipole matrix elements, going beyond simply shifting the exciton energies as is done in Ref.~\cite{schebek_phononmediated_2024}.

In practice, we add a uniform background width of $100$~meV to the state-resolved linewidths from $K^{\rm ph}$ in order to account for residual scattering and finite $\rm{\mathbf{k}}$-point sampling.
We also apply an additional scissor shift $\Delta_{\text{scissor}}$ to all $\tilde{\Omega}^{\tilde{S}}$ used to calculate $\tilde{\varepsilon}_2(\omega)$.
This shift contains two corrections.
The first is the renormalization of the lowest direct band gap at $300$ K due to electron-phonon coupling as discussed in SM S3.2 \cite{sm_see_}; this correction is $-0.933$ eV.
The second correction accounts for the band gap starting point error of $G_0W_0$@PBE; as was shown in Ref.~\cite{ohad_optical_2023}, $G_0W_0$ calculation in BiVO$_4$ that use a non-empirical optimally tuned hybrid functional \cite{wing_band_2021} increases the band gap by $0.30$ eV relative to $G_0W_0$@PBE.
Thus, in total, we apply a red-shift of $\Delta_{\text{scissor}}\,{=}\,0.633$ eV to our spectra.

In order to extend the spectrum computed in Eq.~\ref{eq:eps_2_til_def} to account for more than the first $70$ states, we also plot the conventional clamped-ion BSE spectrum,
\begin{equation}
\label{eq:eps_2_conv}
\varepsilon_2^i(\omega)
=8\pi^2\sum\nolimits_{S>70}|r^i_S|^2\delta(\omega-\Omega^S),
\end{equation}
for the higher-lying states. In this case, the delta functions are replaced with Lorentzian envelopes with broadening of $100$~meV as above.

Finally, because the renormalized spectrum computed for a BZ patch cannot be mapped unambiguously onto the full-grid calculation, we confine our analysis of the effects of $K^{\rm ph}$ in Fig.~\ref{fig:spectra} to the first bright exciton that dominates the clamped-ion full-grid spectrum onset. We therefore only use the spectrum from Eq. \ref{eq:eps_2_til_def} for this first bright peak and then use Eq.~\ref{eq:eps_2_conv} for all the rest of the states. Consequently, the redistribution of exciton oscillator strength to other states in the full grid calculation is neglected. However, this redistribution of weight is relatively uniform for states in the patch.

\nocite{hohenberg_inhomogeneous_1964, kohn_selfconsistent_1965, sleight_crystal_1979, perdew_selfinteraction_1981, godby_metalinsulator_1989, baroni_phonons_2001, dressel_electrodynamics_2002, giustino_electronphonon_2007, giannozzi_quantum_2009, deslippe_berkeleygw_2012, larson_role_2013, hamann_optimized_2013, ponce_temperature_2014, mostofi_updated_2014, cooper_electronic_2014, ponce_temperature_2015, damle_compressed_2015, verdi_frohlich_2015, ponce_epw_2016, giannozzi_advanced_2017, vansetten_pseudodojo_2018, giannozzi_uantum_2020, pizzi_wannier90_2020, vitale_automated_2020, wing_band_2021, sagredo_electronic_2024, gant_optimally_2022, ohad_band_2022, dai_theory_2024, dai_excitonic_2024, lihm_selfconsistent_2024, schebek_phononmediated_2024}

\bibliography{main}

\onecolumngrid

\renewcommand{\thefigure}{S\arabic{figure}}
\renewcommand{\theequation}{S\arabic{equation}}
\renewcommand{\thetable}{S\arabic{table}}
\renewcommand{\thesection}{S\arabic{section}}

\setcounter{secnumdepth}{2}
\setcounter{equation}{0}

\newpage

\centerline{\large{\textbf{Supplementary Materials for:
\textit{Ultrafast Spontaneous Exciton Dissociation via}}}}
\vspace{0.1cm}
\centerline{\large{\textbf{\textit{Phonon Emission in \texorpdfstring{BiVO$_4$}{BiVO4}}}}}
\vspace{0.5cm}

\centerline{
  Stephen E. Gant$^1$, 
  Antonios M. Alvertis$^2$, 
  Christopher J. N. Coveney$^3$, 
}
\centerline{
  Jonah B. Haber$^4$, 
  Marina R. Filip$^3$, 
  Jeffrey B. Neaton$^{1,5,6}$%
}
\vspace{0.5cm}
\centerline{$^1$\textit{Department of Physics, University of California Berkeley, Berkeley, California 94720, USA}}
\centerline{$^2$\textit{KBR, Inc., NASA Ames Research Center, Moffett Field, California 94035, United States}}
\centerline{$^3$\textit{Department of Physics, University of Oxford, Oxford OX1 3PJ, United Kingdom}}
\centerline{$^4$\textit{Materials Science and Engineering, Stanford University, Stanford, CA 94305, USA}}
\centerline{$^5$\textit{Molecular Foundry, Lawrence Berkeley National Laboratory, Berkeley, California 94720, USA}}
\centerline{$^6$\textit{Kavli Energy NanoScience Institute at Berkeley, Berkeley, California 94720, USA}}

\tableofcontents

\section{Calculation Details}
The crystal structure \cite{sleight_crystal_1979} and relaxed atomic positions of the $12$ atom unit cell of m-BiVO$_4$ (space group $15$) can be found in \ref{subsubsec:crystal}. The underlying electronic eigensystem used in our calculations is obtained via density functional theory \cite{hohenberg_inhomogeneous_1964, kohn_selfconsistent_1965} using the PBE functional \cite{perdew_generalized_1996} and the \texttt{Quantum Espresso} code package \cite{giannozzi_quantum_2009, giannozzi_advanced_2017, giannozzi_uantum_2020}, while $GW$ and clamped-ion BSE calculations are carried out using the \texttt{BerkeleyGW} software package \cite{hybertsen_firstprinciples_1985,hybertsen_electron_1986,deslippe_berkeleygw_2012}. Phonon dispersions and electron-phonon matrix elements are obtained via DFPT \cite{baroni_phonons_2001} with \texttt{Quantum Espresso} as well. The electron and phonon eigensystems and electron-phonon matrix elements are interpolated onto finer reciprocal space grids using a combination of the \texttt{Wannier90} \cite{mostofi_updated_2014}, \texttt{EPW} \cite{giustino_electronphonon_2007, verdi_frohlich_2015, ponce_epw_2016}, and \texttt{BerkeleyGW} software packages. $K^{\rm{ph}}$ itself is computed using an in-house code library. Due to the computational cost of converging the $\rm\mathbf{{k}}$- and $\rm\mathbf{{q}}$-grids used in solving the BSE both with and without $K^{\rm{ph}}$, a patch centered around the first bright exciton is used. This has proven to be an effective tool for converging the BSE with little in periodic systems \cite{alvertis_importance_2023}.
\subsection{Density Functional Theory}
Calculations were carried out using fully relativistic optimized norm-conserving Vanderbilt pseudopotentials \cite{hamann_optimized_2013} obtained from the pseudodojo library \cite{vansetten_pseudodojo_2018}. The PBE functional was used for all DFT calculations, and wavefunction plane wave cutoff of 104 Ry was employed for all calculations. A half-shifted $3\times3\times4$ $\rm\mathbf{{k}}$-grid was used to converge the ground state density.

\subsubsection{Crystal Structure}
\label{subsubsec:crystal}
\begin{figure}
    \centering
    \begin{subfigure}[]{0.48\linewidth}
        \begin{overpic}[width=\linewidth]{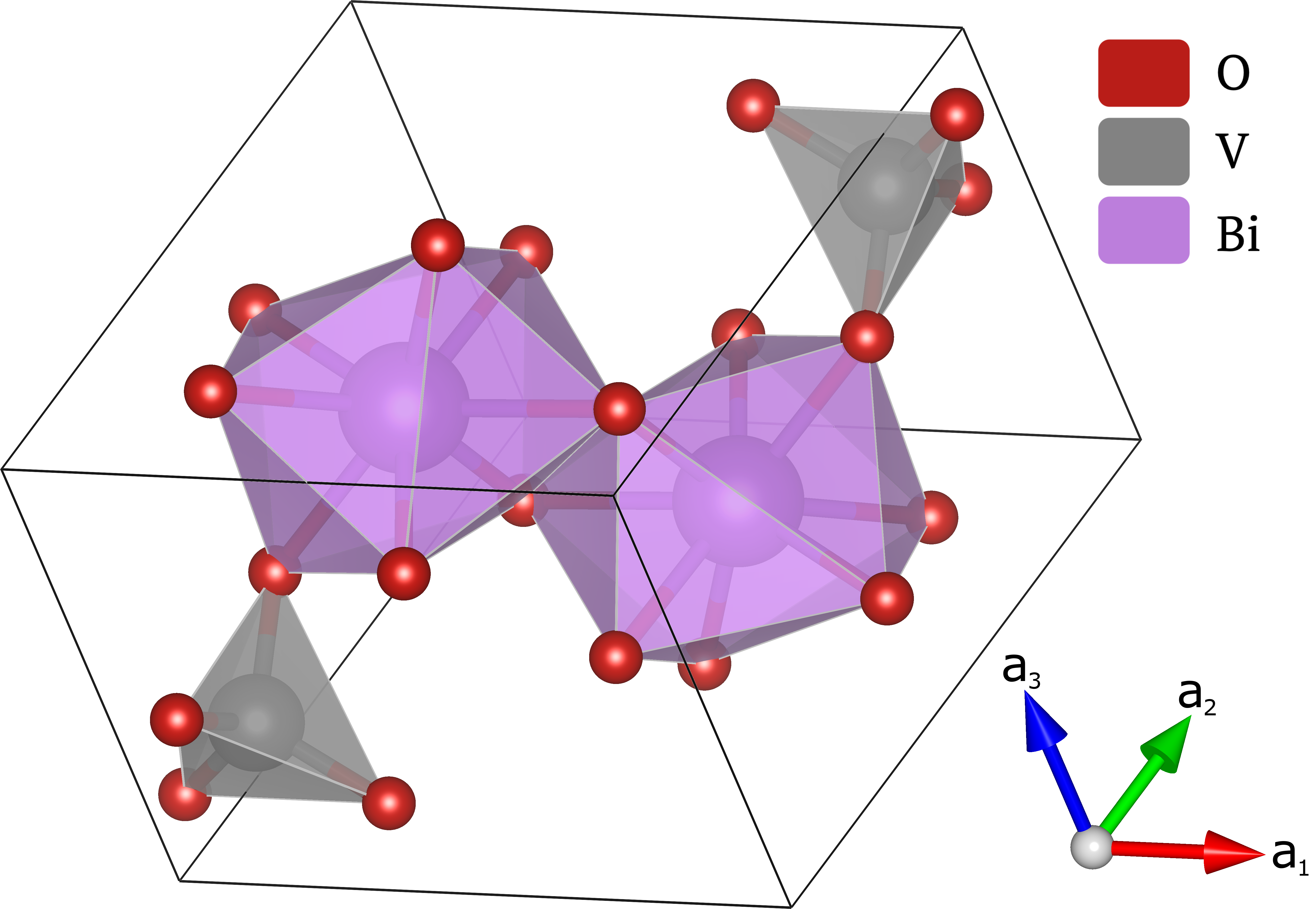}
            \put(-1,70){\small \textbf{(a)}}
        \end{overpic}
    \end{subfigure}
    ~
    \begin{subfigure}[]{0.46\linewidth}
        \begin{overpic}[width=\linewidth]{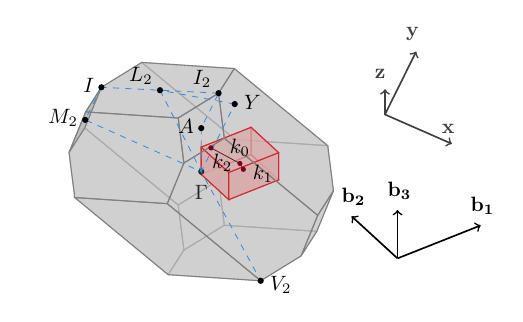}
            \put(-1,66){\small \textbf{(b)}}
        \end{overpic}
    \end{subfigure}
    \\[-0.2cm]
    \caption{a) Unit cell of BiVO$_4$ with lattice vectors $\rm{\mathbf{a}}_1,\rm{\mathbf{a}}_2,\rm{\mathbf{a}}_3$ labeled, and with the atom legend displayed in the top right corner. b) First Brillouin zone of BiVO$_4$ with labeled high symmetry points and reciprocal lattice vectors. The high symmetry paths of the band structure plot in Fig.\ 1(a) of the main text are labeled in blue, and the patch in which the BSE is solved for $K^{\rm{ph}}$ is shown as the red parallelepiped. The in-patch lowest direct gap, conduction band minimum, and valence band maximum are labeled $\rm\mathbf{{k}}_0$, $\rm\mathbf{{k}}_1$, and $\rm\mathbf{{k}}_2$ respectively, and their electronic eigenvalues along this path can be seen in Fig.\ 1(b) of the main text.}
    \label{fig:unit_cell_and_BZ}
\end{figure}

Lattice parameters are fixed in this work and taken from the experimental Ref.~\cite{sleight_crystal_1979} as specified in table \ref{tab:lattice_param}.
\begin{table}[htbp!]
    \centering
    \caption{\centering Lattice vectors in Cartesian coordinates.}
    \vspace{2mm}
    \label{tab:lattice_param}
    \setlength{\tabcolsep}{10pt}
    \begin{tabular}{c|ccc}
    \hline
    \multirow{2}{*}{} & \multicolumn{3}{c}{Cartesian Coordinates ($\angstrom$)} \\
     & x & y & z \\ \hline
    $\rm{\mathbf{a}}_1$ & 3.6235830351 & 5.8485999107 & 0.0000000000 \\
    $\rm{\mathbf{a}}_2$ & -3.6235830351 & 5.8485999107 & 0.0000000000 \\
    $\rm{\mathbf{a}}_3$ & -3.5500108382 & 0.0000000000 & 3.6473943955 \\ \hline
    \end{tabular}
\end{table}
\begin{table}[htbp!]
    \caption{\centering Relaxed (via PBE) atomic positions in crystal coordinates.}
    \vspace{2mm}
    \label{tab:atom_pos}
    \setlength{\tabcolsep}{10pt}
    \begin{tabular}{c|ccc}
    \hline
    \multirow{2}{*}{Atom} & \multicolumn{3}{c}{Position (Crystal Coordinates)} \\
     & $\rm{\mathbf{a}}_1$ & $\rm{\mathbf{a}}_2$ & $\rm{\mathbf{a}}_3$ \\
     \hline
    Bi & 0.627539884 & 0.627539884 & 0.250000000 \\
    Bi & 1.372460116 & 0.372460116 & 0.750000000 \\
    V & 1.125778085 & 0.125778085 & 0.250000000 \\
    V & 0.874221915 & 0.874221915 & 0.750000000 \\
    O & 1.346983574 & 0.062974753 & 0.139765500 \\
    O & 1.653016426 & -0.062974753 & 0.860234500 \\
    O & 1.062974753 & 0.346983574 & 0.360234500 \\
    O & 0.937025247 & 0.653016426 & 0.639765500 \\
    O & 1.706525943 & 0.197757175 & 0.364727435 \\
    O & 1.293474057 & -0.197757175 & 0.635272565 \\
    O & 1.197757175 & 0.706525943 & 0.135272565 \\
    O & 0.802242825 & 0.293474057 & 0.864727435 \\ \hline
    \end{tabular}
\end{table}
The positions of the atoms inside the lattice are relaxed using PBE and are given in table \ref{tab:atom_pos}. The unit cell, its lattice vectors, and the constituent atoms inside are shown in Fig.~\ref{fig:unit_cell_and_BZ}a. Fig.~\ref{fig:unit_cell_and_BZ}b shows the first Brillouin zone (BZ) as well as the reciprocal lattice vectors $\rm{\mathbf{b}}_1,\rm{\mathbf{b}}_2,\rm{\mathbf{b}}_3$. The high symmetry points and the paths between them corresponding to ones taken in the band structure plot of Fig.\ 1(a) in the main text are also shown in blue. Finally, the red parallepiped shows the the location and extent of the patch in which the BSE is solved for $K^{\rm{ph}}$. The additionally labeled $\rm\mathbf{{k}}$ points $\rm\mathbf{{k}}_0,\rm\mathbf{{k}}_1,\rm\mathbf{{k}}_2$ correspond the the lowest optically active direct gap where the patch is centered, the conduction band minimum inside the patch, and the valence band maximum inside the patch respectively (see Fig.\ 1(b) in the main text for their values).

\subsubsection{Spin-Orbit Coupling}
\label{subsubsec:soc}
Though this material contains Bi, which typically exhibits strong spin-orbit coupling (SOC), the orbital character of the conduction and valence bands is dominantly Oxygen $2p$ and Vanadium $3d$ respectively \cite{liu_hole_2020, cooper_electronic_2014}, both of which exhibit significantly less SOC renormalization. Therefore, we neglect SOC for the calculations done in this work. To further test this approximation, we perform calculations at the PBE level on a $12\times12\times16$ $\rm\mathbf{{k}}$-grid and find that with and without SOC turned on, the position of the direct gap where the exciton forms remains stable at its position at $[1/6, -1/6, 1/8]$ (in crystal coordinates). Focusing on a patch centered on this point, and now using a $24\times24\times32$ grid, we also find that upon turning on SOC, the direct gap position still remains fixed at $[1/6, -1/6, 1/8]$, the Brillouin zone locations of the valence band maximum and conduction band minimum do not change, and the direct indirect gap energy difference is reduced by only $2$ meV. Importantly, this means that the favorable energetic conditions which allow for emission-driven exciton dissociation are preserved. The direct gap itself is reduced by $72$ meV with SOC turned on, but given the comparatively small change in the direct-indirect gap difference, we can to a good approximation treat the SOC corrections as a rigid shift and neglect them in our BSE analysis. This does introduce the caveat, however, that the absorption onset will be slightly blue-shifted without SOC effects included.

\subsubsection{Wannierization}
Wannier functions for use in EPW and in the interpolation of the fine grid wavefunctions were obtained using the \texttt{wannier90} code \cite{pizzi_wannier90_2020} with initial projections obtained via the SCDM method \cite{damle_compressed_2015, vitale_automated_2020} using the erfc approach with $\mu_{\text SCDM}=12$ eV and $\sigma_{\text SCDM}=2.5$ eV. A window of $40$ states about the gap was used to form the Wannier functions with a total of $50$ bands being accounted for in the disentanglement procedure. The first 36 eigenstates coming from semi-core electron sates were excluded from the wannierization procedure as they sit sufficiently far from the gap to be neglected in this analysis. A disentanglement frozen window ranging from $13$ to $40$ eV was used to project down the conduction band manifold while preserving the energies of interest around the gap. After 3000 iterations of the minimization routine, sufficient convergence is achieved. The average Wannier function spread is $0.937\; \angstrom^2$, and the minimum and maximum spreads are $0.712\; \angstrom^2$ and $1.990\; \angstrom^2$ respectively.

\subsection{Density Functional Perturbation Theory}
\begin{figure}[htbp!]
    \centering
    \includegraphics[width=0.75\linewidth]{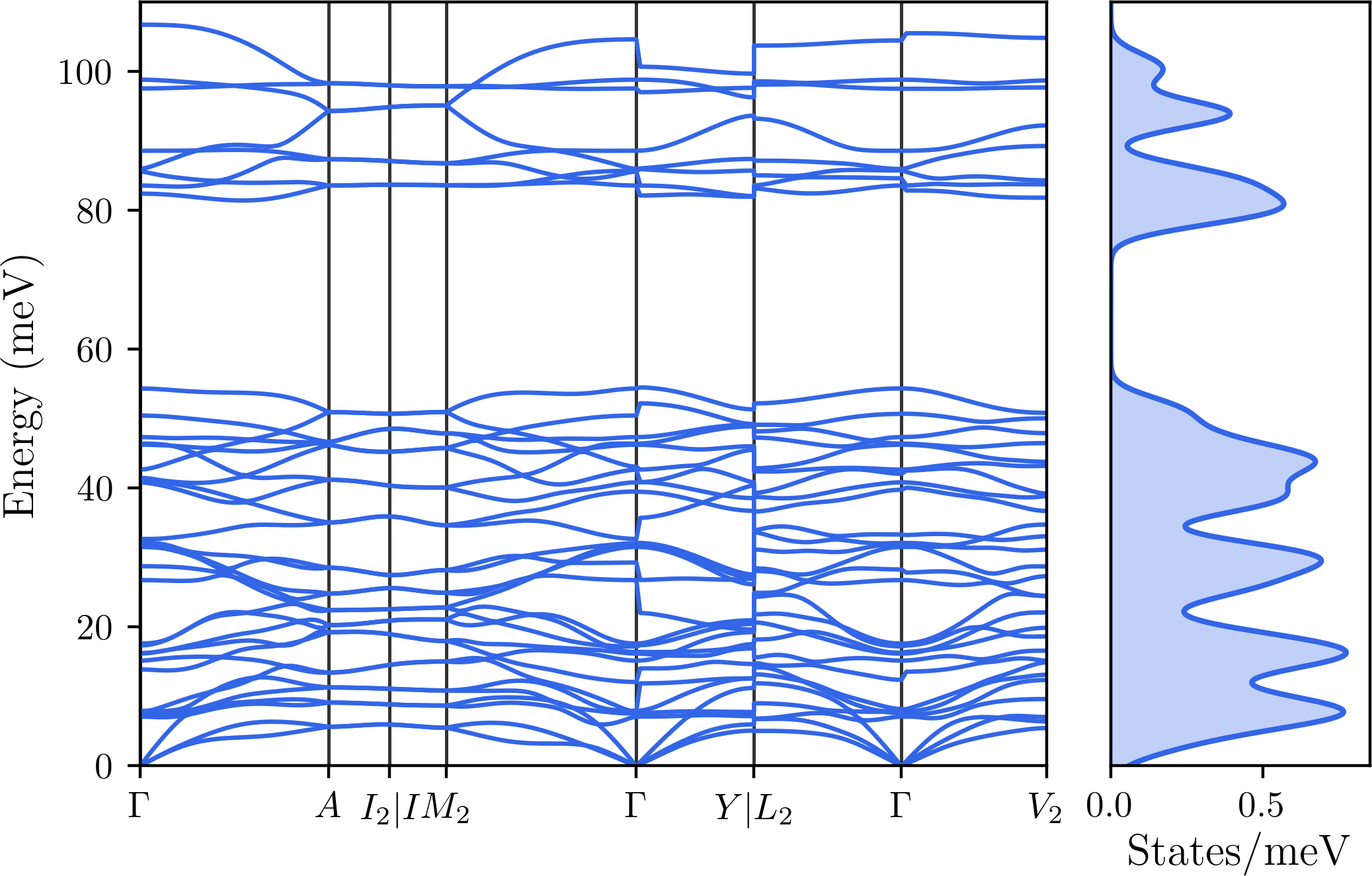}
    \\[-0.2cm]
    \caption{Phonon bands (left) and density of states within the $36\times36\times48$ grid patch using $2$ meV Gaussian smearing (right).}
    \label{fig:phonons}
\end{figure}

Phonon dispersions are computed via density functional perturbation theory in \texttt{Quantum Espresso}. Calculations are carried out on the relaxed structure specified in tables \ref{tab:atom_pos} and \ref{tab:lattice_param} on a $\rm\mathbf{{q}}$-grid of $3\times3\times4$. The resulting phonon band dispersion and density of states are given in Fig.~\ref{fig:phonons}.

Phonon modes and electron-phonon matrix elements are interpolated onto the patch show in Fig.~\ref{fig:unit_cell_and_BZ}b. Specifically, the $\rm\mathbf{{k}}$ points considered in the interpolation are from that patch, while the $\rm\mathbf{{q}}$ points  come from considering the differences between the $\rm\mathbf{{k}}$ points on the patch and exist in a $\Gamma$-centered parallelepiped of the same size as the origin shifted one show in Fig.~\ref{fig:unit_cell_and_BZ}b.

\subsection{\texorpdfstring{\textit{GW}}{GW}-BSE}
Single shot $G_0W_0$ calculations were carried out in the \texttt{BerkeleyGW} code. We use a Godby-Needs generalized plasmon-pole model (GPPM)\cite{godby_metalinsulator_1989} to introduce frequency dependence in the dielectric function, a method that has been shown to accurately reproduce the results of a full frequency calculation \cite{larson_role_2013}. 512 bands are used in the construction of both the dielectric function and the $GW$ self energy, and a cutoff of $30$ Ry was used to truncate the screened coulomb interaction. A half shifted $3\times3\times4$ $\rm\mathbf{{q}}$-grid is used to construct the dielectric function. Convergence data are reported in table \ref{tab:gw_conv}; based on the convergence data in the prior work in Ref.~\cite{ohad_optical_2023}, we estimate that the direct gap at $\Gamma$ is converged to within $\mathord{\sim}30$ meV using these parameters.

\begin{table}[htbp]
    \centering
    \caption{\centering $GW$ calculation convergence data.}
    \vspace{2mm}
    \label{tab:gw_conv}
    \setlength{\tabcolsep}{10pt}
    \begin{tabular}{cccccc}
    \hline
    $N_{\text bands}$ & \begin{tabular}[c]{@{}c@{}}Epsilon\\ Cutoff (Ry)\end{tabular} & GPPM & $\rm\mathbf{{q}}$-grid & \begin{tabular}[c]{@{}c@{}}$\rm\mathbf{{q}}$-shift\\ (crystal)\end{tabular} & \begin{tabular}[c]{@{}c@{}}$E_g^{\Gamma}$ Error\\ (meV)\end{tabular} \\ \hline
    512 & 30 & Godby-Needs & $3\times3\times4$ & $[0.5,0.5,0.5]$ & $+28$ \\ \hline
    \end{tabular}
\end{table}

When solving the BSE in Eq.$\;1$ of the main text, single particle quasiparticle eigenenergies are taken from the $GW$ calculations described in the previous paragraph.
Fine grid wavefunctions for each patch are obtained via a Wannier-based interpolation method in the \texttt{BerkeleyGW} code. We refer to Ref.~\cite{alvertis_phonon_2024} for more details on this.
As described in \ref{subsubsec:soc}, the envelope function for the first bright exciton in reciprocal space is peaked about $k=[1/6,-1/6,1/8]$ in crystal coordinates.
We center all $\rm\mathbf{{k}}$-patches at this location and then build out the patch at the full BZ grid density specified in column 1 of Tab.~\ref{tab:bse_conv}.
The ``mini"-grid inside of this patch is then given in column 2. 

\begin{table}[htbp!]
    \centering
    \caption{BSE convergence data. All patches are centered at $k=[1/6,-1/6,1/8]$ in crystal coordinates and sample at the density in the full Brillouin zone given in the leftmost column.}
    \vspace{2mm}
    \setlength{\tabcolsep}{10pt}
    \begin{tabular}{c|ccc}
    \hline
    BSE Calculation & "Mini" Grid Size & $N_c$ & $N_v$ \\ \hline
    $12\times12\times16$ Patched Grid & $5\times5\times7$ & 2 & 2 \\
    $24\times24\times32$ Patched Grid & $9\times9\times11$ & 2 & 2 \\
    $36\times36\times48$ Patched Grid & $13\times13\times17$ & 2 & 2 \\
    $24\times24\times32$ Full Grid & NA & 4 & 4 \\ \hline
    \end{tabular}
    \label{tab:bse_conv}
\end{table}

We remark here that the size of these patches encompass about $5\%$ of the full BZ, a number which is still small enough to greatly reduce the computational cost of the calculations performed, allowing for greatly improved $\rm\mathbf{{k}}$-space sampling density in the region where the first exciton is located. However, compared to prior recent works which have used this sampling technique, the grids in this work encompass about three times more BZ volume. This is due to two factors. First, the excitons in BiVO$_4$ are not purely Wannier-Mott-like, and are therefore less well localized in reciprocal space. Second, and more importantly, the grid used here needs to encompass the places in the band structure where phonons can scatter the elctron-hole pair. Specifically, as outlined in Appendix A, emission driven processes require a phonon of appreciably finite momentum given by a vector in the BZ which corresponds to an energetically allowed direct-indirect gap transition. In order to encompass the nearest valleys and peaks in the conduction and valence bands respectively where this can occur, the patch used here was further expanded even after converging it to contain the vast majority of the first exciton's wavefunction. If the patches were expanded, more emission channels could be revealed. However, these channels will correspond to larger $\rm\mathbf{{q}}$ phonons, giving them smaller coupling matrix elements. Additionally the exciton wavefunction at either of the $\rm\mathbf{{k}}$ and/or $\rm\mathbf{{k}}'$ points will be farther away from the patch center, further suppressing the size of these channels' contributions to $K^{\rm{ph}}$.

In order to obtain the first bright exciton, only the two highest valence and two lowest conduction bands are used, however, when computing the first few hundred meV of the absorption onset in Fig.\ 4 of the main text, four bands in both the conduction and valence manifolds is are used. For this main text plot, the optical matrix elements are computed along each of the three Cartesian directions using the velocity operator, and the final result averages over these three directions.

\begin{table}[htbp!]
    \centering
    \caption{Band gap data computed via PBE and $G_0W_0$@PBE from within the patch in which the BSE is solved. Calculations are done on a $36\times36\times48$ density grid. $E_g^{\text{dir}}$ is the lowest direct gap, $E_g^{\text{ind}}$ is the lowest indirect band gap within the patch, and $\Delta E_g$ is the direct-indirect gap difference. We note that there exist indirect gaps in the full Brillouin zone which are slightly lower in energy, but these states are farther away in the Brillouin zone from the lowest direct gap, and they are only lower in energy by a few meV.}
    \label{tab:}
    \vspace{2mm}
    \setlength{\tabcolsep}{8pt}
    \begin{tabular}{c|ccc}
     & $E_g^{\text{dir}}$ (eV) & $E_g^{\text{ind}}$ (eV) & $\Delta E_g$ (eV) \\ \hline
    PBE & 2.13 & 2.06 & 0.07 \\
    $G_0W_0$@PBE & 3.45 & 3.35 & 0.10
    \end{tabular}
\end{table}

\subsection{\texorpdfstring{\textit{K\textsuperscript{ph}}}{Kph}}
\subsubsection{\texorpdfstring{$\rm\mathbf{{q}}$}{q}-grid Convergence}
We also check the convergence of the real and imaginary parts of $K^{\rm{ph}}$ with respect to the $\rm\mathbf{{q}}$-grid density.

\begin{figure}[phtb!]
    \centering
    \begin{subfigure}[]{0.485\linewidth}
        \begin{overpic}[width=\linewidth]{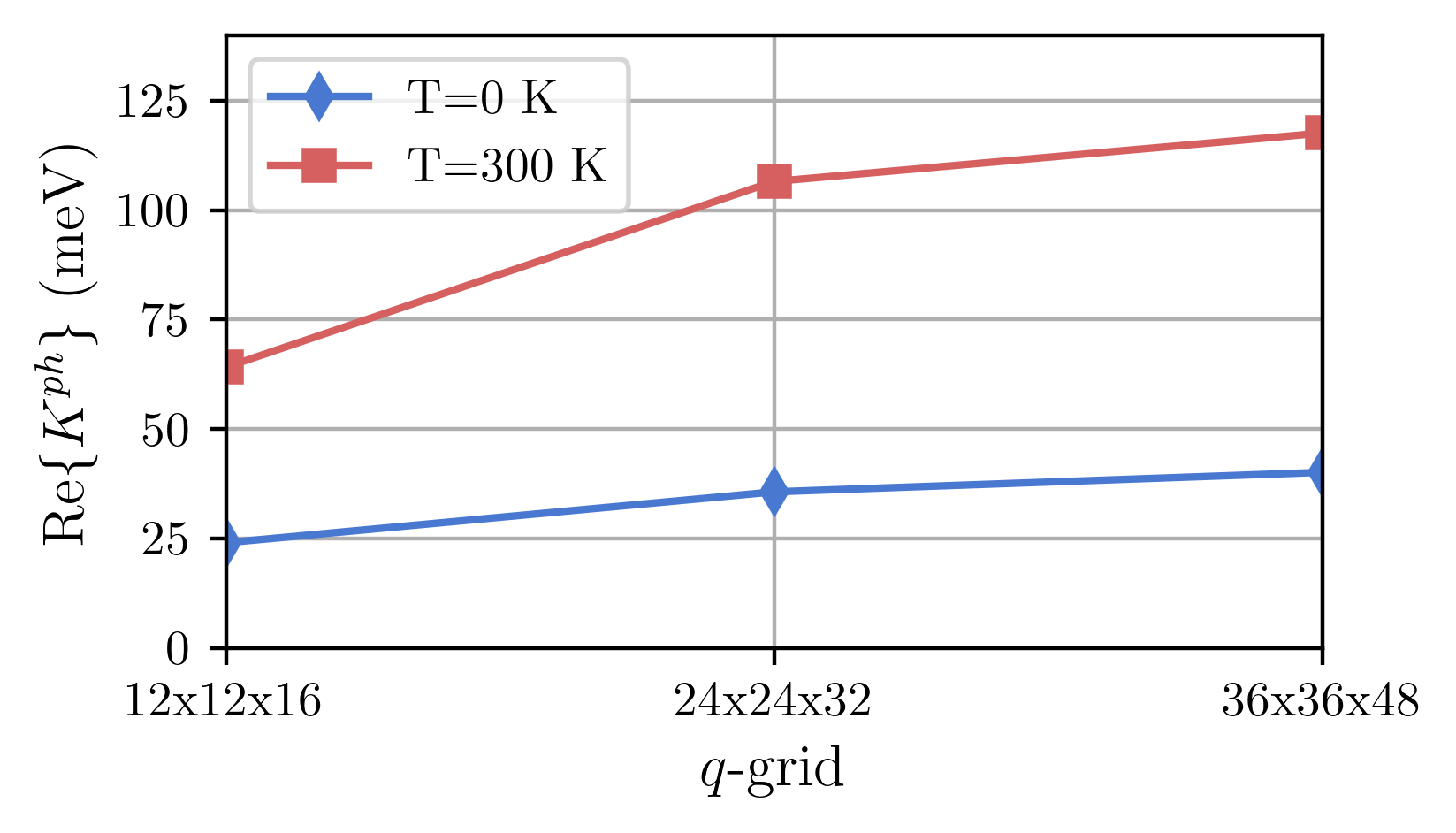}
            \put(1,52){\small \textbf{(a)}}
        \end{overpic}
        \label{fig:Re_Kph_q_conv_grid}
    \end{subfigure}
    \hfill
    \begin{subfigure}[]{0.44\linewidth}
        \begin{overpic}[width=\linewidth]{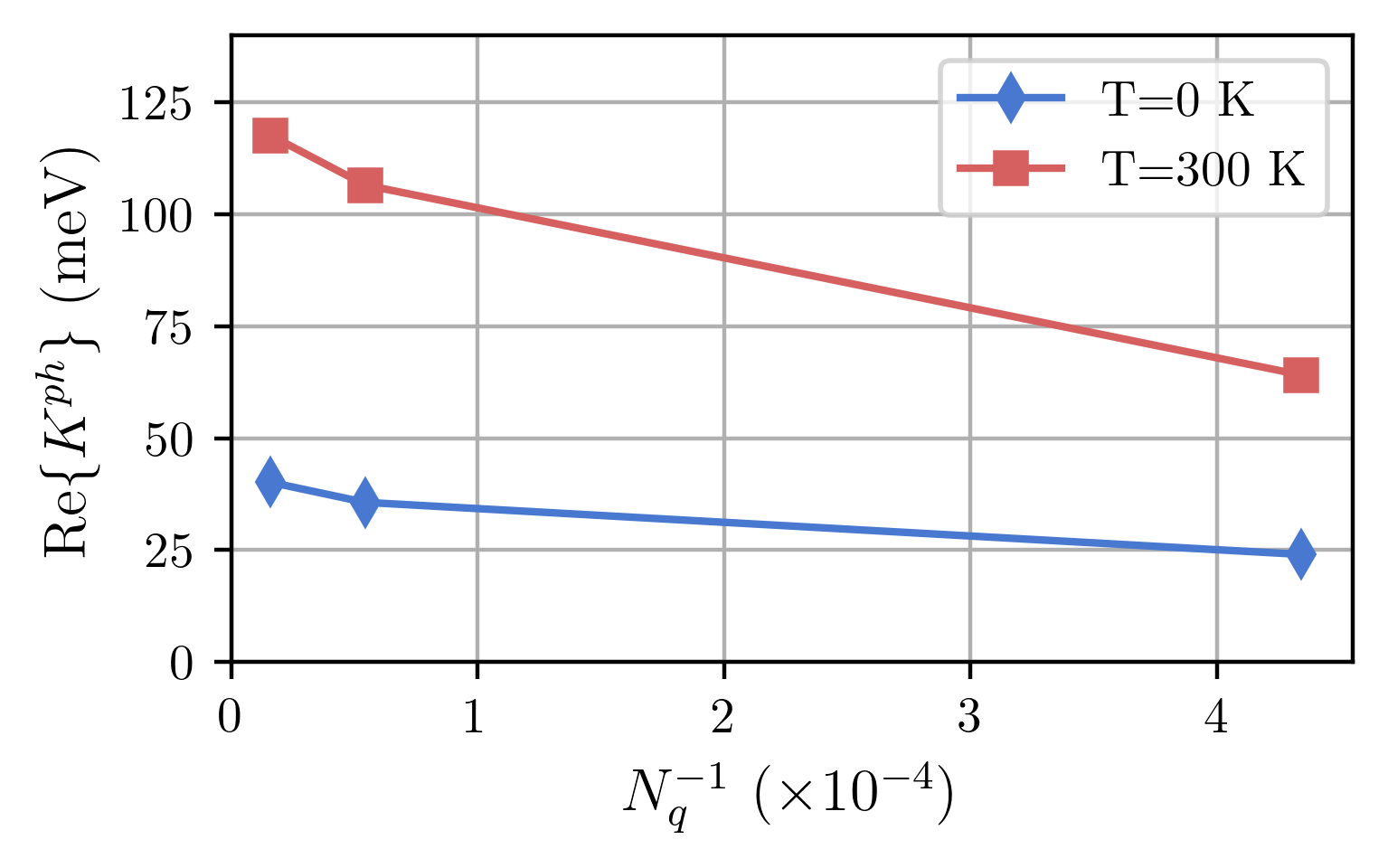}
            \put(1,57){\small \textbf{(b)}}
        \end{overpic}
        \label{fig:Re_Kph_q_conv_Nq}
    \end{subfigure}
    \\[-0.7cm]
    \caption{Convergence of the real part of $K^{\rm{ph}}$ vs the $\rm\mathbf{{q}}$-grid at $T=0$  K and $T=300$ K (a) and vs the number of $\rm\mathbf{{q}}$ points in the full grid $N_q$ that the patch represents (b).}
    \label{fig:Re_Kph_q_conv}
\end{figure}

\begin{figure}[phtb!]
    \centering
    \begin{subfigure}[]{0.485\linewidth}
        \begin{overpic}[width=\linewidth]{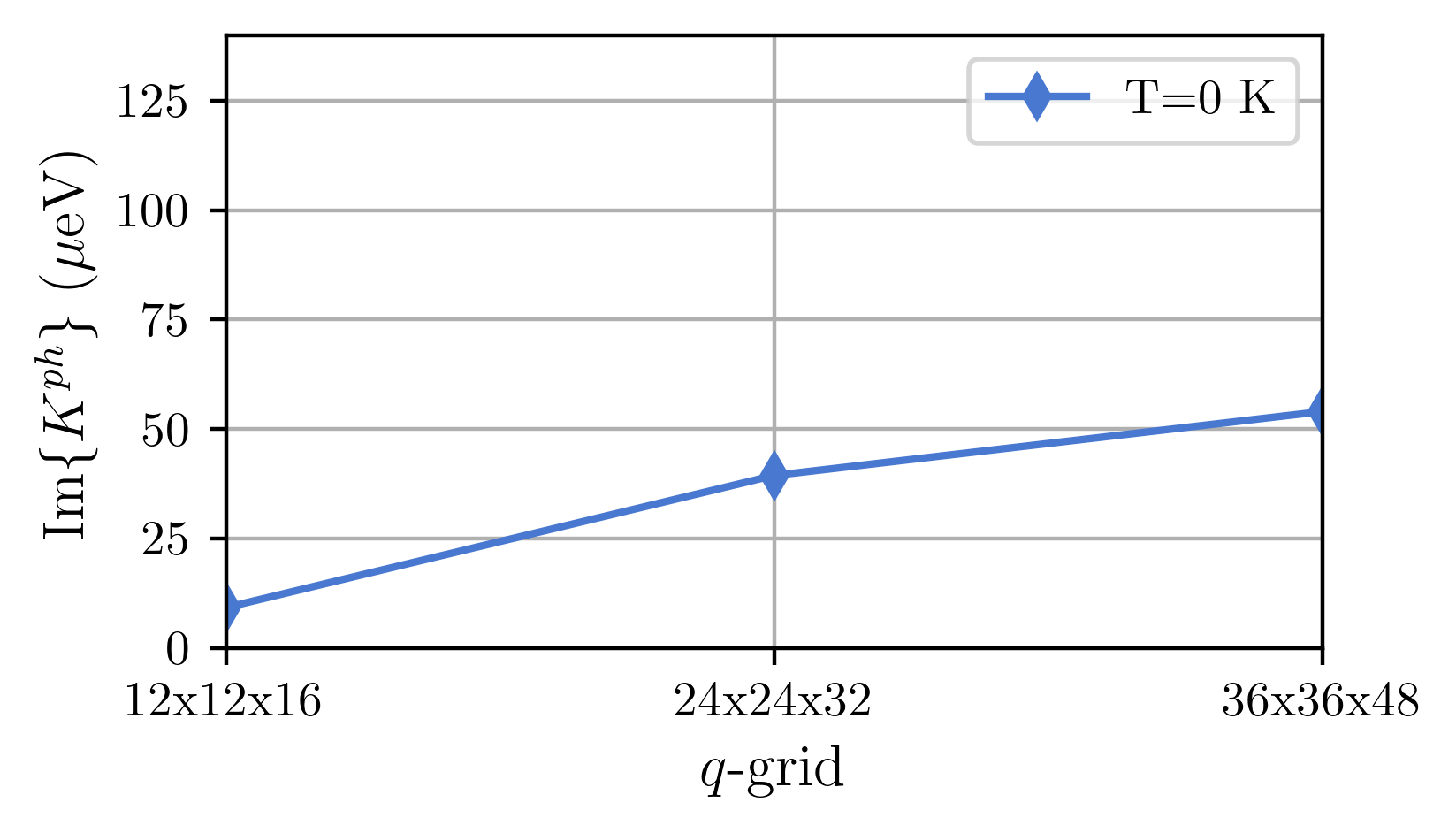}
            \put(1,52){\small \textbf{(a)}}
        \end{overpic}
        \label{fig:Im_Kph_q_conv_0}
    \end{subfigure}
    \hfill
    \begin{subfigure}[]{0.485\linewidth}
        \begin{overpic}[width=\linewidth]{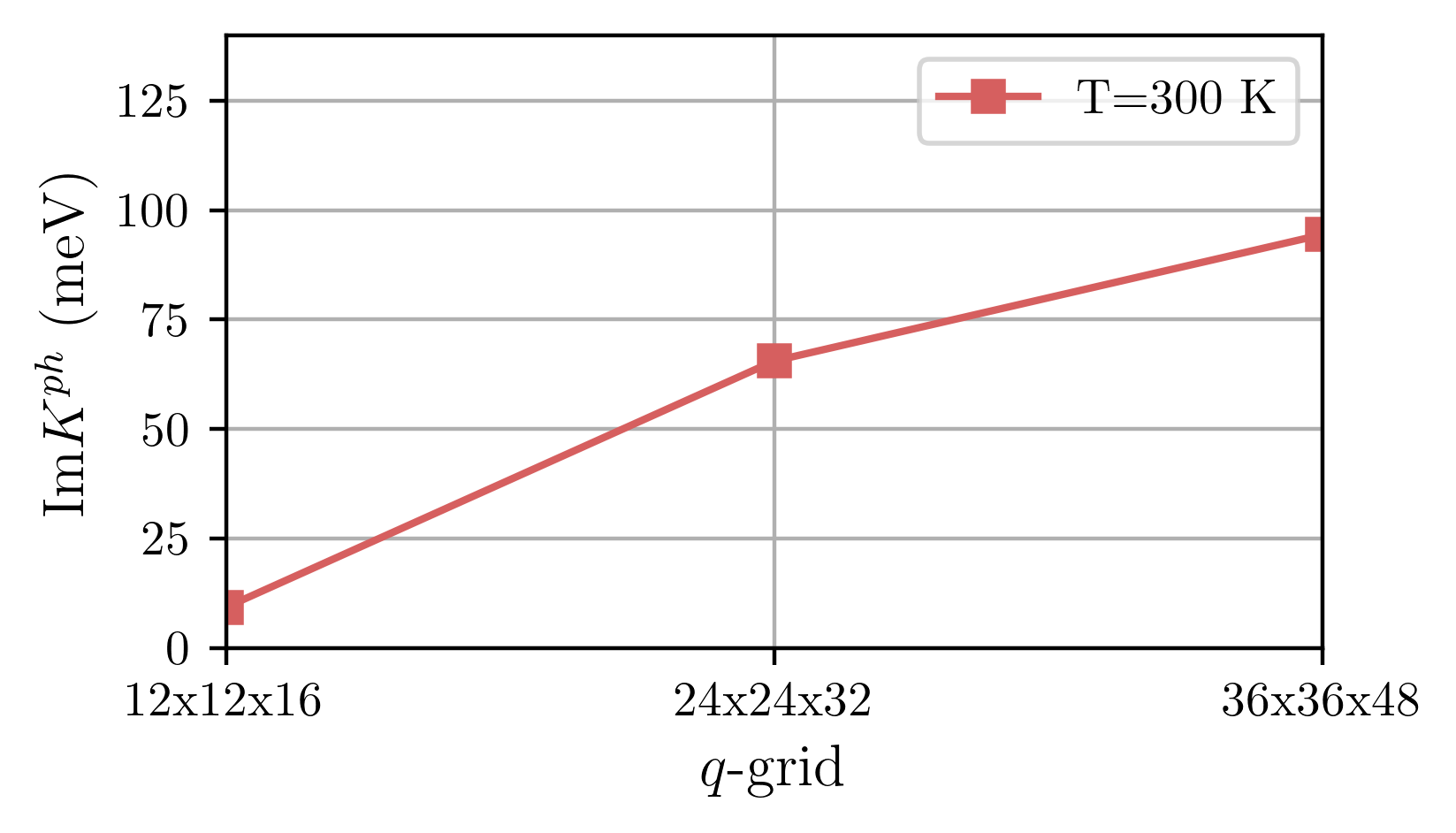}
            \put(1,52){\small \textbf{(b)}}
        \end{overpic}
        \label{fig:Im_Kph_q_conv_300}
    \end{subfigure}
    \\[-0.4cm]
    \begin{subfigure}[]{0.46\linewidth}
        \begin{overpic}[width=\linewidth]{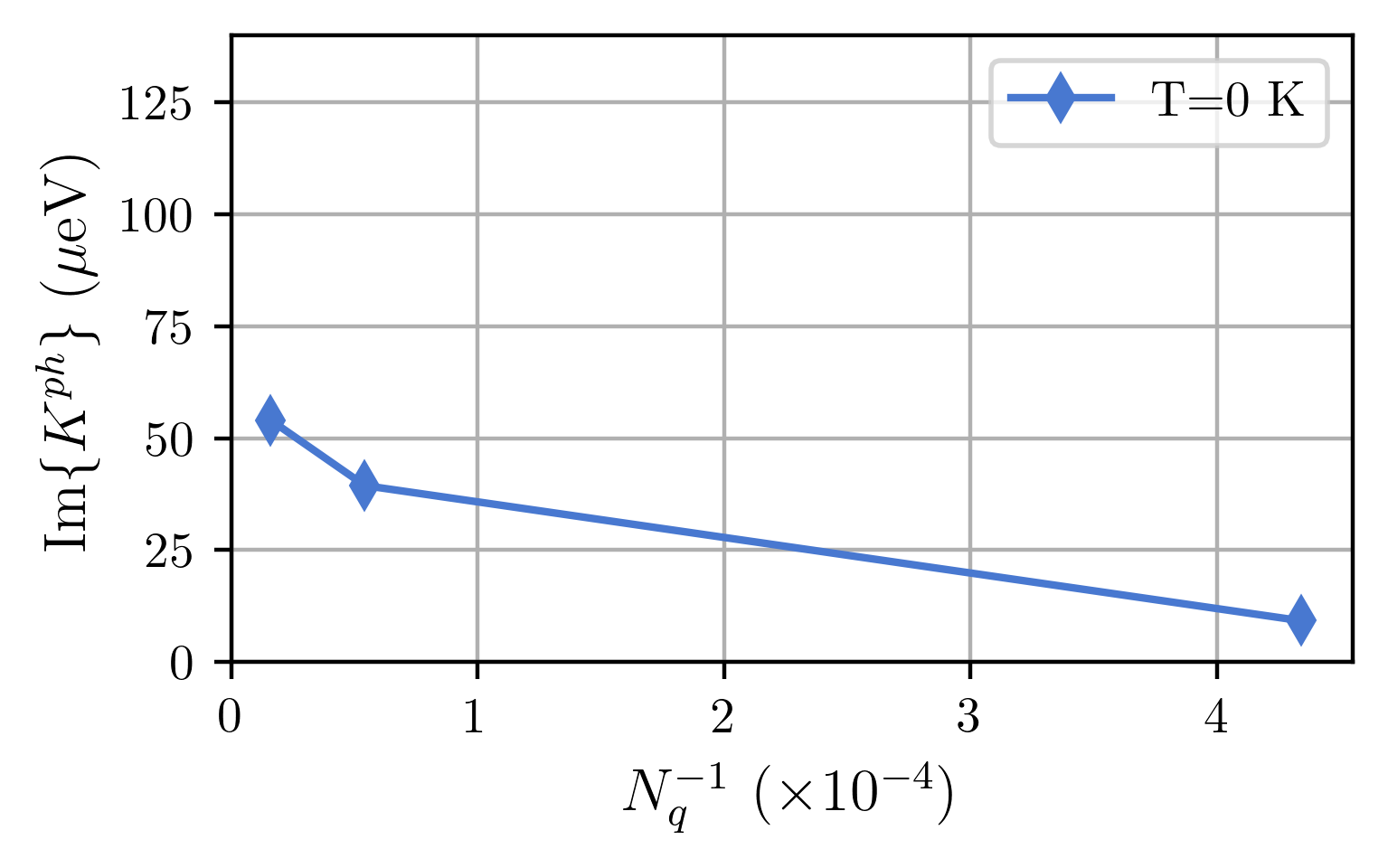}
            \put(1,57){\small \textbf{(c)}}
        \end{overpic}
        \label{fig:Im_Kph_q_conv_inv_0}
    \end{subfigure}
    \hfill
    \begin{subfigure}[]{0.46\linewidth}
        \begin{overpic}[width=\linewidth]{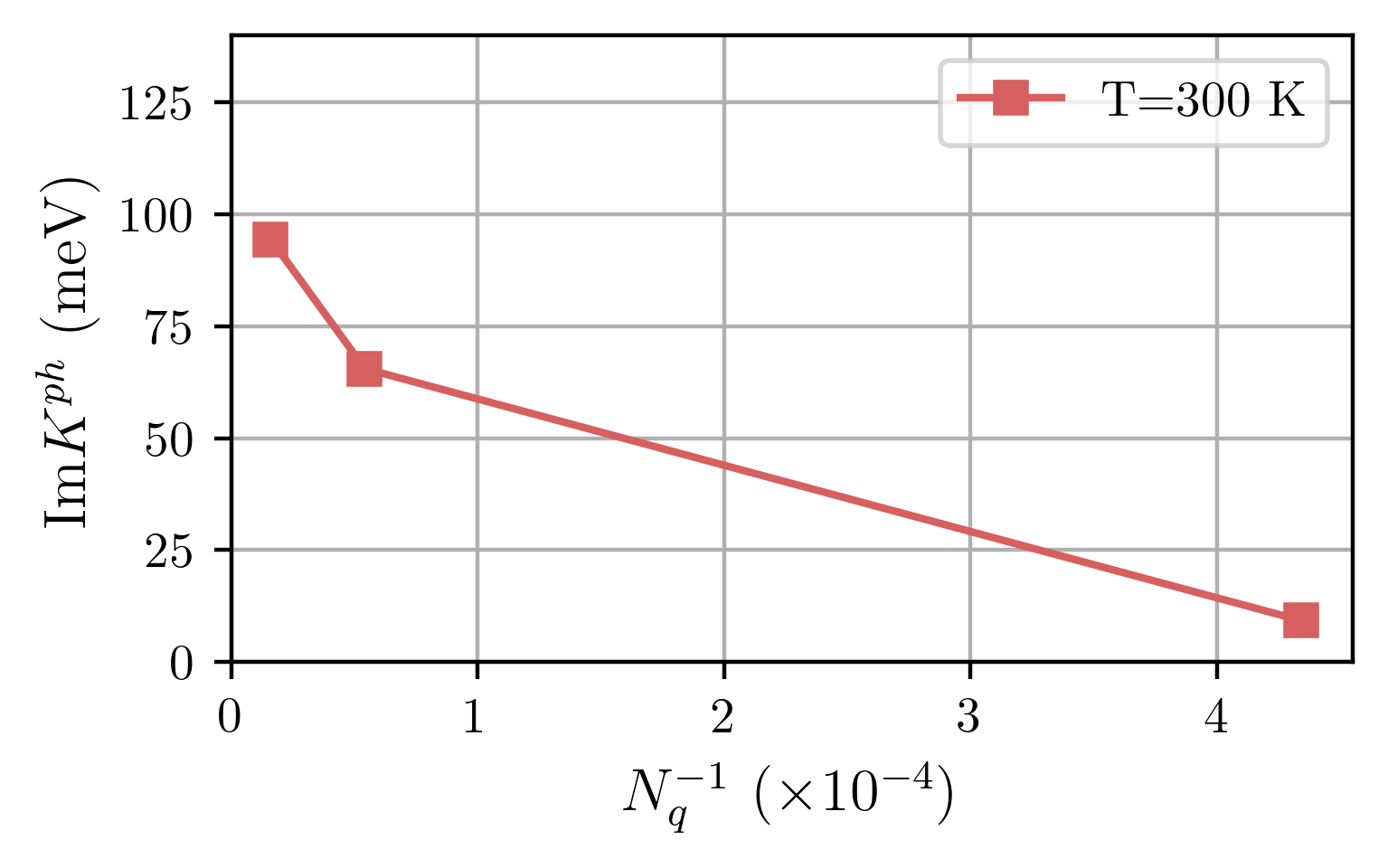}
            \put(1,57){\small \textbf{(d)}}
        \end{overpic}
        \label{fig:Im_Kph_q_conv_inv_300}
    \end{subfigure}
    \\[-0.7cm]
    \caption{Convergence of the imaginary part of $K^{\rm{ph}}$ vs the $\rm\mathbf{{q}}$-grid at $T=0$  K (a) and $T=300$ K (b). (c) and (d) show the same data but vs the number of $\rm\mathbf{{q}}$ points in the full grid $N_q$ that the patch represents.}
    \label{fig:Im_Kph_q_conv}
\end{figure}

Fig.~\ref{fig:Re_Kph_q_conv} shows the convergence of the real part of $K^{\rm{ph}}$ with respect to the $\rm\mathbf{{q}}$-grid density used. As can be seen, even at a $36\times36\times48$ grid, it is not fully converged. Extrapolating from the last two points in (b) for both temperatures would suggest that the real part of $K^{\rm{ph}}$ is underestimated by a few meV. Also of note is the fact that the $300$ K data series appears to converge more slowly.

Fig.~\ref{fig:Im_Kph_q_conv} shows the convergence of the imaginary part of $K^{\rm{ph}}$ with respect to the $\rm\mathbf{{q}}$-grid density used. Even more so than the real part, this quantity is somewhat under-converged even on a $36\times36\times48$ grid. Extrapolating from the last two points in subplots (c) and (d) for both temperatures would suggest that the imaginary part of $K^{\rm{ph}}$ is underestimated by a $\sim10$ meV at $300$ K. As was the case for the real part, the data for $300$ K exhibit slower convergence.

Importantly, both of the above convergence series for the real and imaginary parts of $K^{\rm{ph}}$ suggest that using the grid employed in this work causes an underestimation of the reduction in the exciton binding energy and an overestimation of calculated lifetimes. Though employing the $S$ basis truncation discussed in Appendix B appears to result in a slight overestimation of both aforementioned quantities, the under convergence of the $\rm\mathbf{{q}}$-grid appears to more than cancel out these effects.

\subsubsection{\texorpdfstring{$\eta$}{eta} Convergence}
\label{subsubsec:eta_conv}

\begin{figure}[phtb!]
    \centering
    \begin{subfigure}[]{0.485\linewidth}
        \begin{overpic}[width=\linewidth]{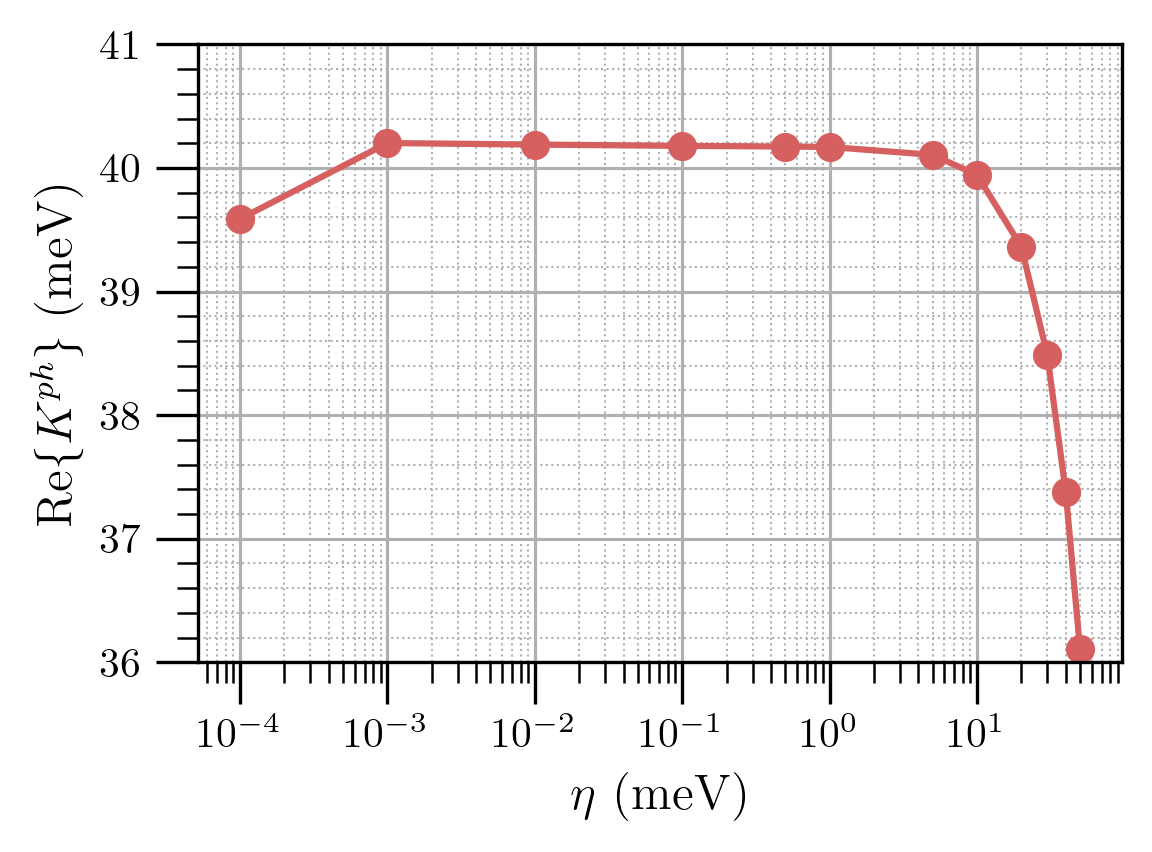}
            \put(1,68){\small \textbf{(a)}}
        \end{overpic}
        \label{fig:eta_conv_re_0}
    \end{subfigure}
    \hfill
    \begin{subfigure}[]{0.485\linewidth}
        \begin{overpic}[width=\linewidth]{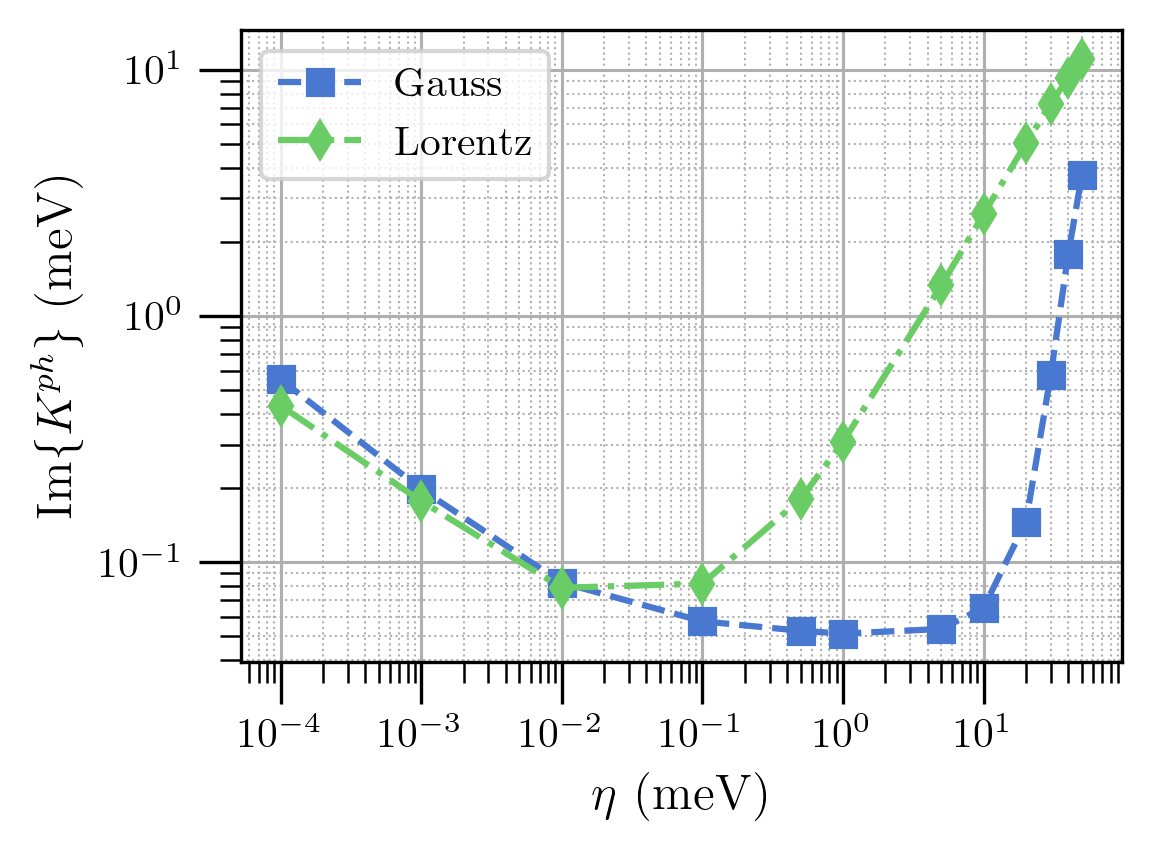}
            \put(1,68){\small \textbf{(b)}}
        \end{overpic}
        \label{fig:eta_conv_im_0}
    \end{subfigure}
    \\[-0.9cm]
    \caption{Convergence of the real and imaginary parts of $K^{\rm{ph}}$ vs. $\eta$ at zero temperature. Note, to aid in readability, logarithmic scales are used for both $x$ axes, but only on the $y$ axis of the imaginary part.}
    \label{fig:eta_conv_comp}
\end{figure}

Fig.~\ref{fig:eta_conv_comp} shows the convergence of the real and imaginary parts of $K^{\rm{ph}}$ with respect to $\eta$ at zero temperature.
Since $T=0$ K here, only the emission channel is active; as can be seen for the imaginary part, Gaussian broadening converges much more stably over three orders of magnitude (from $0.1$ to $10$ meV). The selected broadening value of $\eta=5$ meV is well within the converged plateau of the Gaussian curve. Interestingly, it appears that Lorentzian broadening values in the regime of $0.01$ to $0.1$ meV give comparable results to those of the more well-converged Gaussian broadening. Naively, this choice would appear to be far to small to capture the energy dispersion sampling density. However, it is possible that such a choice could be sensible given the larger tail of a Lorentzian function. 

\begin{figure}[htbp!]
    \centering
    \begin{subfigure}[]{0.485\linewidth}
        \begin{overpic}[width=\linewidth]{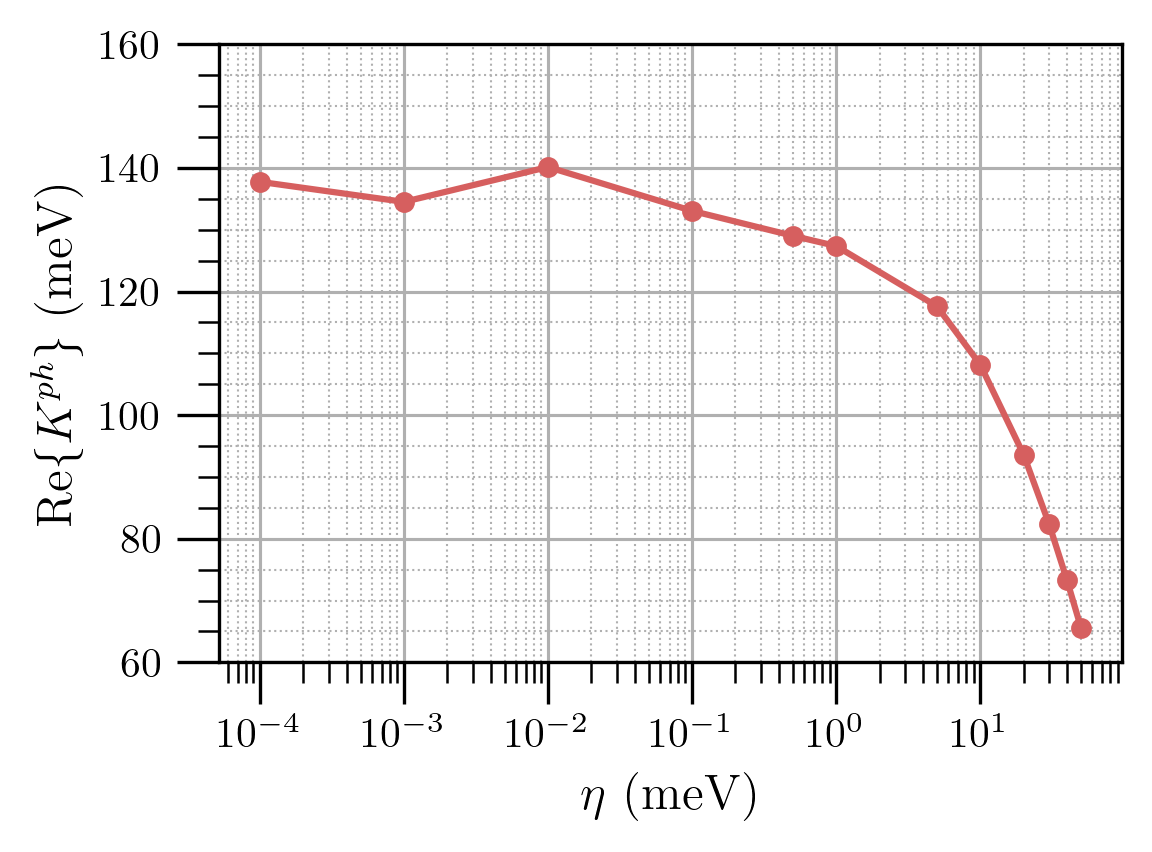}
            \put(1,68){\small \textbf{(a)}}
        \end{overpic}
        \label{fig:eta_conv_re_300}
    \end{subfigure}
    \hfill
    \begin{subfigure}[]{0.485\linewidth}
        \begin{overpic}[width=\linewidth]{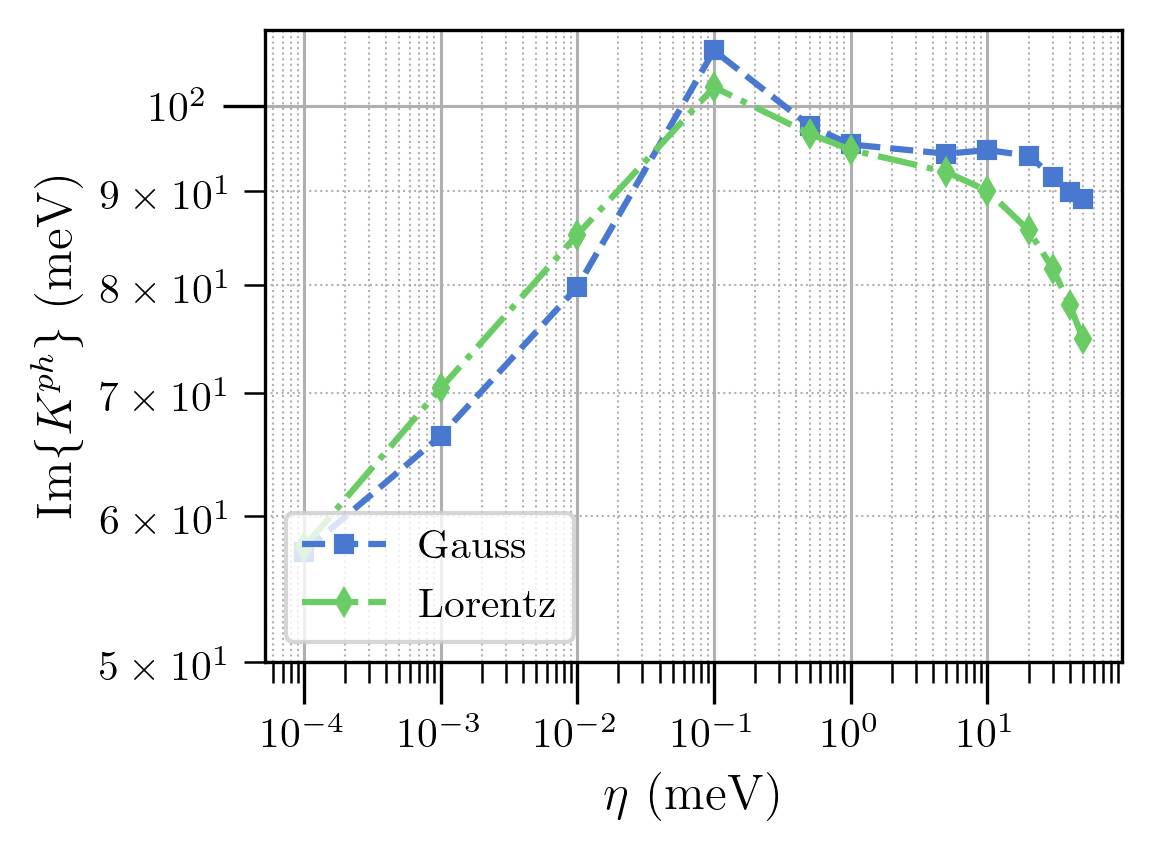}
            \put(1,68){\small \textbf{(b)}}
        \end{overpic}
        \label{fig:eta_conv_im_300}
    \end{subfigure}
    \\[-0.9cm]
    \caption{Convergence of the real and imaginary parts of $K^{\rm{ph}}$ vs. $\eta$ at $300$ K. Note, logarithmic scales are used for both $x$ axes, but only on the $y$ axis of the imaginary part.}
    \label{fig:eta_conv_comp_300}
\end{figure}

\begin{figure}[phtb!]
    \centering
    \begin{subfigure}[]{0.485\linewidth}
        \begin{overpic}[width=\linewidth]{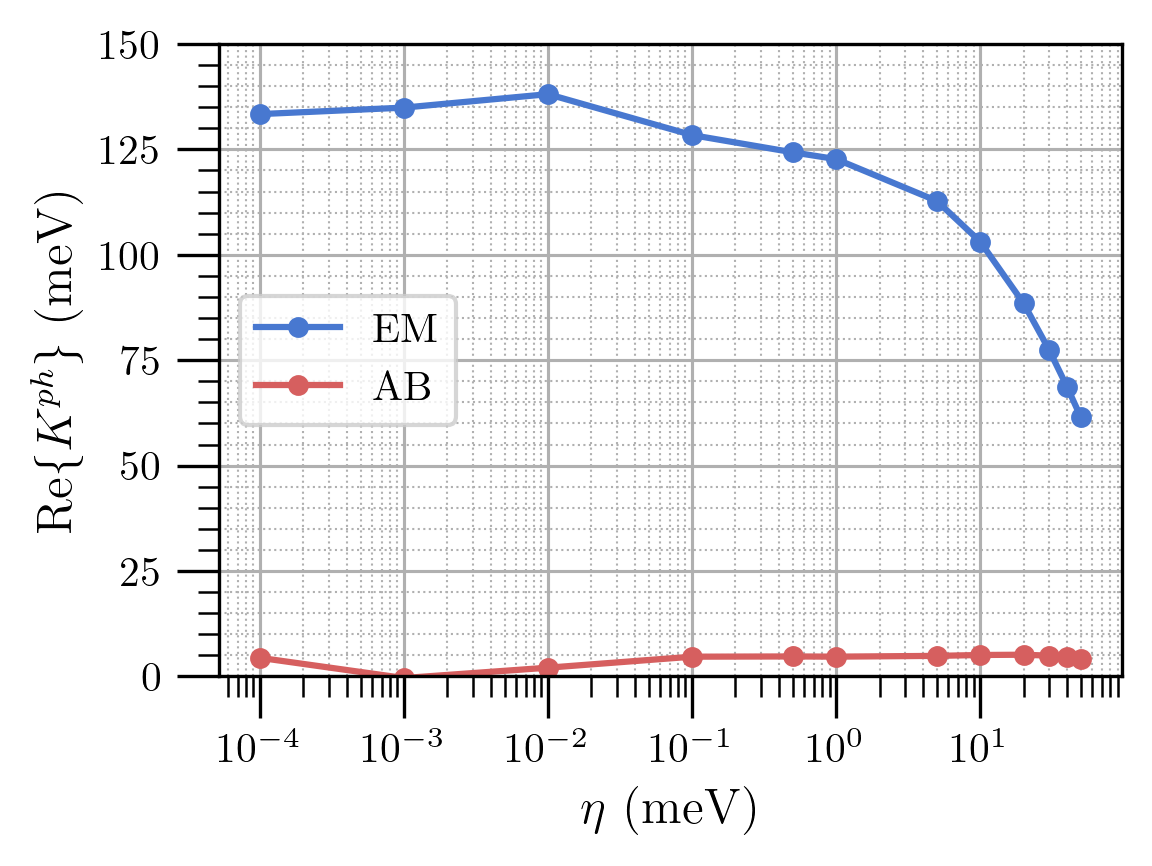}
            \put(1,68){\small \textbf{(a)}}
        \end{overpic}
        \label{fig:eta_conv_re_em_ab_300}
    \end{subfigure}
    \hfill
    \begin{subfigure}[]{0.485\linewidth}
        \begin{overpic}[width=\linewidth]{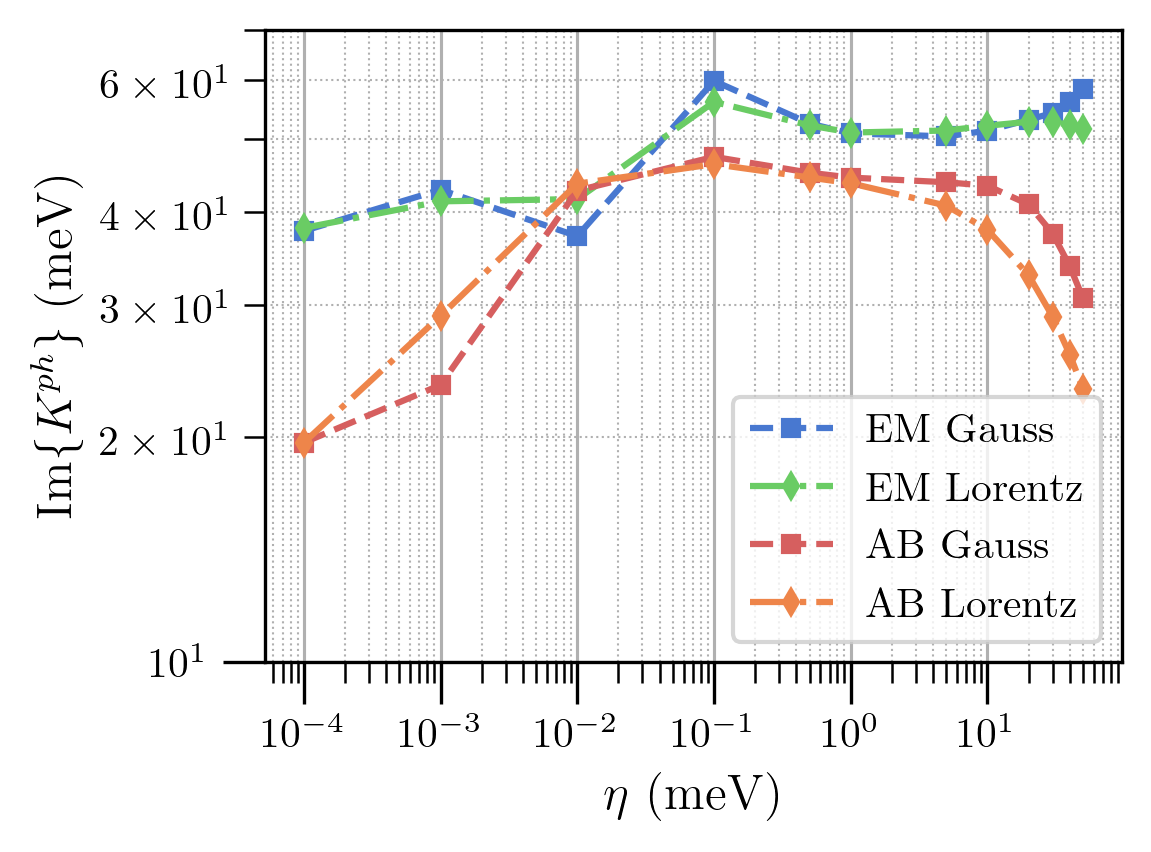}
            \put(1,68){\small \textbf{(b)}}
        \end{overpic}
        \label{fig:eta_conv_im_em_ab_300}
    \end{subfigure}
    \\[-0.9cm]
    \caption{Convergence of the real and imaginary parts of $K^{\rm{ph}}$ split into the emission and absorption channels vs. $\eta$ at $300$ K. Note, logarithmic scales are used for both $x$ axes, but only on the $y$ axis of the imaginary part.}
    \label{fig:eta_conv_em_ab_300}
\end{figure}

Similarly, Fig.~\ref{fig:eta_conv_comp_300} shows the convergence of the real (a) imaginary (b) parts of $K^{\rm{ph}}$ with respect to $\eta$ at $300$ K, and Fig.~\ref{fig:eta_conv_em_ab_300} shows the same data but split into the emission and absorption channels for added detail. Again, Gaussian broadening seems to exhibit a much wider range (from $0.1$ to $10$ meV) of stable convergence with respect to $\eta$. Notably, the real part appears to reach less stable convergence than it did at zero temperature. Comparing the value at $\eta=5$ meV to the approximate asymptote, the real part is underestimated by $\sim20$ meV.

\subsubsection{Perturbation Theory Treatment}
\begin{figure}[htbp!]
    \centering
    \begin{subfigure}[]{0.48\linewidth}
        \begin{overpic}[width=\linewidth]{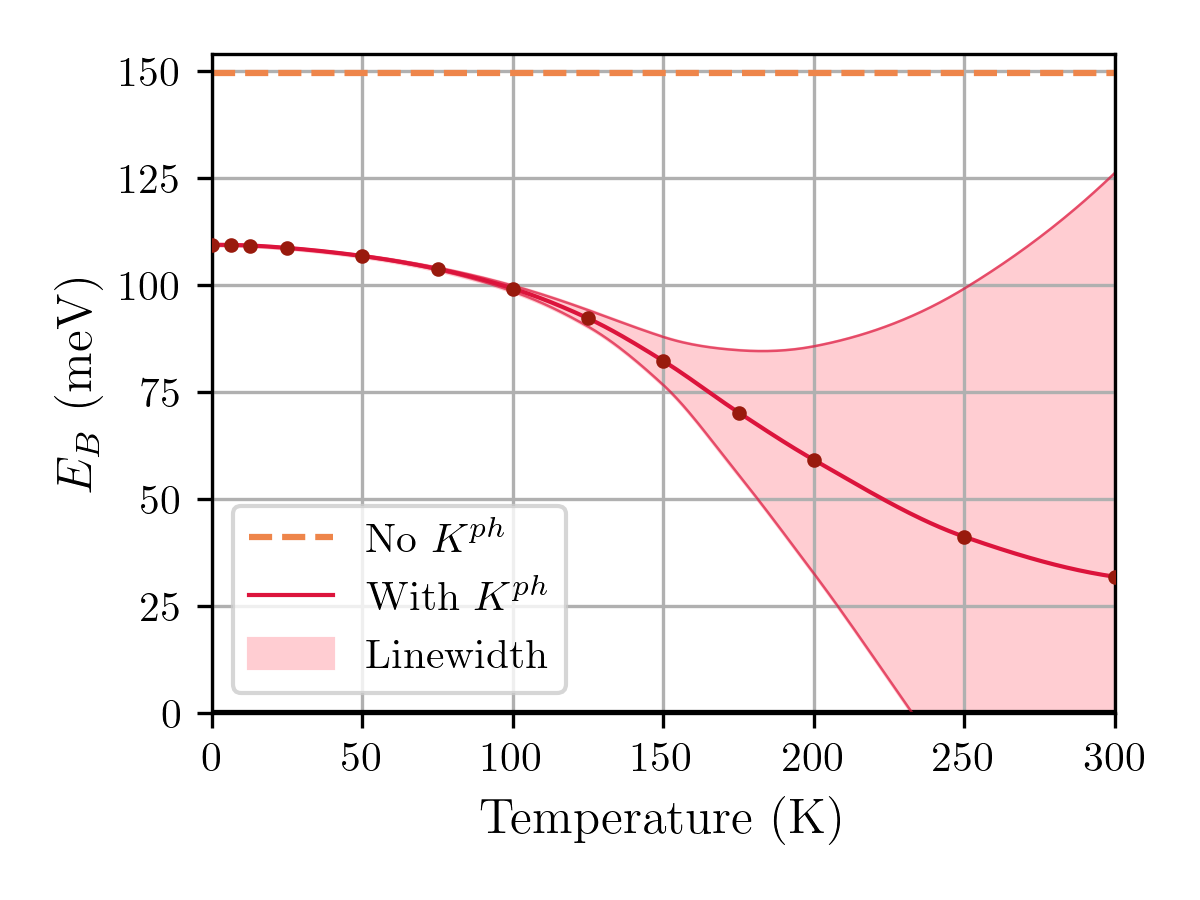}
            \put(-3,70){\small \textbf{(a)}}
        \end{overpic}
        \label{fig:BW_pert}
    \end{subfigure}
    \\[-0.6cm]
    \begin{subfigure}[]{0.48\linewidth}
        \begin{overpic}[width=\linewidth]{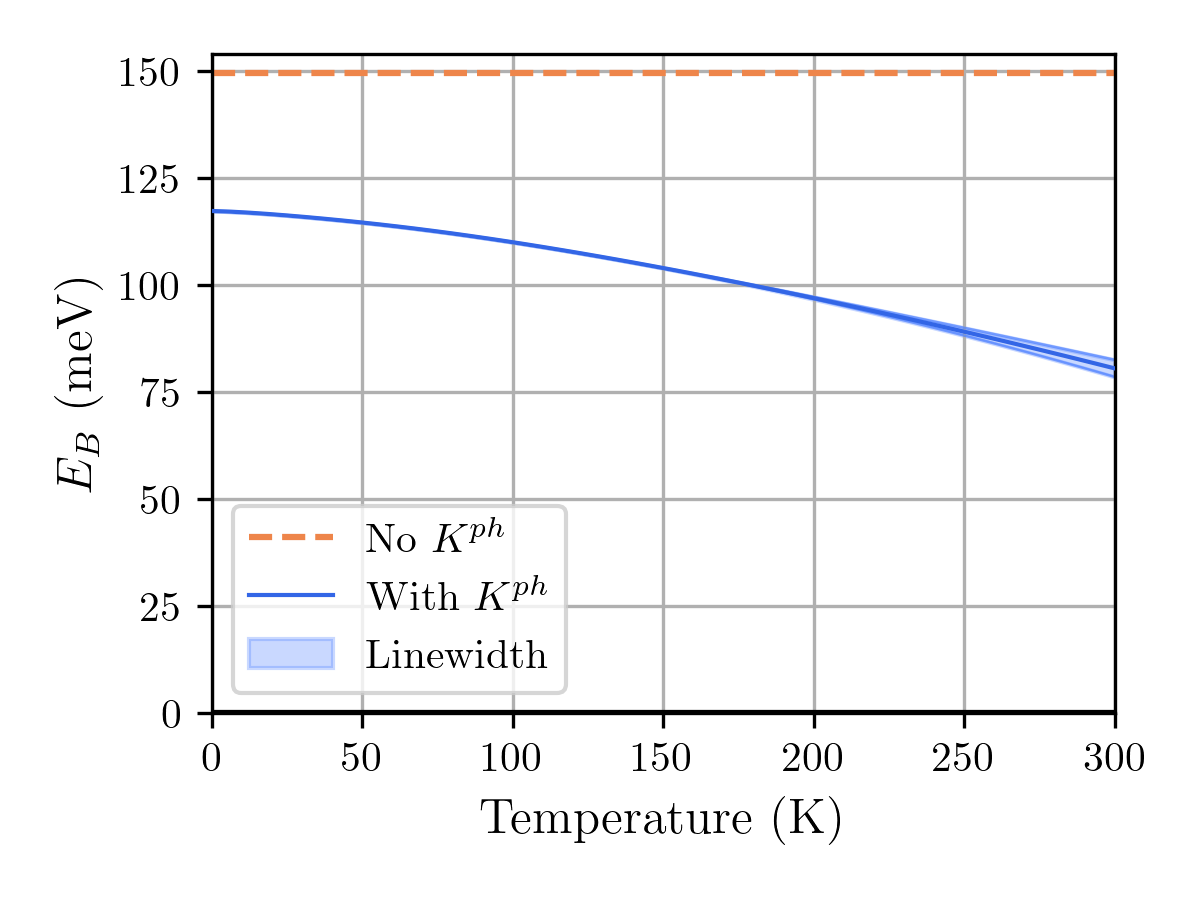}
            \put(-3,70){\small \textbf{(b)}}
        \end{overpic}
        \label{fig:RS_pert_S}
    \end{subfigure}
    ~
    \begin{subfigure}[]{0.48\linewidth}
        \begin{overpic}[width=\linewidth]{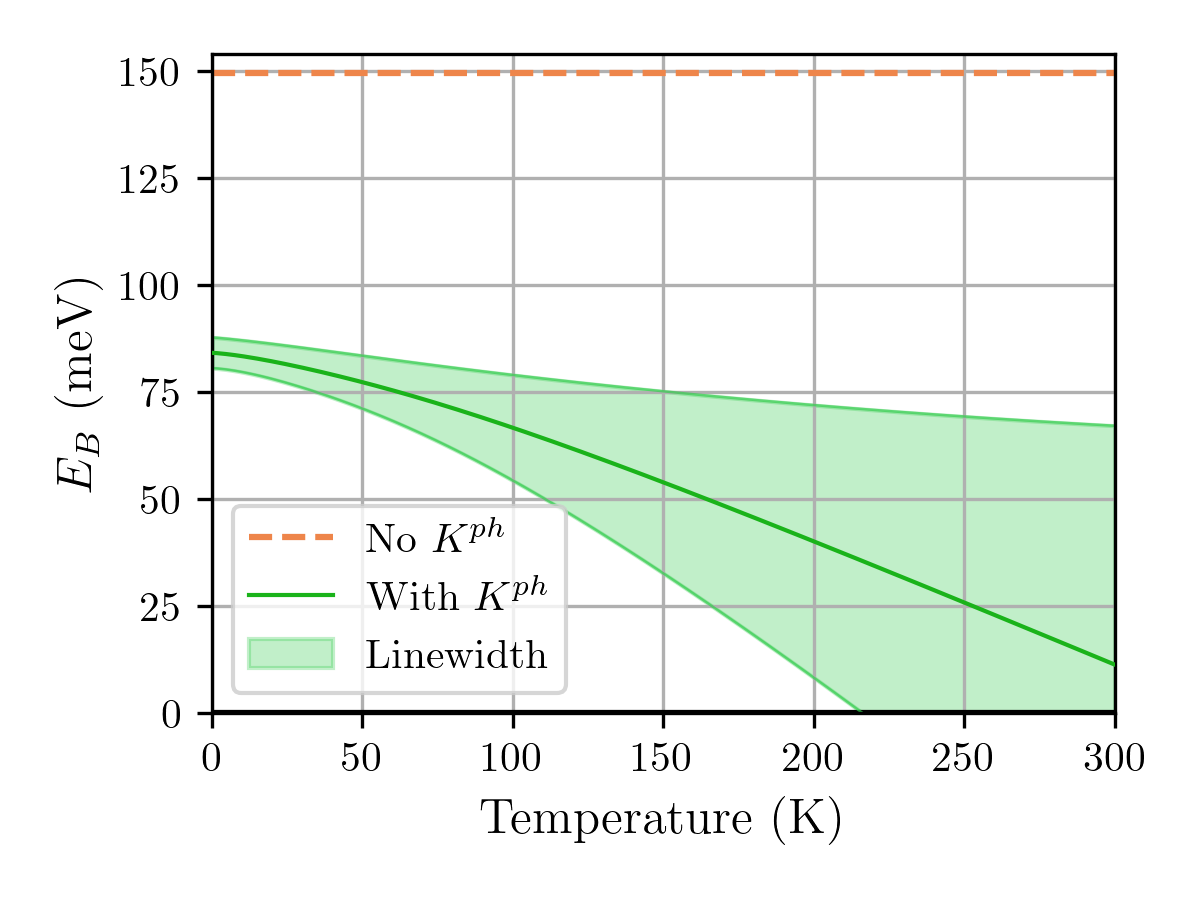}
            \put(-3,70){\small \textbf{(c)}}
        \end{overpic}
        \label{fig:RS_pert_Eg}
    \end{subfigure}
    \\[-0.9cm]
    \caption{Differing perturbative treatments for handling the dynamical nature of $K^{\rm{ph}}$. (a) shows the approach outlined by Eq.~\ref{eq:BW_eq}, (b) shows the approach outlined by Eq.~\ref{eq:RS_S}, and (c) shows the approach outlined by Eq.~\ref{eq:RS_Eg}.}
    \label{fig:RS_vs_BW}
\end{figure}

Fig.~\ref{fig:RS_vs_BW} shows the $K^{\rm{ph}}$ corrected binding energy for the first bright exciton as a function of temperature using three different schemes for solving for the corrected BSE eigenvalue $\tilde{\Omega}^S$. The first scheme corresponding to Fig.~\ref{fig:RS_vs_BW}(a) is that of Eq.\ 4 in the main text. which solves for $\tilde{\Omega}^S$ according to
\begin{equation}
    \label{eq:BW_eq}
    \tilde{\Omega}^S=\Omega^S+K^{\rm{ph}}_{S,S}(\text{Re}\{\tilde{\Omega}^S\},T)
\end{equation}
where it is understood that $\tilde{\Omega}^S$ must be solved for self-consistently at each temperature. This treatment is reminiscent of Brillouin Wigner perturbation theory.

The second scheme corresponding to Fig.~\ref{fig:RS_vs_BW}(b) solves for $\tilde{\Omega}^S$ according to
\begin{equation}
    \label{eq:RS_S}
    \tilde{\Omega}^S=\Omega^S+K^{\rm{ph}}_{S,S}(\Omega^S,T)
\end{equation}
where $\Omega$ is set equal to the ion-clamped exciton binding energy. This treatment is analogous to Raleigh-Schr\"odinger perturbation theory with the ion-clamped exciton solution treated as the unperturbed system. In practice, since $\Omega=\Omega^S$ is independent of temperature, and this approach requires Eq.~\ref{eq:RS_S} to be solved only once for all temperatures since thermal occupation factors can be introduced later for each mode and momentum of phonon for any temperature. This is in contrast to Eq.~\ref{eq:BW_eq} where calculations unavoidably temperature dependent via the self-consistency in $\Omega$.

The third scheme corresponding to Fig.~\ref{fig:RS_vs_BW}(c) solves for $\tilde{\Omega}^S$ according to
\begin{equation}
    \label{eq:RS_Eg}
    \tilde{\Omega}^S=\Omega^S+K^{\rm{ph}}_{S,S}(E_g,T)
\end{equation}
where $\Omega$ is set equal to the direct gap $E_g$ where the exciton forms. This treatment is also analogous to Raleigh-Schr\"odinger perturbation theory but with the the single-particle electron-hole pair treated as the unperturbed system. Like before, since $\Omega=E_g$ is independent of temperature, this approach requires Eq.~\ref{eq:RS_S} be solved only once for all temperatures.

We remark that if either Eq.~\ref{eq:RS_S} or \ref{eq:RS_Eg} where iterated self-consistently for solving $\tilde{\Omega}^S$, then Eq.~\ref{eq:BW_eq} would be recovered. Importantly, for calculating exciton lifetimes, the approach of Eq.~\ref{eq:RS_Eg} amounts to assuming that the exciton has no binding energy and thus significantly finite linewidths are accessible even at zero temperature. Conversely, the approach of Eq.~\ref{eq:RS_S} models an exciton that is overabound relative to Eq.~\ref{eq:BW_eq} or \ref{eq:RS_Eg}, and, as can be seen, the linewidth is much smaller than either approach. We also add that while the approach of Eq.~\ref{eq:RS_S} has been the equation used in both Refs.~\cite{alvertis_phonon_2024, lee_phonon_2024}, \ref{eq:BW_eq} is more well-suited to for computing linewidths and lifetimes because it accurately captures the energy needed to dissociate the exciton that is formed when considering phonon emission and absorption scattering events.

\subsection{Functional Dependence}
In order to establish an increased level of confidence in the results presented in this work, we have investigated the dependence of our results on the underlying DFT exchange correlation approximation by also considering the LDA \cite{kohn_selfconsistent_1965, perdew_selfinteraction_1981} functional. We begin by re-relaxing the structure using LDA and then go on to look at various properties. Specifically, we investigate the magnitude of the real part of $K^{\rm{ph}}$ at zero temperature, defined here as $\Delta E_B$ (i.e. the change in the exciton binding energy for the first exciton at zero temperature). We also investigate the energetics associated with exciton dissociation by considering the difference between the in-patch direct-indirect gap difference ($\Delta E_g$) and the BSE eigenvalue of the first exciton including $K^{\rm{ph}}$ corrections at zero temperature. A positive quantity here indicates that dissociation via phonon emission is energetically allowed. Results are presented in Tab.~\ref{tab:lda_v_pbe}.
\begin{table}[h]
    \centering
    \caption{Functional dependence of $K^{\rm{ph}}$ when using LDA vs PBE. Calculations were carried out on a $9\times9\times11$ grid patch inside of a $24\times24\times32$ grid.}
    \vspace{2mm}
    \setlength{\tabcolsep}{10pt}
    \begin{tabular}{c|cc}
    \hline
    Functional & $\Delta E_B$ (meV) & $\Delta E_g-\tilde{\Omega}^S$ (meV) \\ \hline
    PBE & 36 & 33 \\
    LDA & 36 & 25 \\ \hline
    \end{tabular}
    \label{tab:lda_v_pbe}
\end{table}
We find excellent agreement between LDA and PBE functionals for the exciton binding energy renormalization. Furthermore we find that using both functionals, the direct vs indirect gap energy landscape in the patch is relatively similar, with both functionals permitting exciton dissociation at zero temperature. We note that the allowed energies of emission should grow at higher temperatures and for more converged calculations where the binding energy is more strongly renormalized.

\section{Broadening in \texorpdfstring{$K^{\rm{ph}}$}{Kph}}
\label{sec:eta_role}
As shown in Eq.~$2$ in the main text as well as Eq.~\ref{eq:Kph_T_pm}, the expressions for $K^{\rm{ph}}$ in both the exciton and electron hole bases contain a small positive broadening parameter $\eta$. Specifically, $\eta$ appears in energy denominators of the form $(\Delta E+i\eta)^{-1}$, where $\Delta E$ is an energy difference corresponding to the kernel frequency $\Omega$, electron hole energy differences such as $\Delta_{c\rm\mathbf{{k}}v'\rm\mathbf{{k}}'}$, and a phonon frequency $\omega_{\rm\mathbf{{q}}\nu}$. Formally, $\eta$ is understood to be infinitesimally small, and taking the limit as $\eta\rightarrow0$ for the real and imaginary parts of this denominator we get
\begin{align}
    \label{eq:eta_lim_re}
    \lim_{\eta\rightarrow0}\text{Re}\left\{\frac{1}{\Delta E +i\eta}\right\}
    &=\lim_{\eta\rightarrow0}\frac{\Delta E}{\Delta E^2+\eta^2}
    =\left\{
    \begin{array}{cl}
        \frac{1}{\Delta E} & ,\Delta E\neq 0\\
        0 & ,\Delta E=0
    \end{array}
    \right. \\
    \lim_{\eta\rightarrow0}\text{Im}\left\{\frac{1}{\Delta E +i\eta}\right\}
    &=\lim_{\eta\rightarrow0}\frac{-i\eta}{\Delta E^2+\eta^2}=-\pi\delta(\Delta E).
    \label{eq:eta_lim_im}
\end{align}
Notably, in the imaginary part, we get a delta function that enforces energy conservation.

In numerical calculations, however, $\eta$ cannot be infinitesimally small. For the real part setting $\eta=0$ runs the risk of encountering singularities in the denominator, and for the imaginary part the consequences of using the formal expression on the right side of Eq.~\ref{eq:eta_lim_im} are more severe. For any finite $\Delta E$ in the denominator, no contributions to $\text{Im}\{K^{\rm{ph}}\}$ are picked up, and since these energy differences are sampled on a finite grid, the precise singularities which would give non-zero imaginary contributions are nearly always missed. Thus, some form of broadening must be used to reliably calculate both the real and imaginary part of $K^{\rm{ph}}$.

\begin{figure}[htbp!]
    \centering
    \hspace{13pt}
    \begin{subfigure}[]{0.4\linewidth}
        \begin{overpic}[width=\linewidth]{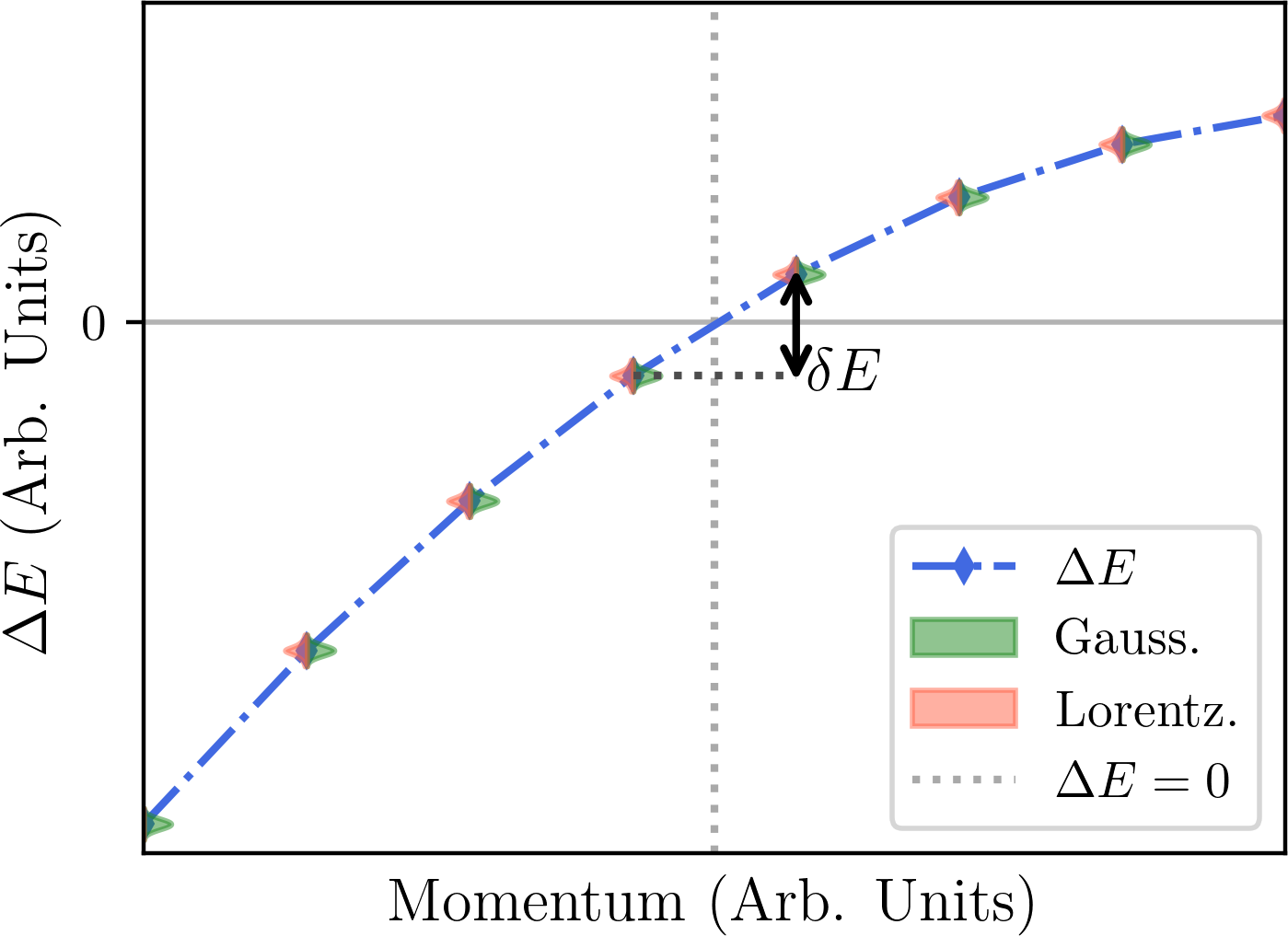}
            \put(-8,70){\small \textbf{(a)}}
        \end{overpic}
        \label{fig:eta_miss}
    \end{subfigure}
    \hspace{16pt}
    \begin{subfigure}[]{0.4\linewidth}
        \begin{overpic}[width=\linewidth]{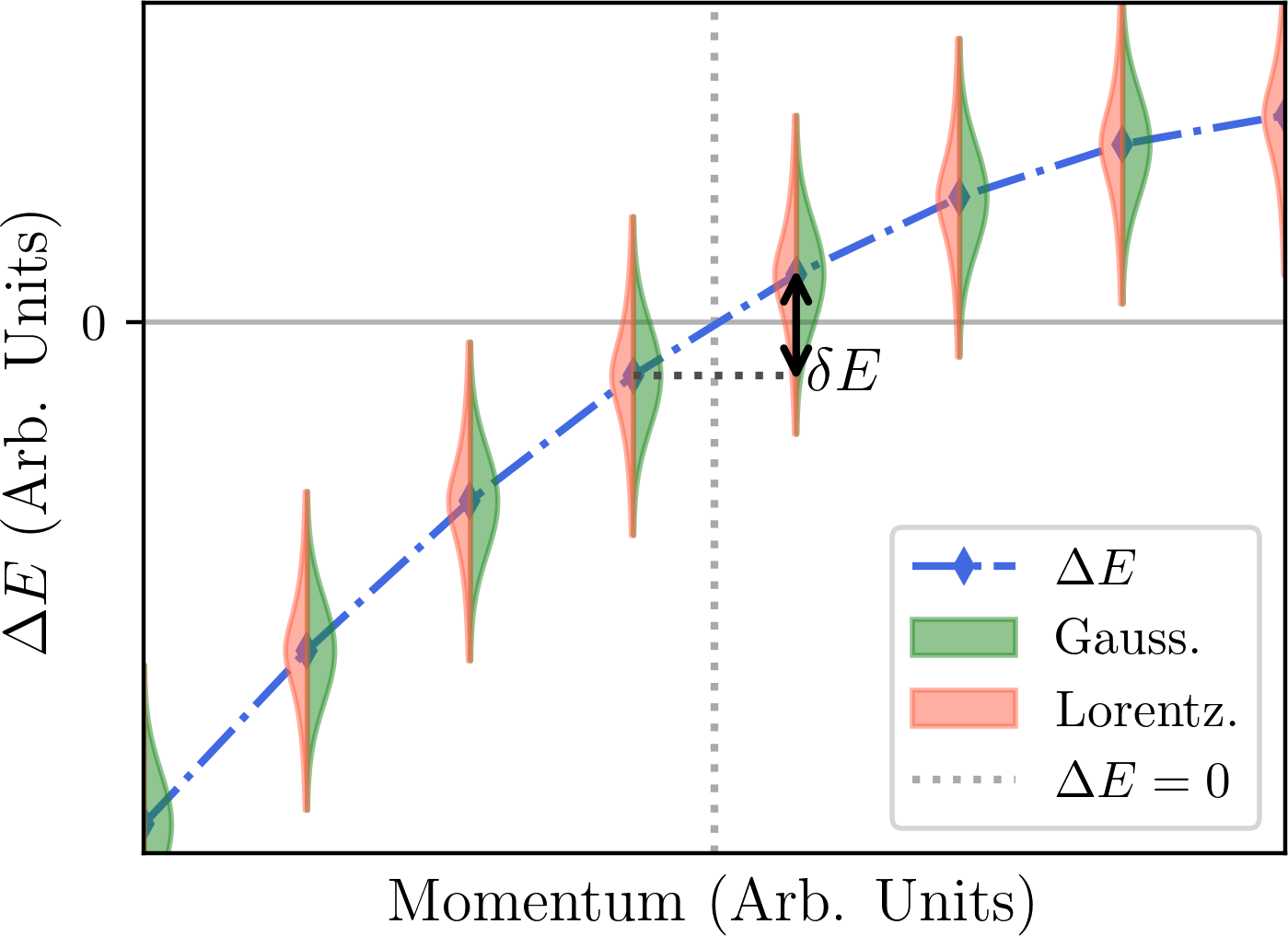}
            \put(-9,70){\small \textbf{(b)}}
        \end{overpic}
        \label{fig:eta_hit}
    \end{subfigure}
    \\[-0.2cm]
    \begin{subfigure}[]{0.415\linewidth}
        \begin{overpic}[width=\linewidth]{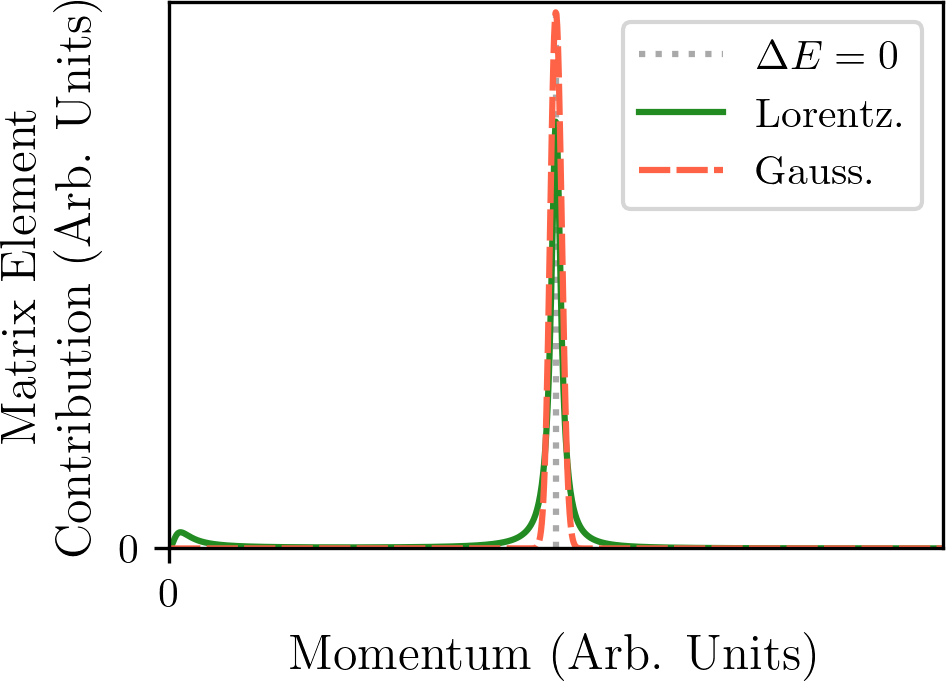}
            \put(-3,69){\small \textbf{(c)}}
        \end{overpic}
        \label{fig:eta_small}
    \end{subfigure}
    \hspace{8pt}
    \begin{subfigure}[]{0.415\linewidth}
        \begin{overpic}[width=\linewidth]{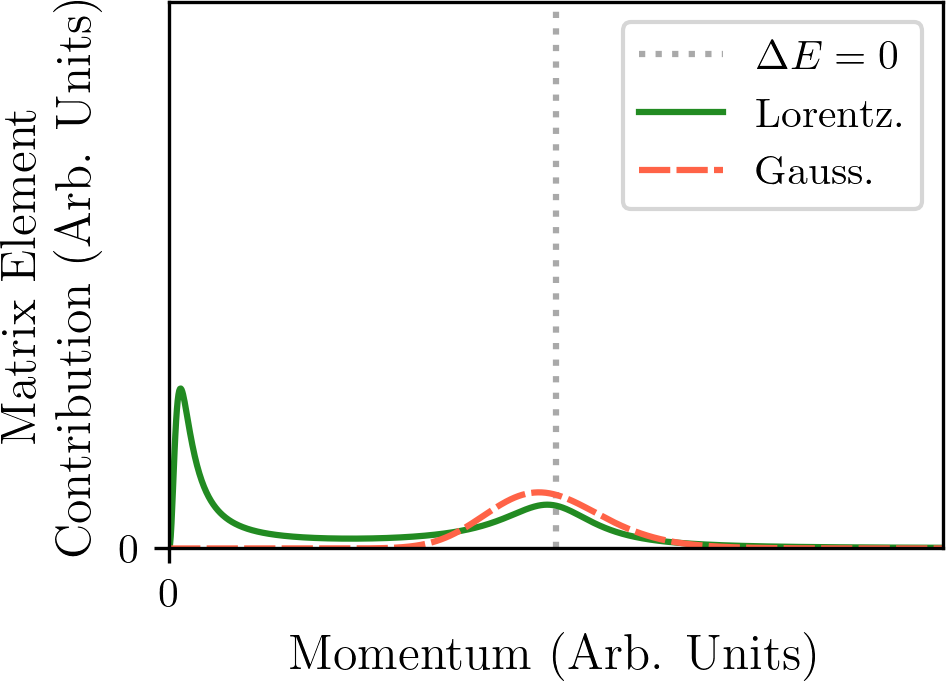}
            \put(-3.5,69){\small \textbf{(d)}}
        \end{overpic}
        \label{fig:eta_big}
    \end{subfigure}
    \\[-0.6cm]
    \caption{Illustration of the differences between of Lorentzian vs Gaussian broadening. (a) Shows a scenario where the choice of $\eta$ relative to the energy spacing $\delta E$ of sampled $\Delta E$ points is small; in the plot, $\delta E$, a proxy for $\eta_{\rm \rm\mathbf{{k}}}$, has a ratio with $\eta$ of $\delta E/\eta\approx19$. In this case the zero crossing is missed. (b) Shows the same setup as (a), except the size of $\eta$ is increased so that $\delta E/\eta\approx1.9$; now the zero crossing is encompassed. (c) shows a Fr{\"o}hlich-like matrix element multiplied by Gaussian and Lorentzian distribution functions, where $\eta$ is chosen to be sufficiently small that the energetically forbidden matrix element peak near $0$ is suppressed. Here, $\eta_M^{\text Lorentz}/\eta\approx5.2$ and $\eta_M^{\text Gauss}/\eta\approx27$. (d) shows the same setup as for (c), but if the size of the smearing is increased by a factor of $10$, as was done for (a) vs (b). In this scenario, $\eta_M^{\text Lorentz}/\eta\approx0.52$ and $\eta_M^{\text Gauss}/\eta\approx2.7$. As can be seen, the Lorentzian envelope erroneously picks up the matrix element peak as the dominant contribution, while the Gaussian envelope does not. Taken together, these figures illustrate a scenario where $\eta_{\rm \rm\mathbf{{k}}}\sim\eta_M$ for Lorentzian broadening, but not for Gaussian broadening.}
    \label{fig:eta_cartoon}
\end{figure}

For the real part, we return to the simple intermediate form given in Eq.~\ref{eq:eta_lim_re} for numerical calculations and find that it converges stably for $\eta\sim10$ meV (see \ref{subsubsec:eta_conv}). For the imaginary part, on the other hand, we find that employing a Lorentzian broadening function (as expressed in the middle of Eq.~\ref{eq:eta_lim_im}) in place of the Dirac delta function on the right side of Eq.~\ref{eq:eta_lim_im} poses serious convergence challenges, especially for emission-driven decay processes. These challenges stem from the fact that the delta function approximations employed in calculating $\text{Im}\{K^{\rm{ph}}\}$ must fulfill two criteria. First, they must be smeared out enough to pick up on zero-crossings in $\Delta E$ sampled on a finite grid. In this sense, there exists a loose lower bound $\eta_{\rm \rm\mathbf{{k}}}$, such that choosing $\eta\ll\eta_{\rm \rm\mathbf{{k}}}$ will miss all zero crossings. This lower bound to choosing $\eta$ is well discussed in Refs.~\cite{giustino_electronphonon_2007, ponce_temperature_2015, giustino_electronphonon_2017, lihm_selfconsistent_2024}, and can be thought of as representing the energy spacing between neighboring $\rm \rm\mathbf{{k}}$ points. Fig.~\ref{fig:eta_cartoon} (a) and (b) show scenarios where $\eta$ is well below and comparable to $\eta_{\rm \rm\mathbf{{k}}}$ respectively.

The second criteria a delta function approximation has to satisfy for these calculations is that it needs to suppress contributions to $\text{Im}\{K^{\rm{ph}}\}$ which come from values of $\Delta E$ that are much greater than $\eta$. This constraint can be more quantitatively established by considering how a matrix element $|M(\rm{\mathbf{q}})|^2$ at some momentum $\rm \rm\mathbf{{q}}$ (with a corresponding energy denominator $\Delta E(\rm{\mathbf{q}})\neq0$) looks when it is multiplied by a delta function approximation function $p(\Delta E,\eta)$ where $\lim_{\eta\rightarrow0}p(\Delta E,\eta)=\delta(\Delta E)$. Specifically, when comparing the matrix element value at a momentum $\rm{\mathbf{q}}_0$ where $\Delta E(\rm{\mathbf{q}}_0)=0$ we require that
\begin{equation}
    p(\Delta E(\rm{\mathbf{q}}),\eta)|M(\rm{\mathbf{q}})|^2\ll p(0,\eta)|M(\rm{\mathbf{q}}_0)|^2,\quad \forall \Delta E(\rm{\mathbf{q}})\gg\eta.
\end{equation}
For many electron-phonon processes, this is trivially satisfied since the matrix elements being selected by the delta function are around $|\rm{\mathbf{q}}|\sim0$ where electron-phonon matrix elements are peaked. However, in the case of indirect gap systems especially when considering phonon emission, the momentum of phonons under consideration is not near $|\rm{\mathbf{q}}|\sim0$ and the coupling matrix elements under consideration are possibly much smaller than those around $|\rm{\mathbf{q}}|\sim0$. In this case, it becomes possible that there exists an $\eta_M$ choice for which
\begin{equation}
    \label{eq:eta_M_def}
    p(\Delta E(\rm{\mathbf{q}}),\eta_M)|M(\rm{\mathbf{q}})|^2=p(0,\eta_M)|M(\rm{\mathbf{q}}_0)|^2,\quad \Delta E(\rm{\mathbf{q}})\gg\eta_M
\end{equation}
In this light, $\eta_M$ provides a loose upper bound on $\eta$. Unless $\eta\ll\eta_M$, matrix elements far away from zero-crossings will erroneously contribute significantly to the imaginary part of $K^{\rm{ph}}$. Fig.~\ref{fig:eta_cartoon} (c) and (d) depict scenarios where $\eta\ll\eta_M$ and $\eta\sim\eta_M$ respectively. Putting these two bounds together, in order for a delta function approximation $p(\Delta E,\eta)$ to work as intended, the value of $\eta$ employed must satisfy the constraint
\begin{equation}
    \eta_{\rm\mathbf{{k}}}\lesssim\eta\lesssim\eta_M.
    \label{eq:eta_constraint}
\end{equation}

Ideally, $\eta_{\rm{\mathbf{k}}}\lesssim\eta_M$, but, in general, it is possible that $\eta_M\ll\eta_{\rm{\mathbf{k}}}$. In this case, there is no value of $\eta$ for which $\text{Im}\{K^{\rm{ph}}\}$ will converge. Thus, careful consideration must be taken to ensure that Eq.~\ref{eq:eta_constraint} is satisfied. Fig.~\ref{fig:eta_cartoon} depicts a scenario where $\eta_{\rm{\mathbf{k}}}/\eta_M\sim4$ for Lorentzian smearing and $\eta_{\rm{\mathbf{k}}}/\eta_M\sim1$ for Gaussian smearing.

Practically speaking, creating conditions where $\eta$ can be converged can be achieved by either reducing $\eta_{\rm{\mathbf{k}}}$ or increasing $\eta_M$. $\eta_{\rm{\mathbf{k}}}$ can be reduced by increasing the $k/q$-grid density, but for a three dimensional grid, increasing the density uniformly significantly increases the cost of solving the BSE and can be impractical. On the other hand, $\eta_M$ can be increased by changing the probability distribution function $p(\Delta E,\eta)$ used in place of $\delta(\Delta E)$. To elucidate this, Eq.~\ref{eq:eta_M_def} can be solved for Lorentzian vs Gaussian broadening. In the case of Lorentzian broadening, this gives
\begin{equation}
    \eta_M^{\text Gauss}=\frac{\Delta E(\rm{\mathbf{q}}_1)}{\sqrt{r_M-1}},
\end{equation}
where $r_M=\nicefrac{|M(\rm{\mathbf{q}}_1)|^2}{|M(\rm{\mathbf{q}}_0)|^2}$ is the ratio of the matrix elements at the peak of the peak of $|M(\rm{\mathbf{q}})|^2$ at $\rm{\mathbf{q}}=\rm\mathbf{{q}}_1$ versus the value at the zero crossing where $\rm{\mathbf{q}}=\rm{\mathbf{q}}_0$. This quantity is well-defined because, as discussed prior, for Eq.~\ref{eq:eta_M_def} to even be satisfied, $r_M$ must be greater than $1$. Similarly, in the case of Gaussian broadening this gives
\begin{equation}
    \eta_M^{\text Lorentz}=\frac{\Delta E(\rm{\mathbf{q}}_1)}{\sqrt{2\ln\left[r_M\right]}}.
\end{equation}

Written this way, $\eta_M^{\text Lorentz}>\eta_M^{\text Gauss}$ for cases where $r_M>3.513$, and therefore instead of increasing $\rm\mathbf{{k}}$-grid density Eq.~\ref{eq:eta_constraint} can potentially be satisfied by switching from Lorentzian to Gaussian broadening, especially when $r_M\gg1$. An example of this scenario can be seen in Fig.~\ref{fig:eta_cartoon} where Lorentzian broadening either fails to register the zero-crossing ((a) and (c)) or erroneously weights matrix contributions far away from the zero crossing ((b) and (d)), while Gaussian broadening is able to pick up on the zero crossing while simultaneously suppressing contributions far from said zero-crossing in (b) and (d).

\section{Single Particle electron-phonon Self-Energy Corrections}
In prior calculations of $K^{\rm{ph}}$ \cite{filip_phonon_2021,alvertis_phonon_2024}, the renormalizing effects of electron-phonon coupling on the single particle energy eigenvalues going into the BSE are neglected. This is because the lattice distortions accompanying and renormalizing the individual electron and hole states can partly or completely cancel each other out when forming an exciton as seen in \textit{ab initio} calculations by Refs.~\cite{dai_theory_2024, dai_excitonic_2024}; this polaron interference effect has also been observed in model calculations when the electron and hole polaron radii are comparable to the exciton radius \cite{mahanti_effective_1972, pollmann_effective_1977}.

\subsection{Exciton and Polaron Radii}
\label{subsec:radii} 
\begin{figure}[htbp!]
    \centering
    \includegraphics[width=0.6\linewidth]{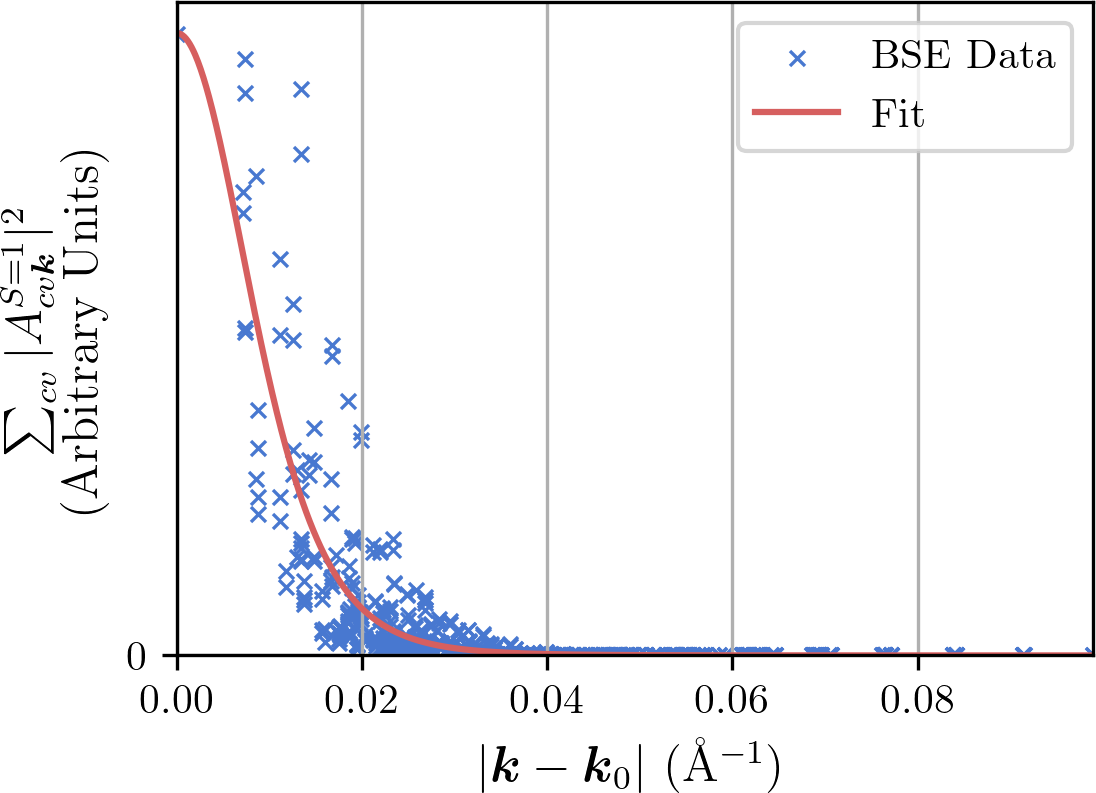}
    \\[-0.3cm]
    \caption{\centering Fit to the norm of the first exciton wavefunction using Eq.~\ref{eq:wan_mot_wfn}.}
    \label{fig:bohr_eff_fit}
\end{figure}

To approximately evaluate the extent to which interference may be present in m-BiVO$_4$, we first calculate the Bohr effective radius $a_{0}^{\text{eff}}$ of the first exciton as well as the polaron radii $r_{e}^p$, $r_h^p$ for the electron and hole. We calculate $a_0^{\text{eff}}$ by fitting the $\rm\mathbf{{k}}$ dependent exciton wavefunction data $\sum_{\rm{cv}}|A_{\rm{cv}\rm\mathbf{{k}}}^{S=1}|^2$ (obtained from solving the BSE) to the modulus squared of the Wannier-Mott model wavefunction
\begin{equation}
    \label{eq:wan_mot_wfn}
    \left|\psi_{\text ex}(\rm\mathbf{{k}})\right|^2=N \left(a_0^{\text{eff}}\right)^3\left(1+\left({a_0^{\text{eff}}}\left|\rm\mathbf{{k}}-\rm\mathbf{{k}}_0\right|\right)^2\right)^{-4}.
\end{equation}
Here, $N$ is a normalization factor, and $\rm\mathbf{{k}}_0$ is the $\rm\mathbf{{k}}$-point where the exciton is centered. The result of this fit for the wavefunction on a $24\times24\times32$ density path grid can be seen in Fig.~\ref{fig:bohr_eff_fit}; a value of $a_0^{\text{eff}}=7.6\;\angstrom$ is obtained.

We estimate polaron radii using the formula of Ref.~\cite{mahanti_effective_1972}:
\begin{equation}
    \label{eq:polaron_radius}
    r^p_{e,h}=\left(2m^*_{e,h}\omega_{LO}\right)^{-1/2}.
\end{equation}
Here, $m^*_{e,h}$ is taken to be the mean effective mass of either the electron or hole at $\rm\mathbf{{k}}_0$, and $\omega_{LO}$ is the highest energy LO phonon frequency. We take $\omega_{LO}$ directly from the phonon calculations needed to compute the el-ph coupling matrix elements, and find $\omega_{LO}=106.8$ meV.

The effective masses for the electron and hole, on the other hand, are obtained from fitting the the $G_0W_0$@PBE bands near $\rm\mathbf{{k}}_0$.
Specifically we fit the bands near $\rm\mathbf{{k}}_0$ to the following second order polynomial:
\begin{equation}
    \label{eq:eff_mass_fit}
    E(k_i)=\sum_i\left(a_i k_i + b_{ii}k_i^2 + \sum_{j>i}b_{ij}k_i k_j\right)+c_0
\end{equation}
with $a_i$, $b_ij$, and $c_0$ being the ten fit parameters. The Hessian matrix $h_{ij}$ corresponding to the inverse of the effective mass tensor is then constructed according to
\begin{equation}
    \label{eq:hessian}
    h_{ij}=\left(m^*\right)^{-1}_{ij}=\frac{\partial^2E}{\partial k_i\partial k_j}=
    \begin{bmatrix}
        2b_{xx} & b_{xy} & b_{xz} \\
        b_{xy} & 2b_{yy} & b_{yz} \\
        b_{xz} & b_{yz} & 2b_{zz} 
    \end{bmatrix}.
\end{equation}
From this, the effective masses are obtained by inverting and diagonalizing $h_{ij}$. These eigenvalues are reported in table \ref{tab:masses}, and the means of the direct gap effective masses for the valence and conduction band are also reported in Tab.~\ref{tab:pol_interfere}.

\begin{table}[htbp!]
    \centering
    \vspace{2mm}
    \setlength{\tabcolsep}{6pt}
    \caption{Effective masses of the electron and hole obtained by diagonalizing the Hessian of the $G_0W_0$@PBE electronic bands.}
    \begin{tabular}{c|ccc|c}
    \hline
    \multicolumn{1}{l|}{} & \multicolumn{3}{c|}{Eigenvalues ($m_e$)} & Mean ($m_e$) \\ \hline
    $m_h^*$ (Direct Gap) & 17.9 & 0.4 & 1.8 & 6.7 \\ \hline
    $m_e^*$ (Direct Gap) & 0.5 & 1.2 & 4.2 & 2.0 \\ \hline
    $m_h^*$ (VBM) & 0.9 & 0.6 & 0.4 & 0.6 \\ \hline
    $m_e^*$ (CBM) & 0.5 & 0.9 & 7.7 & 3.0 \\ \hline
    \end{tabular}
    \label{tab:masses}
\end{table}

\begin{table}[htbp!]
    \centering
    \caption{Effective masses and polaron radii for the electron and hole at $\rm\mathbf{{k}}_0$ (where the exciton forms) vs. the fitted Bohr effective radius for the first exciton.}
    \vspace{2mm}
    \setlength{\tabcolsep}{6pt}
    \begin{tabular}{cccc|c}
    \hline
    $m_e^*$ (a.u.) & $m_h^*$ (a.u.) & $r^p_e$ ($\angstrom$) & $r^p_h$ ($\angstrom$) & $a_0^{\text{eff}}$ ($\angstrom$) \\ \hline
    2.0 & 6.7 & 2.3 & 4.2 & 7.6 \\ \hline
    \end{tabular}
    \label{tab:pol_interfere}
\end{table}

As can be seen in Tab.~\ref{tab:pol_interfere}, the polarons corresponding to the highest valence and lowest conduction bands at $\rm\mathbf{{k}}_0$ are smaller than but comparable in magnitude to the exciton effective radius, suggesting that polaronic interference effects are present but not likely to fully cancel out the renormalizing effects of electron-phonon coupling for the single particle states in the BSE.

\subsection{Electron-Phonon Renormalization}
\label{subsec:el-ph_renorm}
With the results of \ref{subsec:radii} in mind, we also explicitly test the effectiveness of the polaronic interference approximation by computing the renormalization of the highest valence and lowest conduction bands using the \texttt{EPW} code and including these effects in the BSE. For a particular Bloch eigenstate $\ket{nk}$ and energy $E_{nk}$, the electron-phonon corrected eigenenergy $\breve{E}_{nk}$ is given by
\begin{equation}
    \label{eq:fmd_zpr_corr}
    \breve{E}_{n\rm\mathbf{{k}}}=E_{n\rm\mathbf{{k}}}+\braket{n\rm\mathbf{{k}}|\Sigma^{\rm{ph}}(E_{n\rm\mathbf{{k}}})|n\rm\mathbf{{k}}}
\end{equation}
where $\Sigma^{\rm{ph}}$ is the electron-phonon self-energy given by
\begin{equation}
    \label{eq:elph_sigma}
    \Sigma^{\rm{ph}}_{n\rm\mathbf{{k}}}(\omega,T)=\Sigma^{FMD}_{n\rm\mathbf{{k}}}(\omega,T)-\text{Re}\left\{\Sigma^{FMD}_{n\rm\mathbf{{k}}}(\omega=E_F,T)\right\}
\end{equation}
and
\begin{equation}
    \label{eq:elph_FMD_def}
    \begin{aligned}
    \Sigma^{FMD}_{n\rm\mathbf{{k}}}(\omega,T) = \frac{1}{N_q}\sum_{\pm\rm{m\mathbf{q}\nu}}\left|g_{\rm{m,\mathbf{k}+\mathbf{q},n,\mathbf{k},\nu}}\right|^2
    \Bigg[
    \frac{n_B\left(T,\omega_{\rm\mathbf{{q}}\nu}\right)\pm n_F\left(T,E_{m,\rm\mathbf{{k}}+\rm\mathbf{{q}}}\right)-(\pm\nicefrac{1}{2}-\nicefrac{1}{2})}{\omega-(E_{m,\rm\mathbf{{k}}+\rm\mathbf{{q}}}-E_F)\pm\omega_{\rm\mathbf{{q}}\nu}+i\eta}
    \Bigg].
    \end{aligned}
\end{equation}
Eq.~\ref{eq:elph_FMD_def} gives the form of the Fan Migdal (FMD) self-energy ($\Sigma^{FMD}$); in this expression, $E_F$ is the Fermi energy, $n_B(T,\omega)$ and $n_F(T,\omega)$ are the Bose Einstein and Fermi Dirac occupation factors. In Eq.~\ref{eq:elph_sigma}, the real part of the the FMD self-energy at the Fermi level is subtracted from the total self-energy to approximately account for higher order Debye Waller corrections to the electronic self-energy (as outlined in Refs.~\cite{ponce_temperature_2014, ponce_epw_2016}). This approach has been shown to accurately reproduce the actual Debye Waller corrections to within a few meV \cite{ponce_temperature_2014}.

We also compute the effects of including the electron-phonon renormalization of electronic states in the clamped ion BSE as
\begin{equation}
    \label{eq:zpr_BSE}
    \sum_{\rm{cv} \rm{\mathbf{k}}}\left[\breve{\Delta}_{c\rm\mathbf{{k}}v'\rm\mathbf{{k}}'}\delta_{\rm{cv}\rm\mathbf{{k}}\rm{c'v'}\rm\mathbf{{k}}'}+K^{\text{clamped-ion}}_{\rm{cv}\rm\mathbf{{k}}\rm{c'v'}\rm\mathbf{{k}}'}\right]\breve{A}^{\breve{S}}_{\rm{c'v'}\rm\mathbf{{k}}'}=\breve\Omega^{\breve{S}} \breve{A}^{\breve{S}}_{\rm{cv}\rm\mathbf{{k}}}.
\end{equation}
Here, $\breve{\Delta}_{c\rm\mathbf{{k}}v'\rm\mathbf{{k}}'}=\breve{E}_{c\rm\mathbf{{k}}}-\breve{E}_{v'\rm\mathbf{{k}}'}$ is analogous to the single particle energy transition in Eq.\ 1 of the main text, but with the electronic states being renormalized via the electron-phonon self-energy $\Sigma^{\rm{ph}}$.
These effects can all then be incorporated into $K^{\rm{ph}}$ via an expression that is analogous to Eq. 4 of the main text, 
\begin{equation}
\begin{aligned}
\label{eq:Kph_with_FMD}
    \breve{K}^{\rm{ph}}_{\breve{S}\breve{S}'}(\Omega,T)=-\sum_{\substack{\rm{\nu cv}\rm\mathbf{{k}}\\\rm{c'v'}\rm\mathbf{{k}}'}}&\left(\breve{A}^{\breve{S}}_{\rm{cv}\rm\mathbf{{k}}}\right)^*\breve{A}^{\breve{S}'}_{\rm{c'v'}\rm\mathbf{{k}}'} g_{\rm{c\mathbf{k}c'\mathbf{k}'\nu}}g_{\rm{v\mathbf{k}v'\mathbf{k}'\nu}}^*\times\\
    &\left[\frac{1+n_B(T,\omega_{\rm\mathbf{{k}}-\rm\mathbf{{k}}',\nu})}{\Omega-\breve{\Delta}_{c\rm\mathbf{{k}}v'\rm\mathbf{{k}}'}-\omega_{\rm\mathbf{{k}}-\rm\mathbf{{k}}',\nu}+i\eta}\right.
    \left.+\frac{1+n_B(T,\omega_{\rm\mathbf{{k}}-\rm\mathbf{{k}}',\nu})}{\Omega-\breve{\Delta}_{c'\rm\mathbf{{k}}'v\rm\mathbf{{k}}}-\omega_{\rm\mathbf{{k}}-\rm\mathbf{{k}}',\nu}+i\eta}\right.\\
    &\left.+\frac{n_B(T,\omega_{\rm\mathbf{{k}}-\rm\mathbf{{k}}',\nu})}{\Omega-\breve{\Delta}_{c\rm\mathbf{{k}}v'\rm\mathbf{{k}}'}+\omega_{\rm\mathbf{{k}}-\rm\mathbf{{k}}',\nu}+i\eta}
    +\frac{n_B(T,\omega_{\rm\mathbf{{k}}-\rm\mathbf{{k}}',\nu})}{\Omega-\breve{\Delta}_{c'\rm\mathbf{{k}}'v\rm\mathbf{{k}}}+\omega_{\rm\mathbf{{k}}-\rm\mathbf{{k}}',\nu}+i\eta}\right].
\end{aligned}
\end{equation}
This can then be solved self-consistently via
\begin{equation}
    \label{eq:BW_FMD}
    \tilde{\breve{\Omega}}^{\breve{S}}=\breve{\Omega}^{\breve{S}}+\breve{K}^{\rm{ph}}_{\breve{S},\breve{S}}\left(\Omega=\text{Re}\left\{\tilde{\breve{\Omega}}^{\breve{S}}\right\},T\right)
\end{equation}
to find a corrected BSE eigenvalue $\tilde{\breve{\Omega}}^{\breve{S}}$ which includes the effects of both $K^{\rm{ph}}$ and electron-phonon renormalization of the single-particle electronic states.

As seen in Tab.~\ref{tab:el-ph_corr}, at zero temperature at the electronic level, the direct band gap $E_g$ where the exciton forms is reduced by $340$ meV due to the corrections from Eq.~\ref{eq:fmd_zpr_corr}, and at room temperature, the direct band gap is renormalized by $933$ meV, a result that is quite large but also comparable to what was reported in Ref.~\cite{wiktor_comprehensive_2017}.

We still observe a strong exciton peak for the lowest-lying exciton in the solution to Eq.~\ref{eq:zpr_BSE}. The effects of electron-phonon renormalization on exciton energies are shown in in Tab.~\ref{tab:el-ph_corr}. At $0$ K, the size of $K^{\rm{ph}}$ corrections to the real part of the exciton energy are identical, while at $300$ K, they go from $117$ meV to $69$ meV. Importantly, we observe that the effects of electron-phonon corrections to the single-particle states is closely approximated by simply applying a scissor shift based on the direct gap renormalization to the exciton energies.

\begin{table}[htbp!]
    \centering
    \caption{Electron-phonon coupling corrections and their effects on electronic and excitonic energies at $0$ and $300$ K. $\Delta E_g$ denotes the correction to the lowest direct band gap due to electron-phonon interactions. $\Delta \Omega^S$ and $\Delta \tilde{\Omega}^S$ denote the corrections due to electron-phonon coupling in the single particle eigenvalues to the exciton energy with and without $K^{\rm{ph}}$ included respectively. Finally, $\Delta E_B$ and $\Delta \breve{E}_B$ denote the size of $K^{\rm{ph}}$ corrections to the exciton energy with and without electron-phonon coupling included in the single particle eigenvalues respectively.}
    \vspace{2mm}
    \label{tab:el-ph_corr}
    \setlength{\tabcolsep}{5pt}
        \begin{tabular}{c|ccc|cccccc|cc}
        \multirow{2}{*}{T (K)} & \multicolumn{3}{c|}{Direct Band Gap (eV)} & \multicolumn{6}{c|}{Exciton Energy (eV)} & \multicolumn{2}{c}{\begin{tabular}[c]{@{}c@{}}Exciton Binding\\ Energy (meV)\end{tabular}} \\
         & $E_g$ & $\breve{E}_g$ & $\Delta E_g$ & $\Omega^S$ & $\breve{\Omega}^S$ & $\tilde{\Omega}^S$ & $\tilde{\breve{\Omega}}^S$ & $\Delta \Omega^S$ & $\Delta \tilde{\Omega}^S$ & $\Delta E_B$ & $\Delta \breve{E}_B$ \\ \hline
        0 & \multirow{2}{*}{3.448} & 3.108 & -0.340 & \multirow{2}{*}{3.351} & 3.001 & 3.391 & 3.041 & -0.350 & -0.350 & -40 & -40 \\
        300 &  & 2.515 & -0.933 &  & 2.462 & 3.468 & 2.531 & -0.889 & -0.937 & -117 & -69
        \end{tabular}
\end{table}



The imaginary part of $\tilde{\breve{{\Omega}}}^{\breve{S}}$, however, is smaller than that of $\tilde{{\Omega}}^{\breve{S}}$. At $0$ K we calculate it to be $0.0007$ meV, corresponding to a lifetime of $\sim500$ ps. This is two orders of magnitude larger than the value reported in the main text at $0$ K. We attribute this increase to the less favorable energetic conditions for emission-driven dissociation; specifically, the exciton binding energy is increased by an additional $10$ meV. At $300$ K we calculate the imaginary part of $\tilde{\breve{{\Omega}}}^{\breve{S}}$ to be $43$ meV, corresponding to a lifetime of $8$ fs, a value which is slightly larger but still the same order of magnitude as the $3$ fs lifetime reported in the main text.



Electron-phonon calculations were carried out in the \texttt{EPW} code which computes both a Fan-Migdal and an approximate Debeye-Waller correction to the electronic states. Calculations at both $T=0,300$ K were done with a broadening of $15$ meV using a $12\times12\times16$ $\rm\mathbf{{q}}$-grid. An energy window of $10$ eV about the valence band maximum was employed for all calculations.

\section{Dissociation Conditions}
\label{sec:diss_cond_sm}
Here, we discuss the physical constraints of absorption-driven decay processes in the cases of both direct and indirect band gaps.
\begin{figure}[htbp!]
    \centering
    \begin{subfigure}[]{0.45\linewidth}
        \begin{overpic}[width=\linewidth]{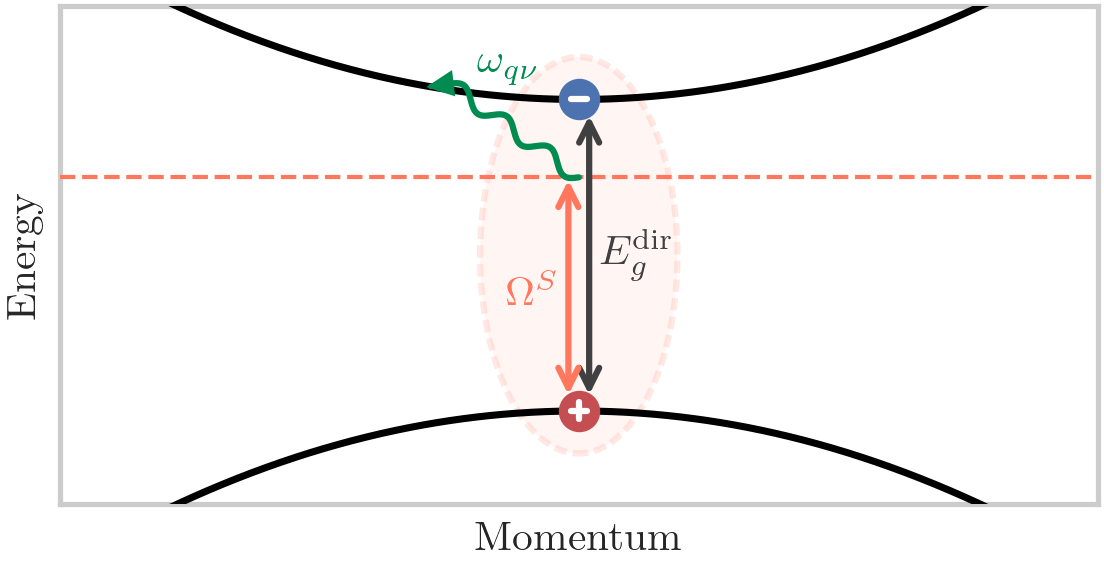}
            \put(-3,48){\small \textbf{(a)}}
        \end{overpic}
        \label{fig:dir_em_cartoon}
    \end{subfigure}
    ~
    \begin{subfigure}[]{0.45\linewidth}
        \begin{overpic}[width=\linewidth]{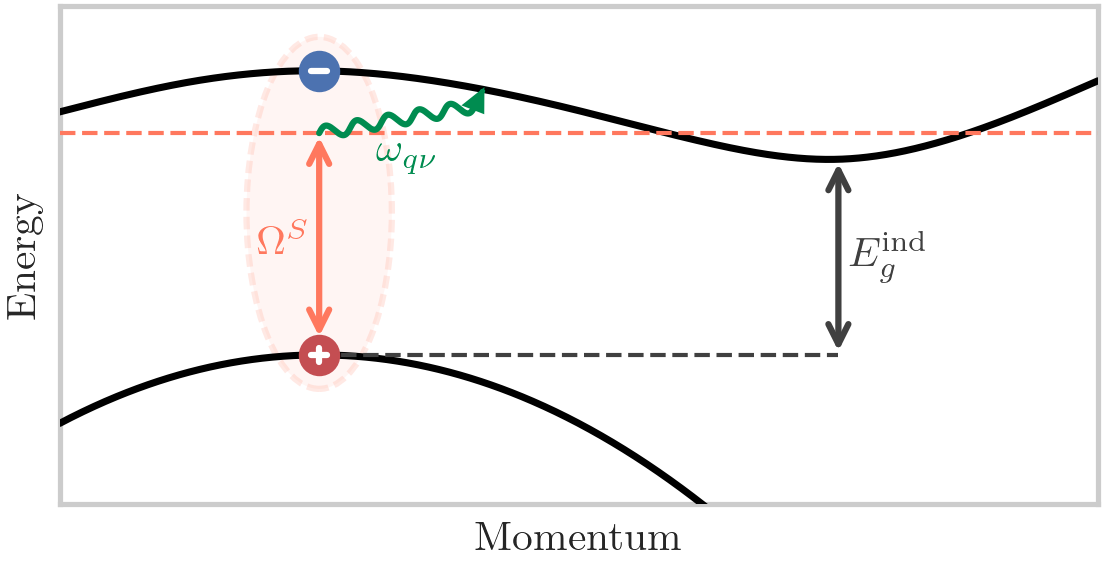}
            \put(-3,48){\small \textbf{(b)}}
        \end{overpic}
        \label{fig:dir_abs_cartoon}
    \end{subfigure}
    \\[-0.7cm]
    \caption{Schematic depictions of dissociation via phonon emission for a direct--\textbf{(a)} and indirect--\textbf{(b)} band gap system.}
    \label{fig:direct_dissociation}
\end{figure}

The linewidths and lifetimes arising out of $K^{\rm{ph}}$ associated with dissociation processes are given by Eq.~$4$ in the main text. Assessing where a finite linewidth can exist can be done by considering the form of the RDOS in Eq.~\ref{eq:rdos_pm} from \ref{sec:DOS}. Focusing first on the absorption channel, the linewidth is $0$ unless the arguments in one of the two delta functions is zero, resulting in the two constraints:
\begin{align}
\label{eq:constr_ab_1}
    \omega_{\rm\mathbf{{k}}-\rm\mathbf{{k}}',\nu}&=(E_{\rm{c}\rm\mathbf{{k}}}-E_{\rm{v}'\rm\mathbf{{k}}'})-\tilde{\Omega}^S\\
\label{eq:constr_ab_2}
    \omega_{\rm\mathbf{{k}}-\rm\mathbf{{k}}',\nu}&=(E_{\rm{c}'\rm\mathbf{{k}}'}-E_{\rm{v}\rm\mathbf{{k}}})-\tilde{\Omega}^S.
\end{align}
Focusing on the Eq.~\ref{eq:constr_ab_1}, because $\omega_{\rm\mathbf{{k}}-\rm\mathbf{{k}}',\nu}$ is bounded from below by $0$ and from above by $\omega_{LO}^{\text max}$, there can only be absorption from this channel if
\begin{equation}
\label{eq:constr_ab_3}
    \omega_{LO}^{\text max}\geq(E_{\rm{c}\rm\mathbf{{k}}}-E_{\rm{v}'\rm\mathbf{{k}}'})-\tilde{\Omega}^S\geq0.
\end{equation}

In the event a system's fundamental electronic band gap is direct, and the state $\tilde{\Omega}^S$ lies below the gap, the rightmost constraint in Eq.~\ref{eq:constr_ab_3} is satisfied, and even in the event a BSE eigenvalue lies above the direct gap, there exist higher energy single particle states to scatter to. Addressing the left inequality, the minimum value of $(E_{\rm{c}\rm\mathbf{{k}}}-E_{\rm{v}'\rm\mathbf{{k}}'})-\tilde{\Omega}^S$ for a particular state $S$ in a direct gap system with a band gap of $E_g^{\text{dir}}$ is simply $E_g^{\text{dir}}-\tilde{\Omega}^S$. Putting this all together we arrive at the heuristic that absorption-driven decay in direct gap systems is only allowed if $\omega_{LO}^{\text{max}}\geq E_g^{\text{dir}}-\tilde{\Omega}^S$. This scattering process will first only be allowed for $\rm\mathbf{{q}}=0$ phonons when $E_g^{\text{dir}}-\tilde{\Omega}^S=\omega_{LO}^{\text{max}}$, and if $E_g^{\text{dir}}-\tilde{\Omega}^S$ is strictly less than $\omega_{LO}^{\text{max}}$, scattering involving phonons of finite $\rm\mathbf{{q}}$ becomes allowed too. Likewise, when the difference between $\omega_{LO}^{\text{max}}$ and $E_g^{\text{dir}}-\tilde{\Omega}^S$ is small, only high-energy phonons can be absorbed, but as the difference grows, the allowed phonon energy range does so as well.
Of particular note is that corrections from $K^{\rm{ph}}$ typically reduce the exciton binding energy vs. the clamped-ion regime, so it is possible for this channel to open up once phonon screening is introduced especially at higher temperatures where the binding energy is reduced by more.

If gap is indirect, the minimum value of $(E_{\rm{c}\rm\mathbf{{k}}}-E_{\rm{v}'\rm\mathbf{{k}}'})-\tilde{\Omega}^S$ for a particular state $S$ is $E_g^{\text{ind}}-\tilde{\Omega}^S$, where $E_g^{\text{ind}}$ is the indirect band gap. Thus, the previous heuristic for absorption-driven decay in direct gap systems can easily be extended to indirect gap systems with the constraint that $\omega_{LO}^{\text{max}}\geq E_g^{\text{dir}}-\tilde{\Omega}^S$.

\newpage
\section{Momentum-Resolved \texorpdfstring{$K^{\rm{ph}}$}{Kph}}
For completeness, we also compute and plot the momentum-resolved form of $K^{\rm{ph}}$ on a $36\times36\times48$ density grid for both the real and imaginary parts at $0$ and $300$ K. In the following plots, each pixel shows the contribution to $K^{\rm{ph}}$ corresponding to the all phonon modes with momentum $\rm\mathbf{{q}}=\rm\mathbf{{q}}_1+\rm\mathbf{{q}}_2+\rm\mathbf{{q}}_3$. Each panel shows the variation along the reciprocal lattice vectors $\rm{\mathbf{b}}_1$ and $\rm{\mathbf{b}}_2$ with $\rm{\mathbf{b}}_3$ held fixed.
\subsection{Re[\texorpdfstring{$K^{\rm{ph}}$}{Kph}]}
The contributions to the real part of $K^{\rm{ph}}$ are well contained inside of the chosen patch, with the strongest contributions coming from near $|\rm\mathbf{{q}}|=0$.

\begin{figure}[h!]
    \centering
    \includegraphics[width=0.78\linewidth]{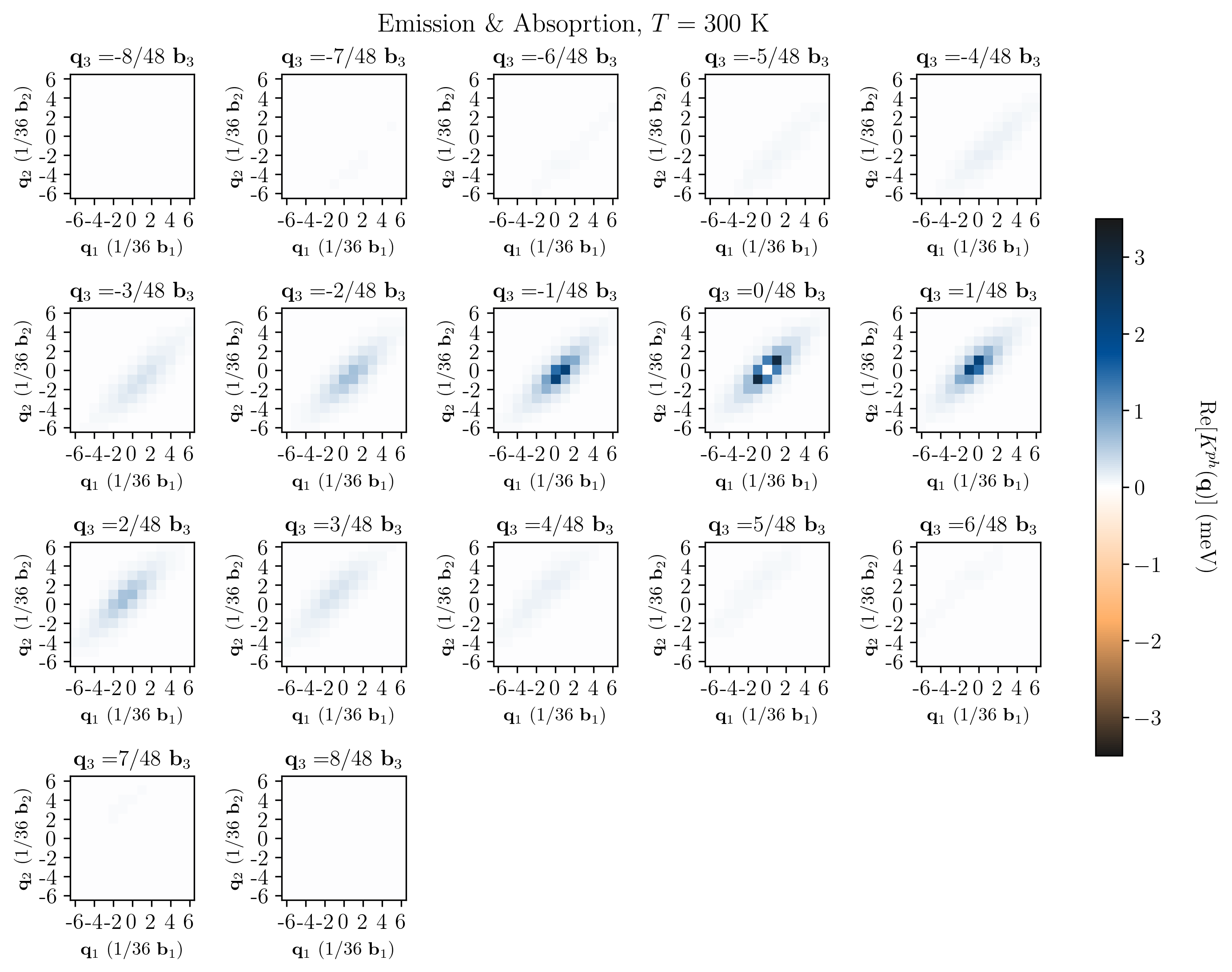}
    \\[-0.4cm]
    \caption{Momentum-resolved contributions to the real part of $K^{\rm{ph}}$ coming from both emission and absorption channels at room temperature}
    \label{fig:q_res_re_kph_tot_300}
\end{figure}

\begin{figure}[p!]
    \centering
    \includegraphics[width=0.78\linewidth]{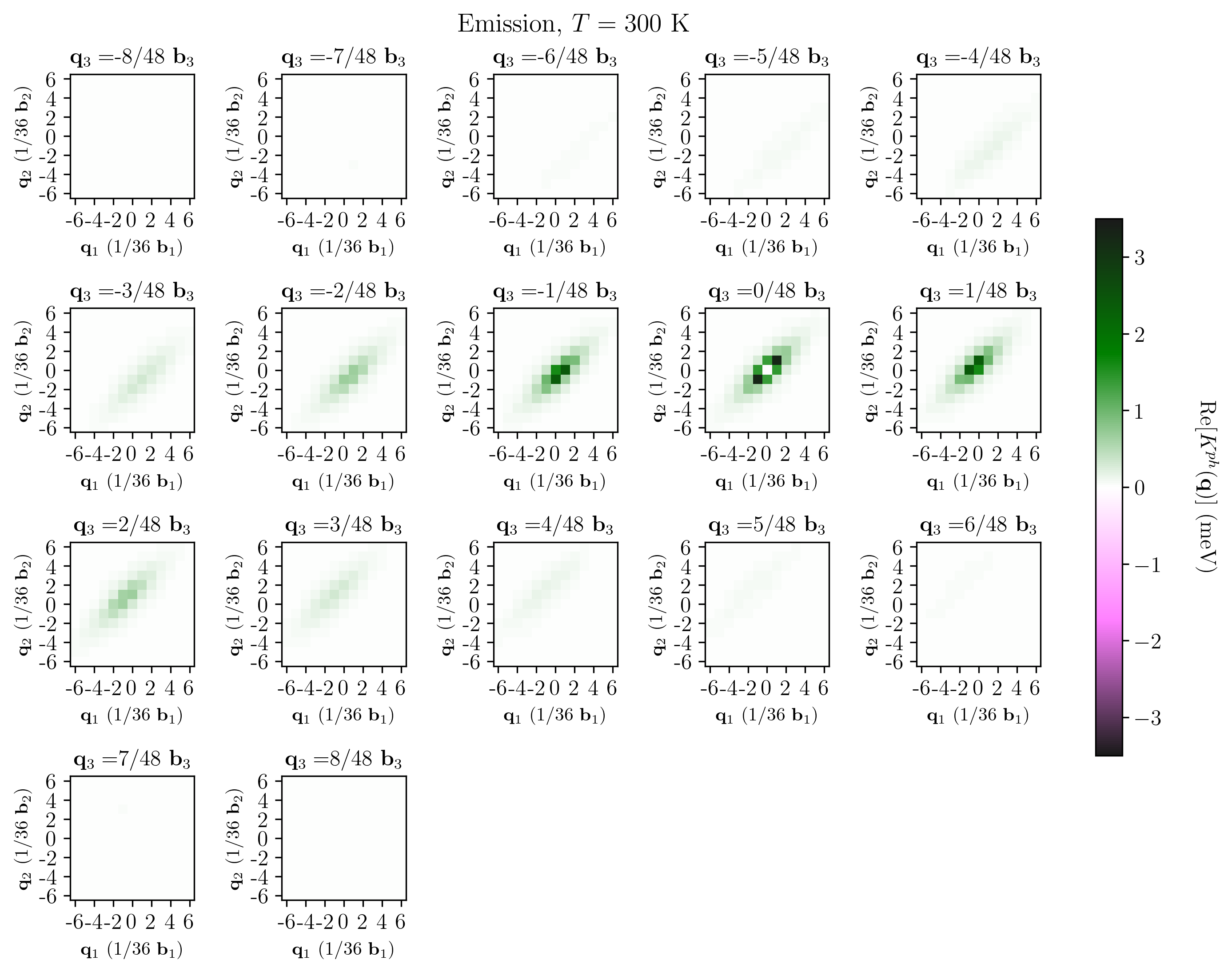}
    \\[-0.4cm]
    \caption{Momentum-resolved contributions to the real part of $K^{\rm{ph}}$ coming from only the emission channel at room temperature}
    \label{fig:q_res_re_kph_em_300}
\end{figure}

\begin{figure}[p!]
    \centering
    \includegraphics[width=0.78\linewidth]{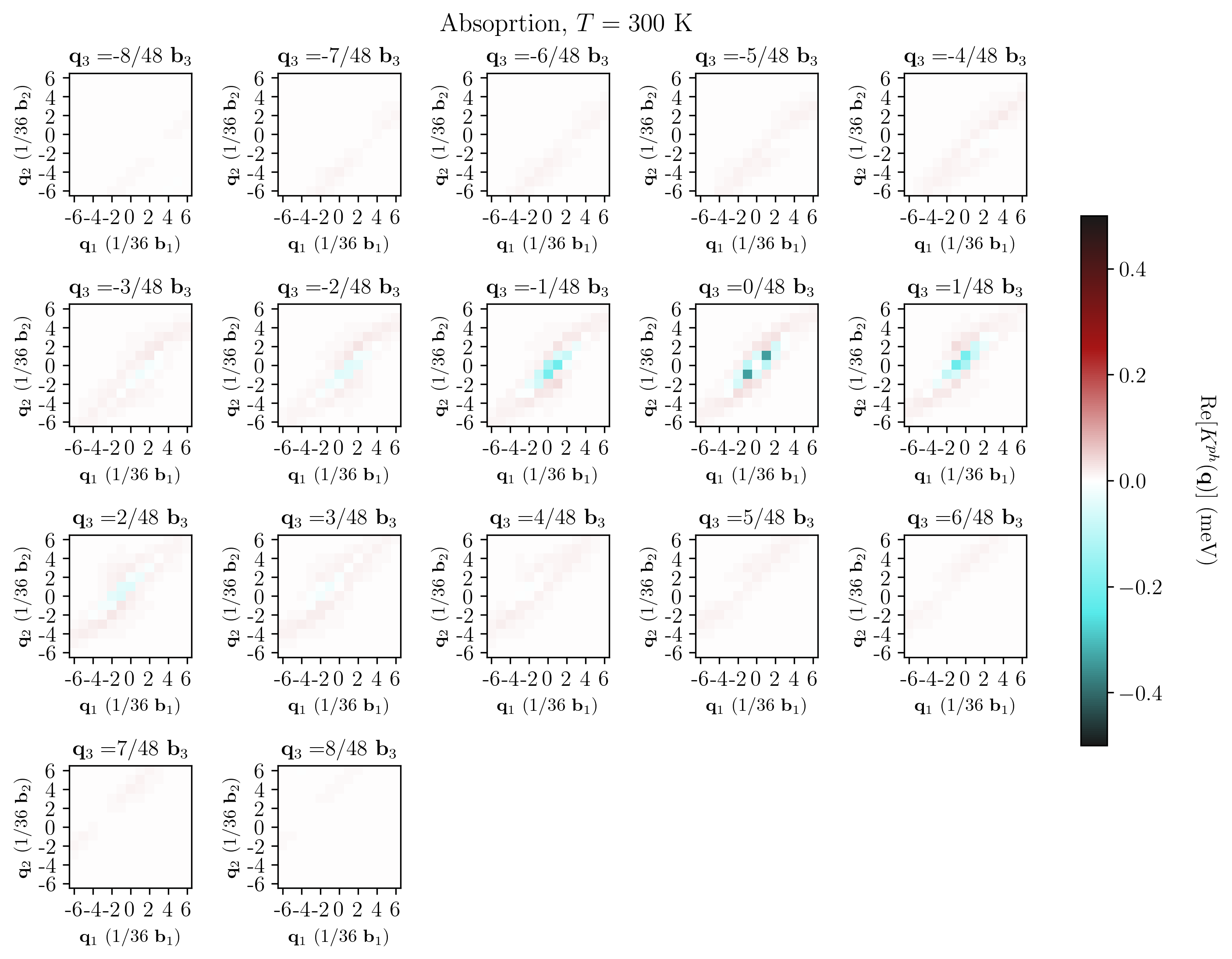}
    \\[-0.4cm]
    \caption{Momentum-resolved contributions to the real part of $K^{\rm{ph}}$ coming from only the absorption channel at room temperature}
    \label{fig:q_res_re_kph_ab_300}
\end{figure}

\begin{figure}[p!]
    \centering
    \includegraphics[width=0.78\linewidth]{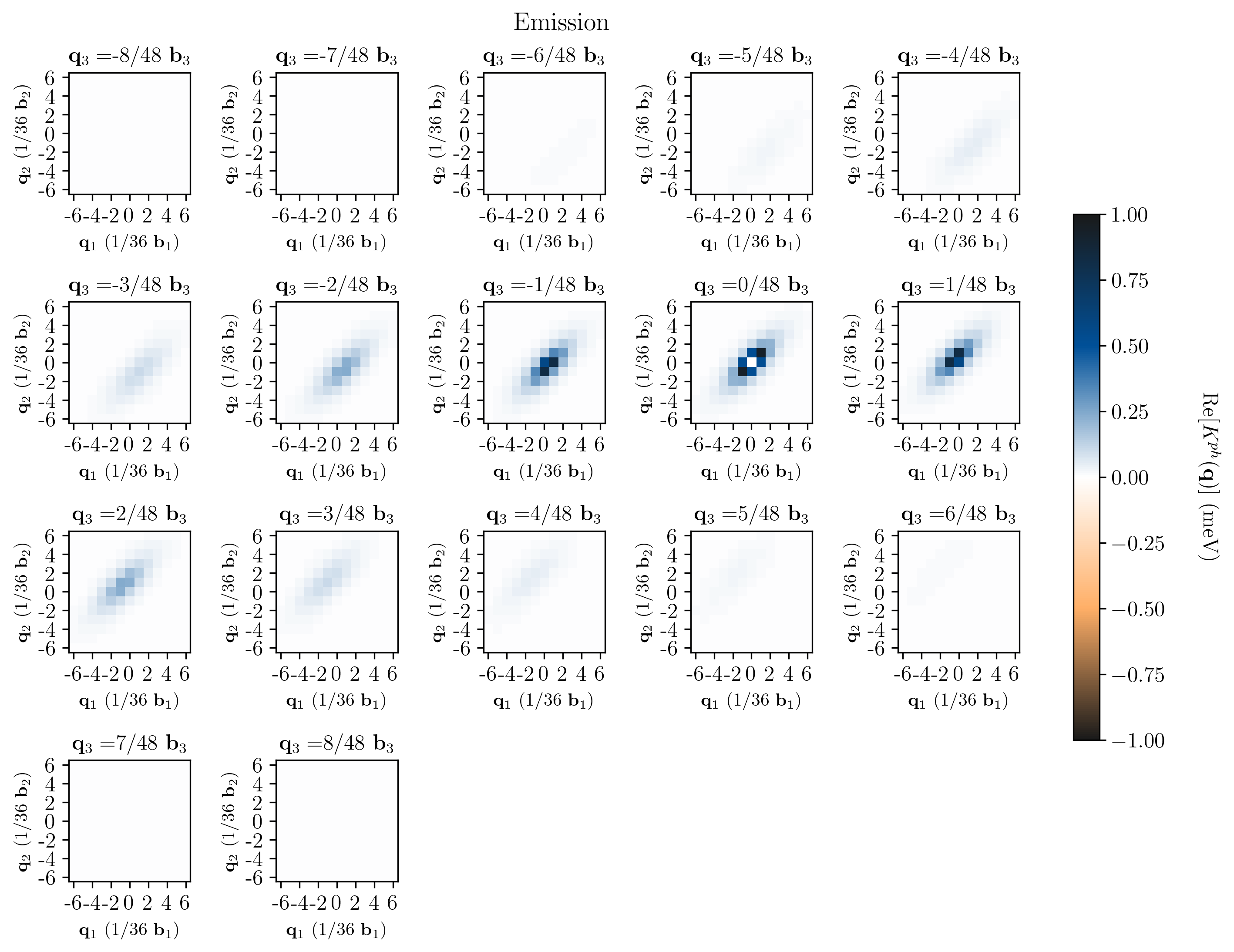}
    \\[-0.4cm]
    \caption{Momentum-resolved contributions to the real part of $K^{\rm{ph}}$ at $0$ K. Note that due to the temperature dependence of the absorption channel, only emission is present here.}
    \label{fig:q_res_re_kph_tot_0}
\end{figure}

\newpage
\subsection{Im[\texorpdfstring{$K^{\rm{ph}}$}{Kph}]}
\begin{figure}[p!]
    \centering
    \includegraphics[width=0.78\linewidth]{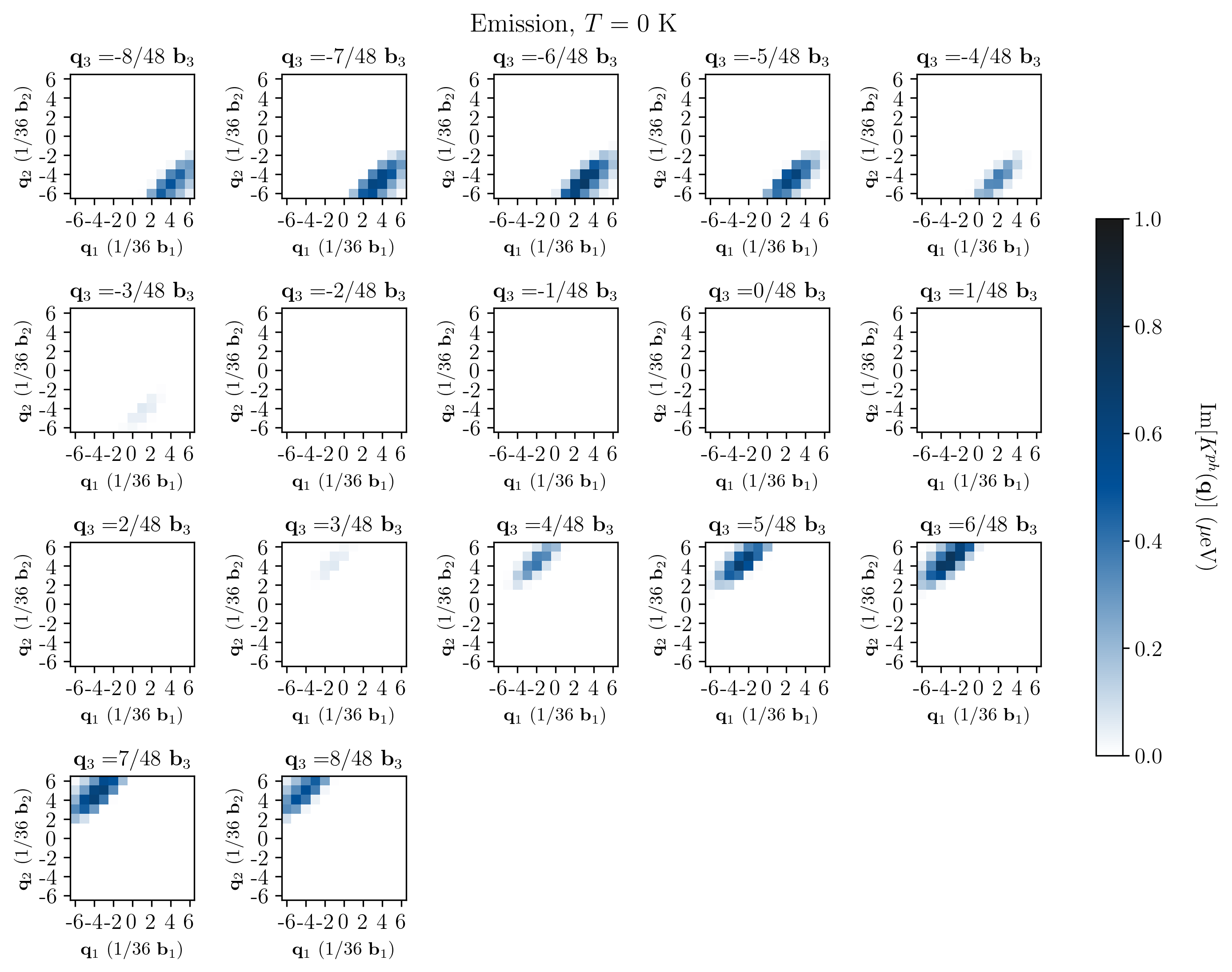}
    \\[-0.4cm]
    \caption{Momentum-resolved contributions to the imaginary part of $K^{\rm{ph}}$ at $0$ K. Note that due to the temperature dependence of the absorption channel, only emission is present here.}
    \label{fig:q_res_im_kph_tot_0}
\end{figure}

The contributions to the imaginary part of $K^{\rm{ph}}$ are well contained inside of the chosen patch at room temperature, with the strongest contributions coming from near $|\rm\mathbf{{q}}|=0$. Interestingly, the emission channel at room temperature is dominated by nearly direct transitions with $|\rm\mathbf{{q}}|\sim0$. As discussed in \ref{sec:diss_cond_sm}, this only occurs when ${\tilde{\Omega}^S}$ is greater than the direct gap as occurs for the $S=1$ exciton in BiVO$_4$ at room temperature. At zero temperature, only the emission channel is active, and it's strength is approximately three orders of magnitude weaker than at room temperature. This is due to indirect gap scattering with necessarily non-zero $|\rm\mathbf{{q}}|$ dominating. Fig.~\ref{fig:q_res_im_kph_tot_0} shows that these contributions are not fully contained in the patch, suggesting that the low-temperature dissociation lifetime may be slightly overestimated with respect to the size of the patch.

\begin{figure}[h!]
    \centering
    \includegraphics[width=0.78\linewidth]{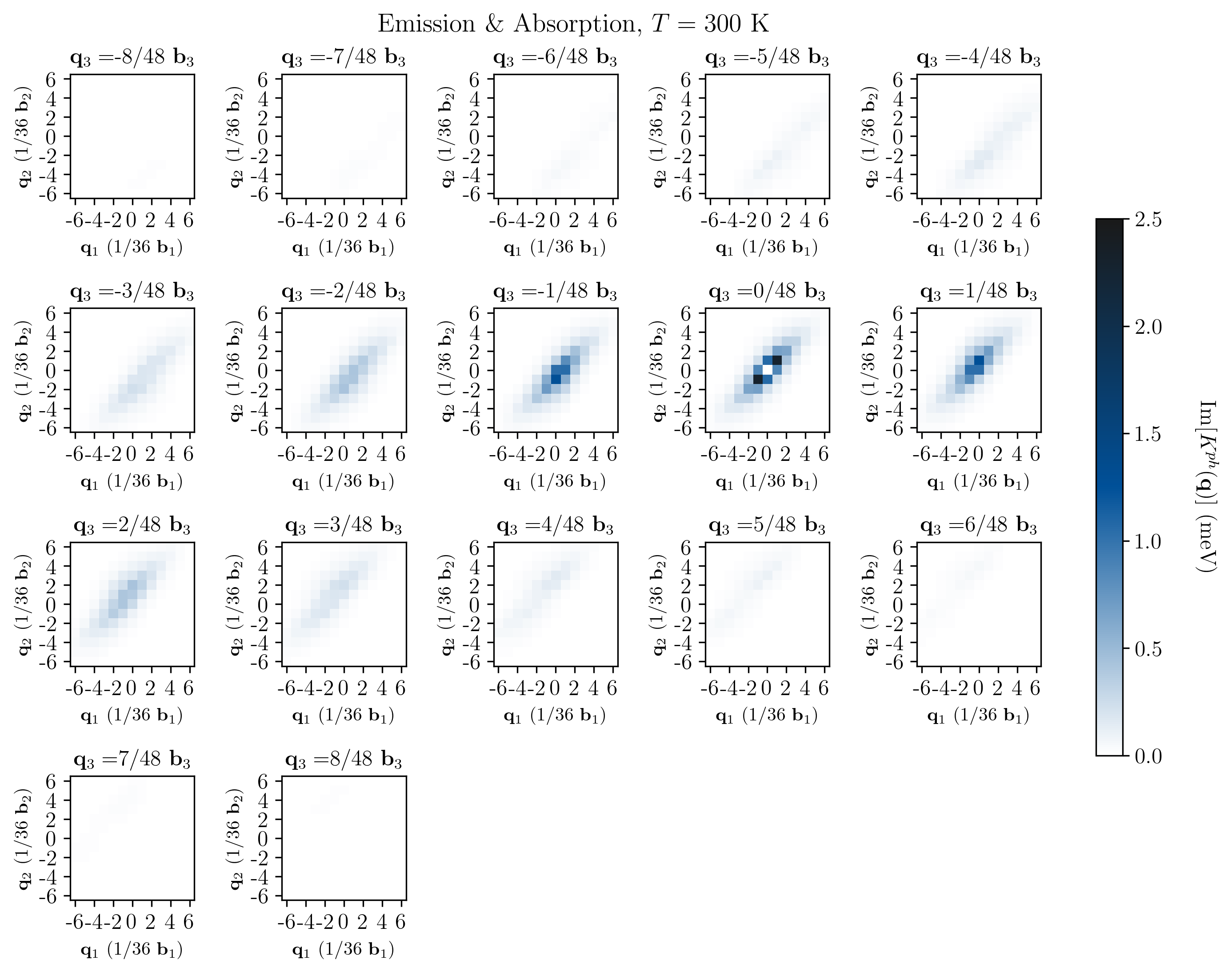}
    \\[-0.4cm]
    \caption{Momentum-resolved contributions to the imaginary part of $K^{\rm{ph}}$ coming from both the emission absorption channels at room temperature}
    \label{fig:q_res_im_kph_tot_300}
\end{figure}

\begin{figure}[p!]
    \centering
    \includegraphics[width=0.78\linewidth]{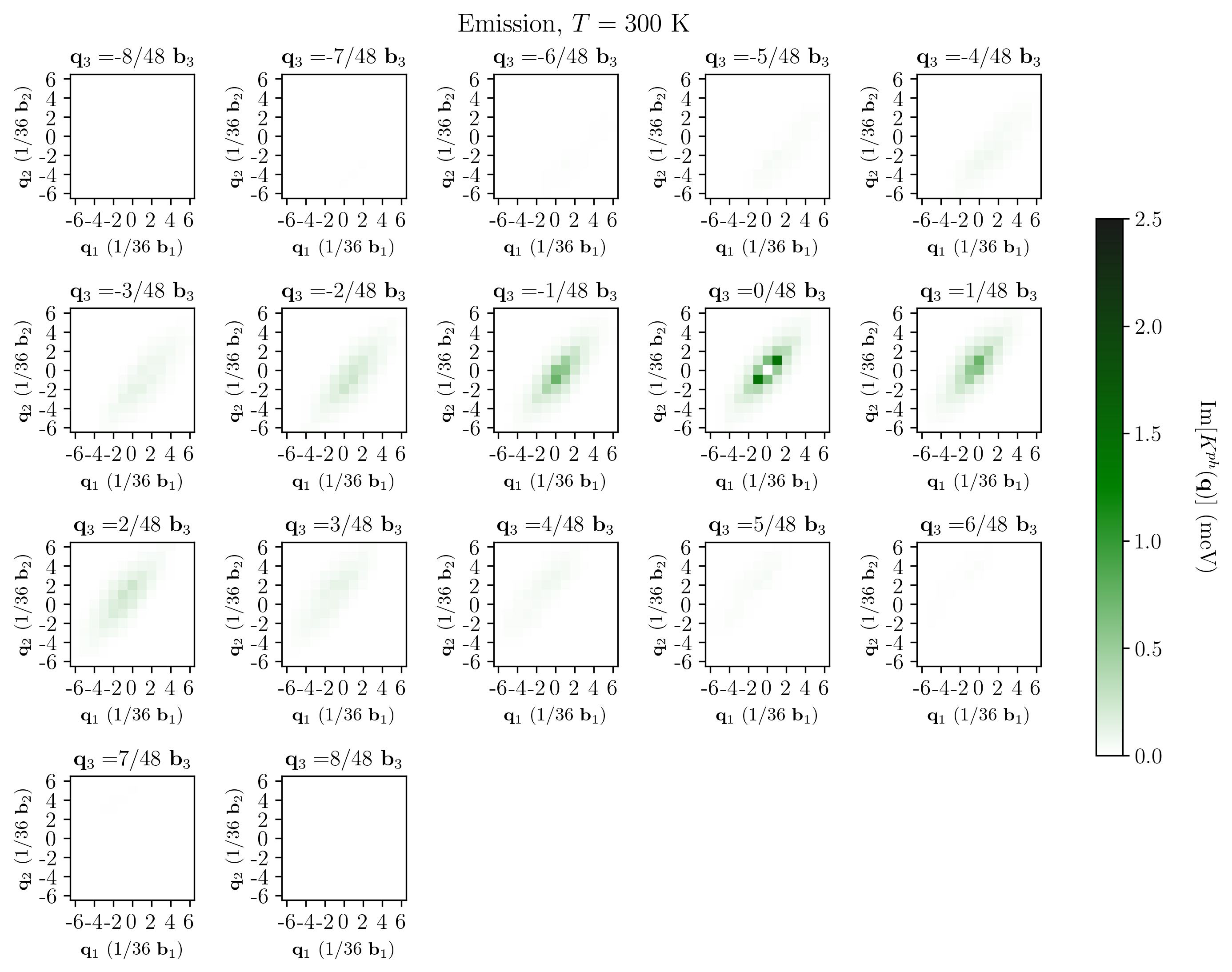}
    \\[-0.4cm]
    \caption{Momentum-resolved contributions to the real part of $K^{\rm{ph}}$ coming from only the absorption channel at room temperature}
    \label{fig:q_res_im_kph_em_300}
\end{figure}

\begin{figure}[p!]
    \centering
    \includegraphics[width=0.78\linewidth]{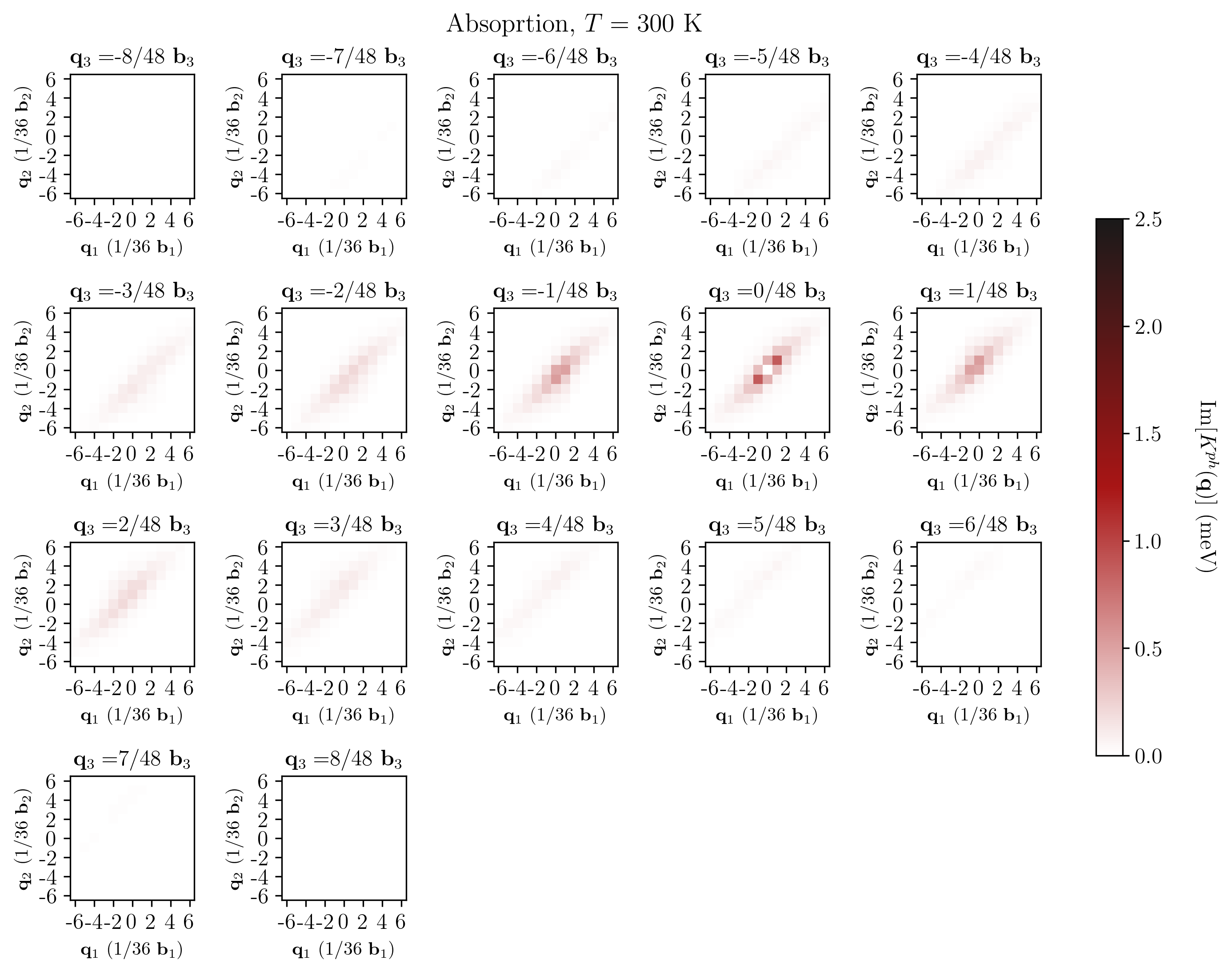}
    \\[-0.4cm]
    \caption{Momentum-resolved contributions to the imaginary part of $K^{\rm{ph}}$ coming from only the absorption channel at room temperature}
    \label{fig:q_res_im_kph_ab_300}
\end{figure}

\newpage
\section{Phonon Mode Contributions to \texorpdfstring{$K^{\rm{ph}}$}{Kph}}
\label{sec:DOS}
In order to offer greater insight into the individual phonon mode contributions to $K^{\rm{ph}}$ and to provide further analysis of the separated emission (EM) and absorption (AB) channels, we compute the spectral density $D$ of the real and imaginary parts of $K^{\rm{ph}}$ separately for each channel. Specifically, these spectral densities are given by
\begin{equation}
\label{eq:D_Re_Kph_T_pm}
    D^{\pm}_{\text{Re}\{K^{\rm{ph}}_{S,S'}\}}(\omega,\Omega,T)=
    \sum_{\rm{\nu}}
    \text{Re}\left\{\sum_{\rm{ cv}\rm\mathbf{{k}}\rm{c'v'}\rm\mathbf{{k}}'}
    \left(A^S_{\rm{cv}\rm\mathbf{{k}}}\right)^*A^{S'}_{\rm{c'v'}\rm\mathbf{{k}}'}K^{\rm{ph},\pm}_{\rm{\nu cv}\rm\mathbf{{k}}\rm{c'v'}\rm\mathbf{{k}}'}(\Omega,T)
    \delta(\omega-\omega_{\rm\mathbf{{k}}-\rm\mathbf{{k}}',\nu})
    \right\}
\end{equation}
where the $+$ and $-$ channels denote absorption and emission channels respectively and where $K^{\rm{ph},\pm}_{\rm{\nu cv}\rm\mathbf{{k}}\rm{c'v'}\rm\mathbf{{k}}'}(\Omega,T)$ are the mode-and channel-resolved phonon kernels defined as
\begin{equation}
\label{eq:Kph_T_pm}
    K^{\rm{ph},\pm}_{\rm{\nu cv}\rm\mathbf{{k}}\rm{c'v'}\rm\mathbf{{k}}'}(\Omega,T)=
    g_{\rm{c\mathbf{k}c'\mathbf{k}'\nu}}g_{\rm{v\mathbf{k}v'\mathbf{k}'\nu}}^*
    \left[\frac{\pm\nicefrac{1}{2}-\nicefrac{1}{2}-n_B(T,\omega_{\rm\mathbf{{k}}-\rm\mathbf{{k}}',\nu})}{\Omega-\Delta_{c\rm\mathbf{{k}}v'\rm\mathbf{{k}}'}\pm\omega_{\rm\mathbf{{k}}-\rm\mathbf{{k}}',\nu}+i\eta}
    +\frac{\pm\nicefrac{1}{2}-\nicefrac{1}{2}-n_B(T,\omega_{\rm\mathbf{{k}}-\rm\mathbf{{k}}',\nu})}{\Omega-\Delta_{c'\rm\mathbf{{k}}'v\rm\mathbf{{k}}}\pm\omega_{\rm\mathbf{{k}}-\rm\mathbf{{k}}',\nu}+i\eta}\right]
\end{equation}
The combination of these two terms, summed over all modes gives the standard electron-hole basis representation of the temperature-dependent phonon kernel according to
\begin{equation}
\label{eq:Kph_T_mode}
    K^{\rm{ph}}_{\rm{cv}\rm\mathbf{{k}}\rm{c'v'}\rm\mathbf{{k}}'}(\Omega,T)=\sum_{\nu} \left[K^{\rm{ph},+}_{\rm{\nu cv}\rm\mathbf{{k}}\rm{c'v'}\rm\mathbf{{k}}'}(\Omega,T)+K^{\rm{ph},-}_{\rm{\nu cv}\rm\mathbf{{k}}\rm{c'v'}\rm\mathbf{{k}}'}(\Omega,T)\right].
\end{equation}
By design, adding both quantities and integrating over $\omega$ yields back the original total value of $\text{Re}\{K^{\rm{ph}}_{S,S'}\}(\Omega,T)$,
\begin{equation}
    \int_0^\infty \left[D^{+}_{\text{Re}\{K^{\rm{ph}}_{S,S'}\}}(\omega,\Omega,T)+D^{-}_{\text{Re}\{K^{\rm{ph}}_{S,S'}\}}(\omega,\Omega,T)\right] d\omega=\text{Re}\{K^{\rm{ph}}_{S,S'}\}(\Omega,T).
\end{equation}

\begin{figure}[htbp!]
    \centering
    \includegraphics[width=0.6\linewidth]{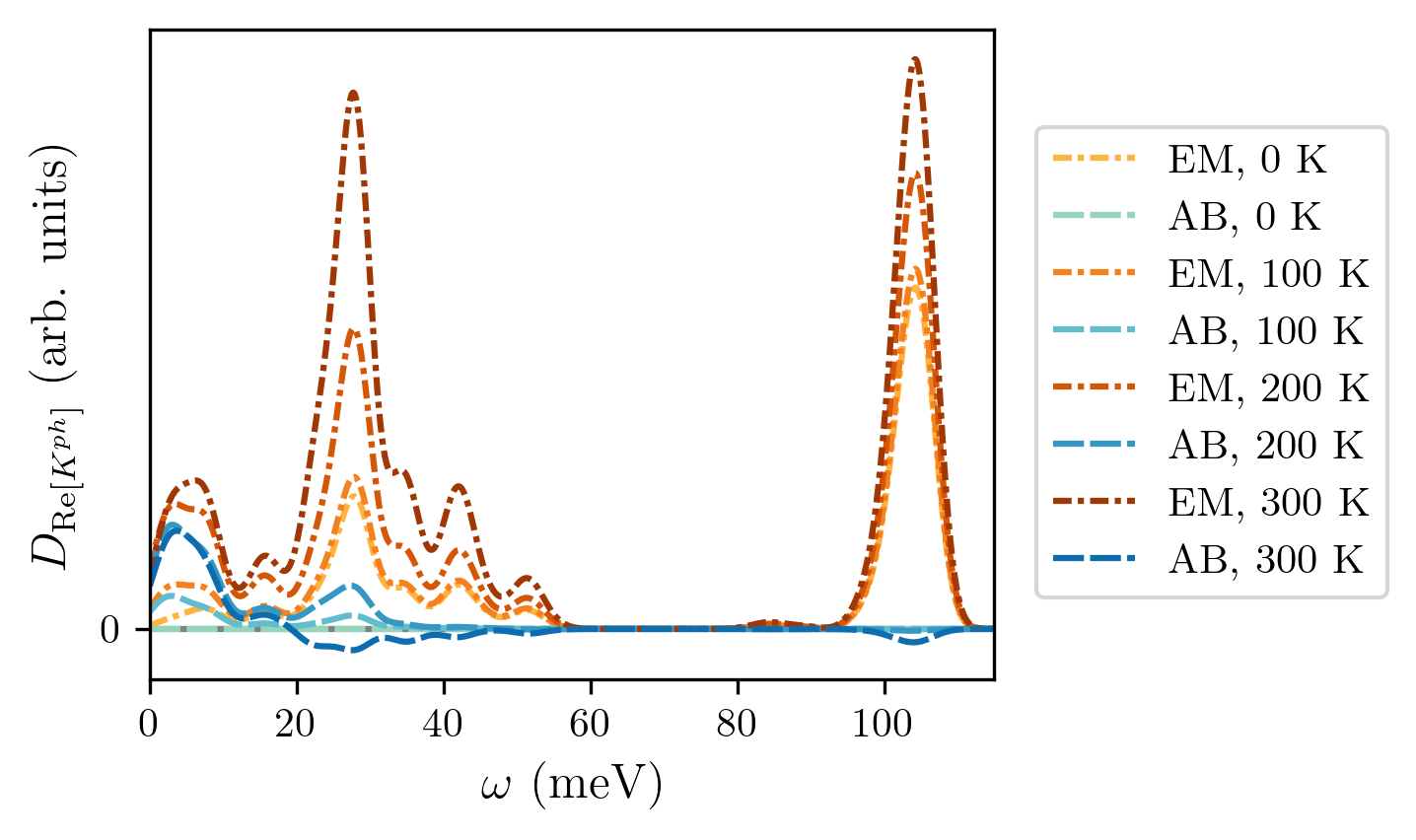}
    \caption{Phonon frequency resolved contributions to the real part of $K^{\rm{ph}}$ for the first exciton. Lower energy modes as well as the highest LO modes can be seen playing an important role in the renormalization of the exciton binding energy.}
    \label{fig:dos_kph_re}
\end{figure}

The emission and absorption spectral densities of the real part for the first exciton (i.e. $S=S'=1$ and $\Omega=\tilde{\Omega}^{S=1}$) are plotted in Fig.~\ref{fig:dos_kph_re}. At zero temperature, the contributions from the absorption channel can be seen in Fig.~\ref{fig:dos_kph_re} to be $0$ as expected, and they begin to become non-zero at finite temperatures as the Bose-Einstein occupation factors for each phonon grow. Interestingly, we observe that while the sign of the emission driven processes remains positive for all frequencies, the sign of the contributions from absorption driven processes flips for higher frequency modes at higher temperatures. This trend corresponds with what one would expect as $\Omega$ increases for the absorption terms in Eq.~\ref{eq:Kph_T_pm}, where the sign in front of $\omega_{\rm\mathbf{{k}}-\rm\mathbf{{k}}',\nu}$ flips. Additionally, as previous models have suggested \cite{filip_phonon_2021}, the highest LO modes, which are just above $100$ meV for BiVO$_4$, make up a substantial fraction of the total weight of $D^{-}_{\text{Re}\{K^{\rm{ph}}\}}$. However, we observe here that many phonons between $20$ to $60$ meV also couple strongly and play a significant role in the screening of excitons such that by room temperature, these modes contribute just as much as the higher LO ones. This highlights the importance of \textit{ab initio} calculations for understanding the full contribution to the phonon kernel in real materials.

\begin{figure}[htbp!]
    \centering
    \begin{subfigure}[]{0.48\linewidth}
        \begin{overpic}[width=\linewidth]{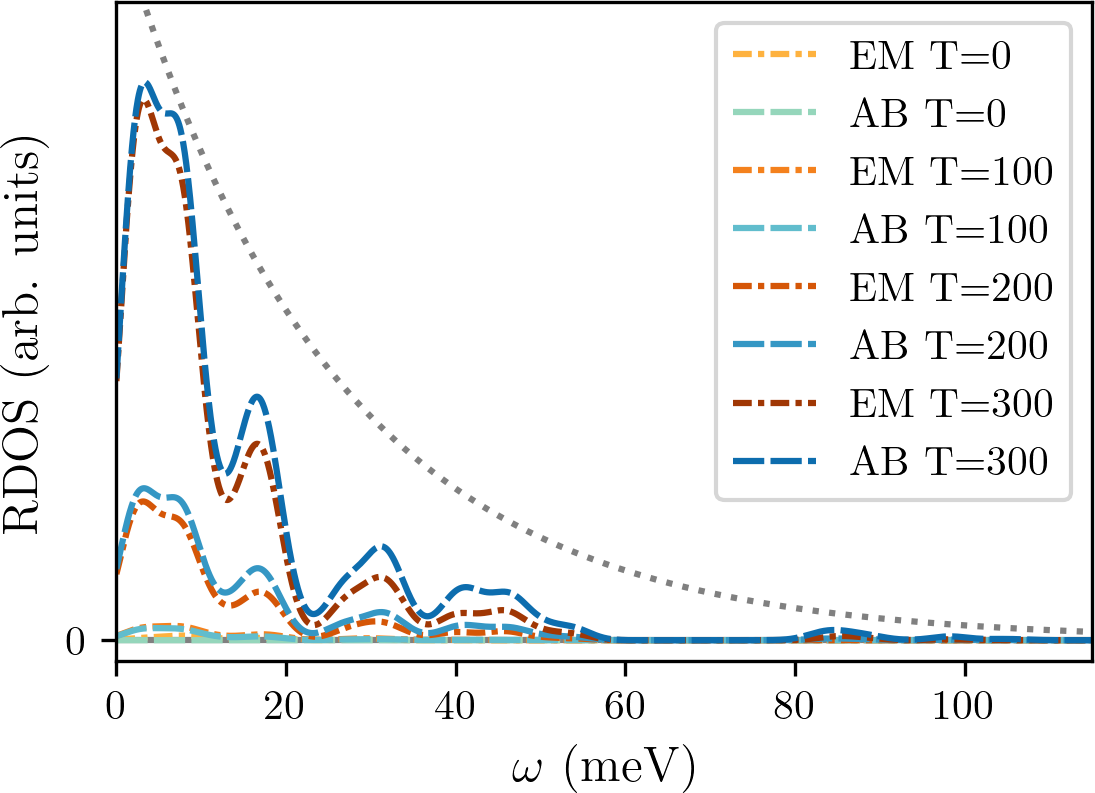}
            \put(-1,70){\small \textbf{(a)}}
        \end{overpic}
        \label{fig:rdos_1}
    \end{subfigure}
    ~
    \begin{subfigure}[]{0.48\linewidth}
        \begin{overpic}[width=\linewidth]{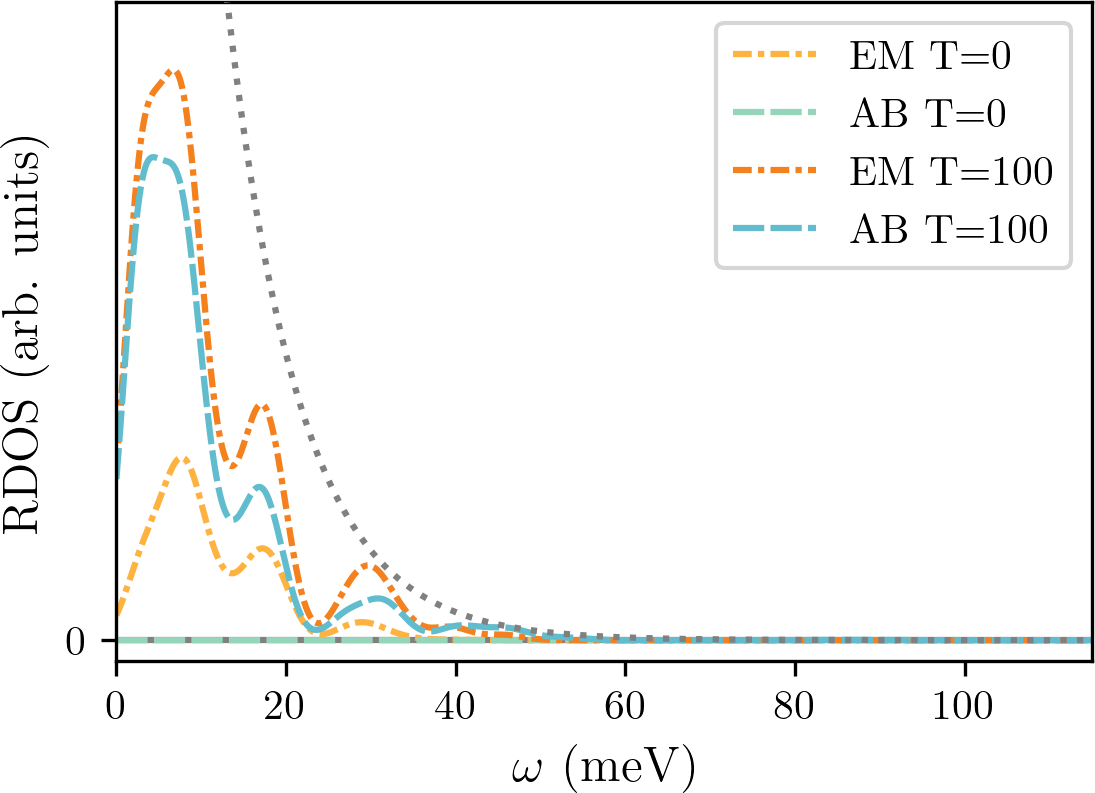}
            \put(-1,70){\small \textbf{(b)}}
        \end{overpic}
        \label{fig:rdos_2}
    \end{subfigure}
    \\[-0.7cm]
    \caption{Phonon restricted density of states for the emission and absorption channels as a function of temperature. These data are equivalent to $ D_{\text{Re}\{K^{\rm{ph}}_{S,S'}\}}(\omega,T)$ if the matrix element for each allowed transition is set to $1$. A scaled Bose-Einstein occupation envelope is also overlaid for $300$ and $100$ K in plots (a) and (b) respectively.}
    \label{fig:rdos}
\end{figure}
\begin{figure}[htbp!]
    \centering
    \begin{subfigure}[]{0.48\linewidth}
        \begin{overpic}[width=\linewidth]{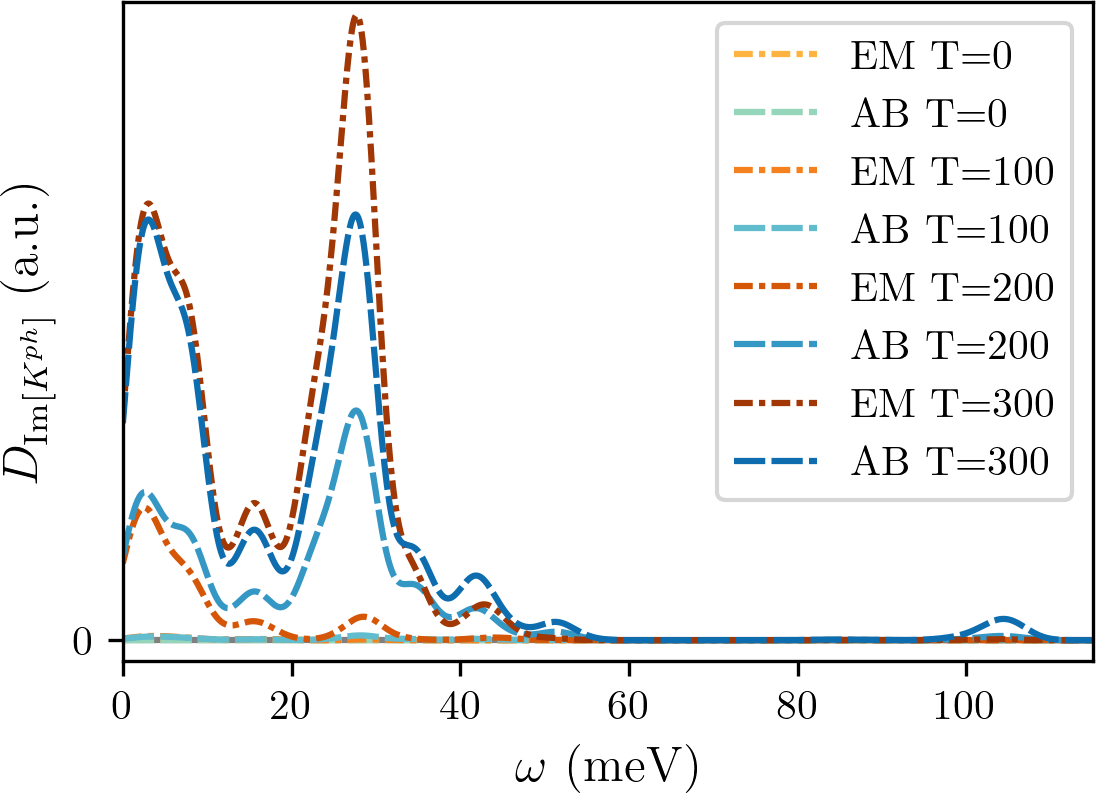}
            \put(-1,70){\small \textbf{(a)}}
        \end{overpic}
        \label{fig:mrdos_1}
    \end{subfigure}
    ~
    \begin{subfigure}[]{0.48\linewidth}
        \begin{overpic}[width=\linewidth]{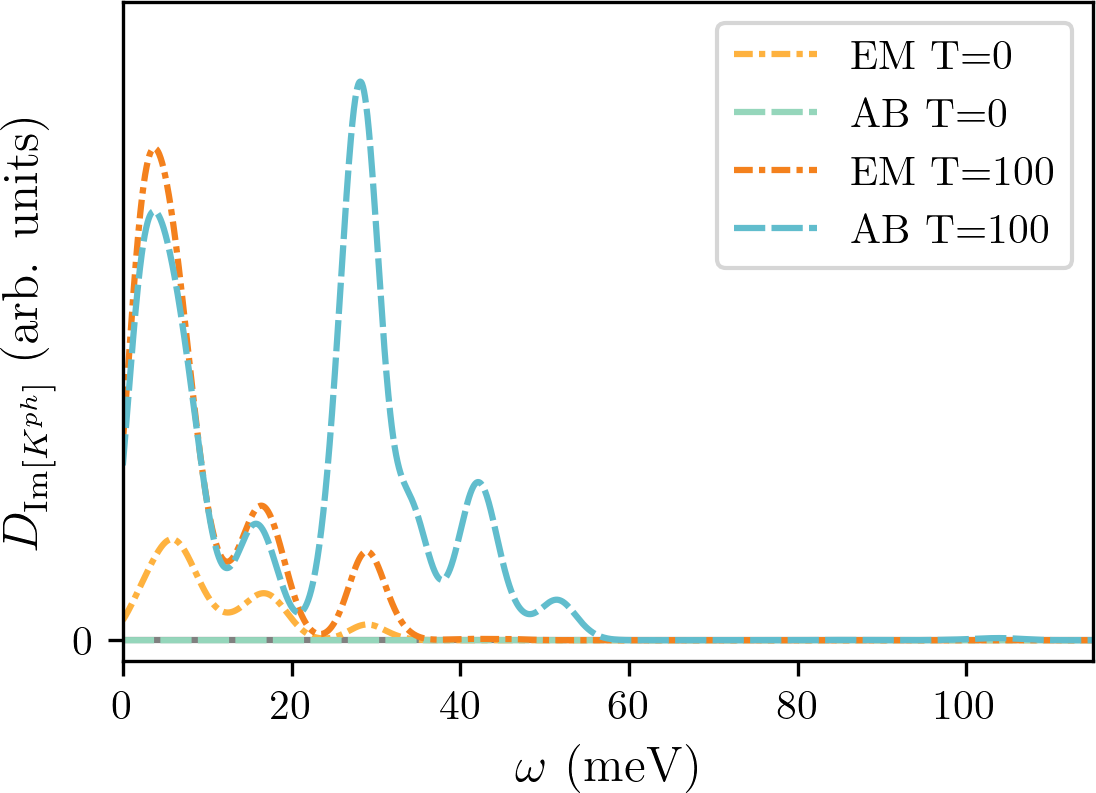}
            \put(-1,70){\small \textbf{(b)}}
        \end{overpic}
        \label{fig:mrdos_2}
    \end{subfigure}
    \\[-0.7cm]
    \caption{Phonon frequency resolved contributions to the imaginary part of $K^{\rm{ph}}$ for the first exciton. Lower energy modes as opposed to the LO mode play a dominant role here. Compared to the RDOS, the contributions from modes around $20$ to $40$ meV are also enhanced.}
    \label{fig:dos_kph_im}
\end{figure}

Similar to the real part, the spectral density of $\text{Im}\{K^{\rm{ph}}\}$ is calculated according to
\begin{equation}
\label{eq:D_Im_Kph_T_pm}
    D^{\pm}_{\text{Im}\{K^{\rm{ph}}_{S,S'}\}}(\omega,\Omega,T)=
    \sum_{\rm{\nu}}
    \text{Im}\left\{\sum_{\rm{ cv}\rm\mathbf{{k}}\rm{c'v'}\rm\mathbf{{k}}'}
    \left(A^S_{\rm{cv}\rm\mathbf{{k}}}\right)^*A^{S'}_{\rm{c'v'}\rm\mathbf{{k}}'}K^{\rm{ph},\pm}_{\rm{\nu cv}\rm\mathbf{{k}}\rm{c'v'}\rm\mathbf{{k}}'}(\Omega,T)
    \delta(\omega-\omega_{\rm\mathbf{{k}}-\rm\mathbf{{k}}',\nu})
    \right\}
\end{equation}
and as was the case for the real part, adding both channels' spectral densities and integrating over $\omega$ recovers the full imaginary part $\text{Im}\{K^{\rm{ph}}_{S,S'}\}(\Omega,T)$.

When considering the spectral density of the imaginary part of $K^{\rm{ph}}$ it is helpful to first consider the restricted density of states RDOS associated with said spectral density. For the EM and AB channels, the RDOS is calculated according to
\begin{equation}
\begin{aligned}
\label{eq:rdos_pm}
    RDOS^{\pm}(\omega,\Omega,T)=
    \sum_{\rm{\nu vc}\rm\mathbf{{k}}\rm{v'c'}\rm\mathbf{{k}}'}&
    \left|\pm\nicefrac{1}{2}-\nicefrac{1}{2}-n_B(T,\omega_{\rm\mathbf{{k}}-\rm\mathbf{{k}}',\nu})\right|\delta(\omega-\omega_{\rm\mathbf{{k}}-\rm\mathbf{{k}}',\nu})\times\\
    &\Big[\delta(\Omega-\Delta_{c\rm\mathbf{{k}}v'\rm\mathbf{{k}}'}\pm\omega_{\rm\mathbf{{k}}-\rm\mathbf{{k}}',\nu})+\delta(\Omega-\Delta_{c'\rm\mathbf{{k}}'v\rm\mathbf{{k}}}\pm\omega_{\rm\mathbf{{k}}-\rm\mathbf{{k}}',\nu})\Big].
\end{aligned}
\end{equation}
In short, the $+$ and $-$ RDOS functions are equivalent to $D^{\pm}_{\text{Im}\{K^{\rm{ph}}_{S,S'}\}}(\omega,T)$ if the matrix element contributions coming from $A^S_{\rm{cv}\rm\mathbf{{k}}}$ and $g_{\rm{n\mathbf{k}m\mathbf{k}'\nu}}$ to the expression in Eq.~\ref{eq:D_Im_Kph_T_pm} are all set to $1$. As such, it elucidates the possibility for dissociation to occur after thermal occupation as well as energy and momentum conservation are taken into account.

Fig.~\ref{fig:rdos} shows the RDOS for the first exciton as a function of temperature; it demonstrates a few notable trends. First, as expected, the absorption RDOS is zero at zero temperature due to the Bose-Einstein occupation factor. Second, because Eq.~$3$ from the main text is solved self-consistently, the allowed phonon frequencies grow as a function of temperature as the exciton binding energy is reduced; this follows the expected trend discussed in \ref{sec:diss_cond_sm}. Third, Due to both thermal occupation as well as the higher number of allowed transitions for lower energy phonons, the RDOS is heavily weighted towards lower energy phonons even when higher energy transitions become allowed.

Returning to the full imaginary spectral density, Fig.~\ref{fig:dos_kph_im} shows the phonon frequency resolved spectral density of $\text{Im}\{K^{\rm{ph}}\}$ for the first exciton. Compared to the RDOS plots, there is an enhanced contribution of the phonons around $20$ to $40$ meV, and although it's overall size is small, the contribution from the highest LO mode via the absorption channel is also greatly enhanced. These data suggest that especially when examining lifetimes and/or linewidths, modeling exciton dissociation via just the most strongly coupled LO modes will miss almost all of the contributions to such quantities as the lower frequency modes dominate the imaginary part of $K^{\rm{ph}}$. We also note that while at zero temperature, only the emission channel contributes, and at room temperature, both channels contribute to the imaginary part of $K^{\rm{ph}}$ almost equally.

All of the plots in this section use a Gaussian function with a smearing of $2$ meV to represent all Dirac delta functions. A $36\times36\times48$-density $\rm\mathbf{{q}}$-grid (with a $13\times13\times17$ mini-grid in the patch) was used to calculate $K^{\rm{ph}}$.

\section{Absorption Spectra Renormalization}
\label{sec:abs_spec_sm}
\subsection{Dipole Matrix Element Renormalization}
\begin{figure}[htbp!]
    \centering
    \includegraphics[width=0.47\linewidth]{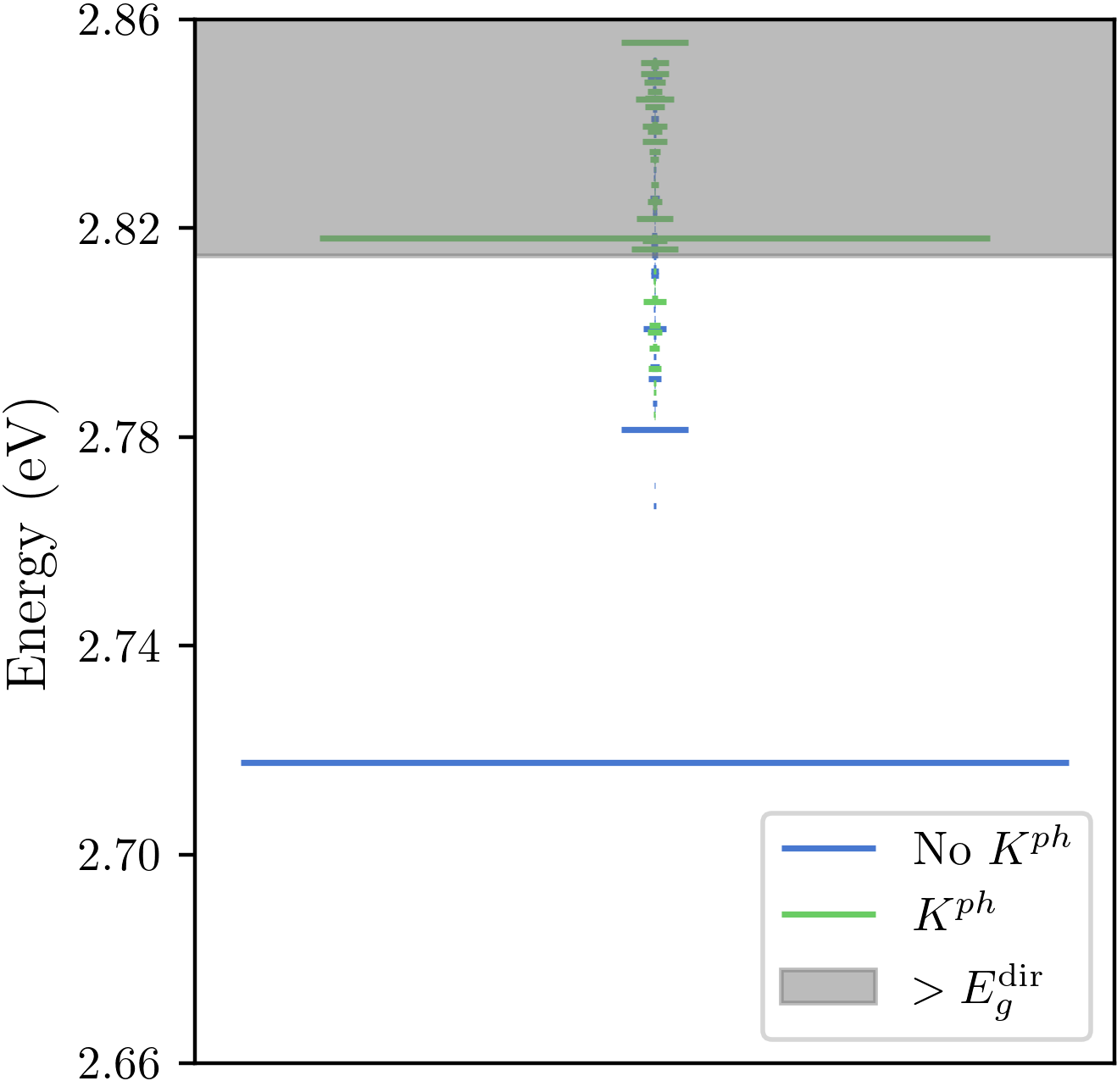}
    \caption{Stick plot of how the directionally averaged dipole matrix elements squared are renormalized by diagonalizing Eq.~B1 for the first $70$ $N_S$ states. The relative widths of each line are all scaled to the strongest element, the lowest-lying exciton in the clamped-ion limit.}
    \label{fig:osc_str_renorm}
\end{figure}

The real parts of the renormalized directionally averaged dipole matrix elements $D^i_{\tilde{S}}$ are compared to those obtained in the clamped-ion limit in Fig.~\ref{fig:osc_str_renorm} (for the first 70 states in each case).
Though there is some degree of re-ordering, there is still a dominantly strong oscillator in both limits.
After state mixing induced by $K^{\rm ph}$, this dominantly strong state still have $\mathord{\sim}80\%$ overlap with the camped-ion exciton wavefunction.
The same scissor shift discussed in Appendix D is also applied here.

\subsection{Wavefunction Renormalization}
\begin{figure}[htbp!]
    \centering
    \includegraphics[width=0.56\linewidth]{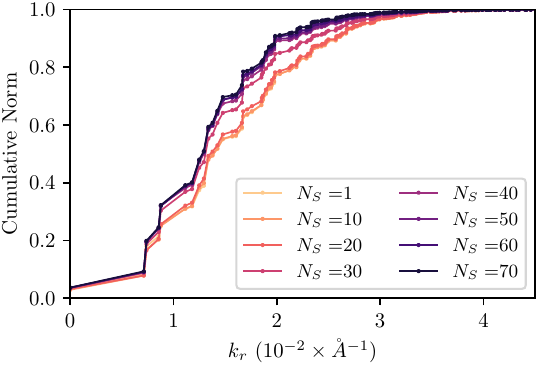}
    \caption{Cumulative norm of the first exciton wave-function versus distance $k_r$ in reciprocal space, computed while $K^{\rm ph}_{SS'}$ mixes the lowest $N_S$ states (off-diagonal coupling).
    The curve converges to unity slightly more rapidly for larger $N_S$, signifying modestly more ${\rm\mathbf{k}}$-space localization and increased real-space delocalization.
    Each exciton is centered at ${\rm\mathbf{k}}_0$.}
    \label{fig:wfn_loc}
\end{figure}
We also assess the extent to which wavefunction mixing affects the localization of the wavefunction of the lowest exciton. Because Eq.~B1 is a non-hermitian eigen-problem, the left and right eigenvectors can be expressed in the electron-hole basis as
\begin{align}
    \tilde{A}_{\rm cv\mathbf{k}}^{\tilde{S}}
    &=\sum_SA^{S*}_{\rm cv\mathbf{k}}\tilde{A}_S^{\tilde{S}}\\
    \tilde{B}^{\tilde{S}}_{\rm cv\mathbf{k}}
    &=\sum_S \tilde{B}^{\tilde{S}}_{S'}A^{S}_{\rm cv\mathbf{k}}.
\end{align}
In this way the norm of this state is expressed as
\begin{equation}
    w^{\tilde{S}}_{\rm cv\mathbf{k}}=
    \tilde{B}^{\tilde{S}}_{\rm cv\mathbf{k}}
    \tilde{A}_{\rm cv\mathbf{k}}^{\tilde{S}},
\end{equation}
and it satisfies the normality constraint,
\begin{equation}
    \sum_{\rm cv\mathbf{k}} w^{\tilde{S}}_{\rm cv\mathbf{k}}=1,
\end{equation}
even though $w^{\tilde{S}}_{\rm cv\mathbf{k}}$ is in general complex.
In other words, whatever remaining imaginary parts there are in $w^{\tilde{S}}_{\rm cv\mathbf{k}}$ will exactly cancel in the sum, while the real parts will add up to 1.

Thus, to assess the localization of the $K^{\rm ph}$ modified exciton wavefunction for the lowest-lying state (centered at $\rm{\mathbf{k}_0}$), we plot in Fig.~\ref{fig:wfn_loc} the cumulative normalization of the real part of $w^{\tilde{S}}_{\rm cv\mathbf{k}}$
\begin{equation}
    C(k_r)=
    \frac{\sum_{|{\rm\mathbf{k}}-{\rm\mathbf{k}}_0|\leq k_r}\sum_{\rm cv}
    \text{Re}\{w^{\tilde{S}}_{\rm cv\mathbf{k}}\}}
    {\sum_{\rm cv\mathbf{k}}w^{\tilde{S}}_{\rm cv\mathbf{k}}}.
\end{equation}
As higher-lying states that are more localized in ${\rm \mathbf{k}}$-space mix into the state, it can be seen becoming more localized in reciprocal space and subsequently more delocalized in real-space.
By $N_S=50$, the new exciton wavefunction appears well-converged, and is only modestly more localized than in the clamped-ion limit.

\end{document}